# The Economic & Sustainability Future of Cellular Networks

Year 3 – Individual Project

**University ID: 1016557,**
**[Pick the date]**

# Acknowledgements

This project would not have been possible without the support of Dr Weisi Guo, my project supervisor. The constructive feedback and valuable guidance provided by Dr Guo throughout the project was extremely helpful and encouraging.

# 1.0 Authors assessment of the project

The Engineering contribution of this project is the analysis of economic and environmental stability of 4th generation wireless communication LTE in the face of increased consumer demand, rising energy consumption and falling profitability. The project gauges an understanding of the potential impact of 4G in the UK.

The contribution is both important and relevant as major operators have undergone or undergoing national system upgrades from 3rd generation to 4th generation networks. This is an expensive process and the cost projections made can help network operators devise the most suitable tariffs to ensure profitability can be achieved and maintained.

Network operators can use this research to gain a better understanding of the tariff strategy required to be both profitable and environmentally friendly. While researchers can benefit from extending the analysis further, using the project results to determine the impact of specific wireless technologies on economic and environmental sustainability. There are very few integrated studies that consider wireless performance, power consumption and profitability of wireless networks. This holistic analysis is therefore beneficial to both industry and academic research.

An area of weakness in the project is profit calculation, which is a pre-retail-expenditure value calculated per $km^2$ rather than an absolute value. This was due to the information related to network operators and industry being commercial sensitivity and therefore limited. As a result the project focused on developed city areas, which are the most densely populated areas and therefore responsible for most of the wireless data demand.

## 2.0 Summary

*Global data traffic is expected to grow exponentially in the next few years with video and smartphone applications driving data growth. Many mobile network providers in the UK have either deployed or planning to deploy 4$^{th}$ generation Long-Term-Evolution (LTE) mobile technology as the solution to meet capacity demands. This study evaluates the technological improvements in 4G LTE in comparison to 3G High Speed Packet Access (HSPA) and further conducts a techno-economic analysis using primary researched tariff data to determine network operator profitability and mobile tariff strategy to meet user demand. To ensure holistic analysis, the study also considers the environmental impacts of LTE by determining the annual carbon emission for a network operator. The study results shows LTE will prove profitable; however a trade-off has to be made by network operators between meeting consumer tariff demands or increasing profitability. Analysis also shows a 63% reduced in carbon emissions is possible with migration to 4G services with implication of further financial benefits for network operators as a result.*

## Table of Contents

## 3.0 Introduction

There are many contributing factors to why the market needs to move towards 4G; however the key driver is to meet the demand of 'data hungry' users. Current 3G HSPA network offer theoretical throughput speeds of 14.4 Mbps; however 4G LTE offers theoretical downlink speeds of 100Mbps. (1) As a result users can support application such as HDTV, skype and video conferencing, all applications that previously required fixed broadband services. User experience is also enhanced in 4G, as consumers can experience low latency, improved system capacity, enhanced coverage, and greater spectral efficiency. The project investigates the technological improvements in 4G and further evaluates the economical and environmental benefits.

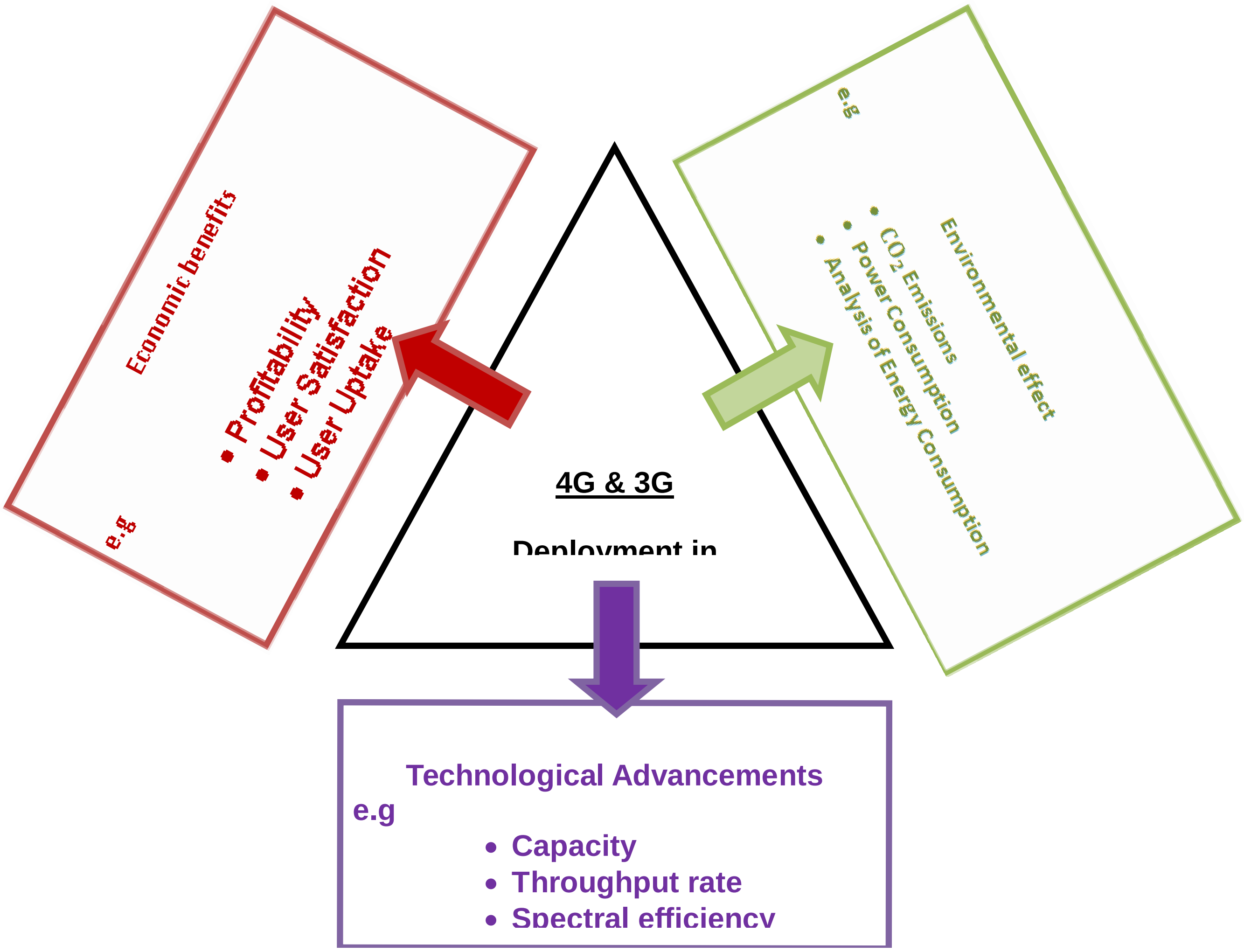

Figure 1 Example of the intended project outcomes

### 3.1 Project Aim

The aim of the project is to gain an understanding into the mechanisms that link three important wireless metrics together: capacity, energy consumption and profit margins. In doing so, one can begin to consider how to optimize the relationship, subject to consumer demands and other obligations.

### 3.2 Project Objectives

1. Understand the theoretical capacity gains achieved by upgrading the multiple-access system from 3G HSPA to 4G LTE, and verify findings with literature review and theoretical results.
2. Obtain the resulting energy consumption savings from the capacity increase due to 4G upgrade. Convert energy savings to operational expenditure (OPEX) savings.
3. Obtain data on capital expenditure costs (CAPEX) for implementing the 4G upgrade, include spectrum purchase, and engineer training and equipment costs.
4. Examine the 3G and proposed 4G tariffs to see how they can help to increase the profit margins. How do these tariffs reflect the capacity improvements found in Objective 1? How can understanding of the previous objectives, help to formulate better tariffs.

## 4.0 Research Methodology

### 4.1 Research Strategy

The project conducted was purely theoretical consisting of both primary and secondary data research. Primary data research involved collecting mobile tariff data from various mobile operators worldwide. Secondary data research involved analysis of mobile user trends and literature review.

## 4.2 Research Method

Several milestones were set throughout the project, with the ultimate goal to achieve the project aim. The following shows an outline of the project progress through the year.

**Week 6 - 7**[1] Basic theory and inter relationship between capacity, energy and cost were explored. The concept of frequency reuse by network operators and its application in multi cell cellular network was understood. Concepts related to purchasing of spectrum, allocation of bandwidth and frequency reuse to avoid interference were investigated.

**Week 7-10** Literature relating to the capacity gains, power consumptions and methods of calculating OPEX costs were reviewed. 3G & 4G tariff database was created. Matlab code was developed to produce the regression fit for collected tariff data.

**Week 10 - 13** Power consumption model was developed with understanding of capacity gains from 3G to 4G. Model used to calculate 3G & 4G power consumption per cell and per $km^2$.

**Week 14 - 16** CAPEX and OPEX costs associated with 3G & 4G networks researched to produce costing for proposed 4G network per $km^2$. CAPEX transformed to OPEX costs to produce OPEX model and OPEX saving calculated.

**Week 17 - 19** Tariff data regression model and OPEX cost model combined to form profit analysis. Profit for 3G & 4G calculated for different percentage

1 The project commenced in week 6 as a switch was made from the original project allocated with Professor Mias, to this project with Dr Guo.

| | |
|---|---|
| | uptake in each technology. |
| **Week 19 - 20** | Power consumption model used with application of literature theory to produce carbon emission analysis. |
| **Week 20+** | Project findings compiled and project report produced. |

# 5.0 Research Theory

## 5.1 Literature Review

To understand the implication of deploying 4G in the UK it was important to conduct a literature review. The review was conducted to gain an understanding of the project topic and to use pre existing research to either validate results or determine areas of further investigation. The literature reviewed in this section includes subject areas of: environmental implications; technological advancement; and financial analysis of cellular networks.

Mobile network operators have invested heavily in the deployment of 4G, with many expected to launch 4G services in the UK by mid 2013 (2). The heavy investment in 4G networks is a result of global network leaders such as Cisco predicting exponential growth in mobile data traffic. Research by Cisco (3), concludes the average smartphone user will generate 2.7GB of traffic per month in 2017, an 8 fold increase from the 2012 average of 342MB. The predicted growth in smartphone data traffic is driven by an increase in usage of mobile applications, where video/communication applications account for 45% of all smartphone data consumption. To support data intensive applications such as video calling, live streaming and HDTV streaming large data throughput is required. Research comparing HSPA+ and LTE (1) measured a 56% throughput gain in the downlink for LTE using 5MHz bandwidth, while throughput performance using 10MHz bandwidth showed a 42% gain in favour of LTE. The research incorporated data throughput measurements in good, medium and bad radio conditions and concluded that LTE spectral efficiency is greater than HSPA+ over a wide variety of radio conditions. Paper (4) and source (5) outlined a theoretical approach using stochastic geometry, to calculate cell throughput. The sources demonstrated how network capacity could be modelled as a function of fading and user distance from cell. The technique provided a method to determine the spectral efficiency of LTE and HSPA with consideration to SINR inefficiencies losses. The spectral efficiency of LTE and HSPA could then be used to calculate data throughput per cell. Although the spectral efficiency gains measured in (1) differ from the theoretical spectral efficiency gains in (5), the conclusion derived was the same; LTE spectral efficiency is greater than that of HSPA. However the methodology to calculate throughput per cell in (4) (5) assumes only on active user, yet no evidence is presented to suggest that assuming one active user is feasible.

Base-station (BS) power consumption has been of interest in recent years for both industry and academic research. Prior research showed that outdoor base-stations are responsible for 70% of the total cellular network energy consumption (6) and contribute to two thirds of the total $CO_2$ emitted by wireless access networks (7). Furthermore, several UK network operators have pledged to reduce their $CO_2$ emissions; with network operator Vodafone aiming to reduce $CO_2$ emissions by 50% in developed markets by 2020 (8). Research conducted in (9) relates BS power consumption to varying traffic demand. The research defines an LTE BS model where the equipment contributing to power consumption is clear. A data traffic model is then defined considering population density per $km^2$ for low, medium and high data traffic profiles. Results show that macro BS's consume the most power while providing greatest coverage and femto BS's consume least power while providing a small coverage area. The research also shows power consumption from voice and SMS traffic is negligible in comparison to mobile data traffic. The fundamental equation used to calculate power consumption in (9) is also used in papers (10) and (11). The equation shows the power consumption of a base station (cell) is a function of transmitted power, traffic load over capacity and overhead power. In (12) the same fundamental equation is used and a comparison between LTE and HSPA shows that, for a coverage area of $100km^2$ LTE consumes 39.5% less power and requires 81 less base stations compared to HSPA. The fundamental equation used in all four papers is a sound method to calculate power consumption as the same underlining results and conclusions were achieved, showing power consumption increase with traffic load. Although a comparison of power consumption was made in paper (12), there was no comparison made between energy consumption of HSPA and LTE, which is directly related to both power consumption and operational expenditure (10) (11).

The migration from 3G to 4G is primarily due to the increase in throughput and higher spectral efficiency gains (13). However for network operators the migration from 3G to 4G has economic cost implications. Published research (14) (15) (16) have all shown the economic costs associated with various 3G and 4G technologies. Papers (10) (11) define capital expenditure costs (CAPEX) which can then be converted to operational expenditure (OPEX) costs via a loan and used to model the costs associated with BS. All papers have concluded that BS's energy bill is a large contributor to the OPEX costs, with energy bills contributing up to 16% of the total

OPEX costs for a 3G network. Although a comparison between 3G LTE and 3G HSPA (17) revealed 3G LTE can deliver lower data access cost per megabyte, no operational cost comparison between 3G and 4G has been made, an assessment which could reveal if the migration from 3G to 4G provides any operational costs benefits for network operators in the UK.

The importance of reducing carbon emissions has been stressed by network operators (18) and in published research (9) (10) (11) (12). In (9) the equation which focuses on total network energy is shown to be closely related to the $CO_2$ emissions; however a method to calculate $CO_2$ emissions from LTE base stations is not discussed. Research in (11) shows how an estimation of the $CO_2$ produced by macro BS can be achieved using the 'Carbon Emission Ratio factor' however no case study showing results is presented. An estimate of the amount of $CO_2$ produced by 4G and 3G BS's is however presented in report (19) and is the only report to provide detailed findings on the $CO_2$ emission of each user. Although a good indicator of the consequent effect on the environment, the methodology used to obtain $CO_2$ emission per user considers 'reverse auction bidding' to obtain the data traffic demand. This is not a method used in the UK nor in Europe and was not approved by a credible publishing institute and therefore the methodology must be rejected. However the research does indicate the importance of conducting a comparison between 3G and 4G carbon emissions, the result could show the environmental impact in technology migration as well as the increase/decrease in user carbon footprint.

The rise in traffic data demand and required increase in capacity has resulted in network operator investing heavily in 4G technology. As a result, profitability is a key issue for network operators; research by Ofcom (20) has shown mobile operator voice revenues have fallen in the last decade while mobile data revenues have increased 22%. Mobile operators are therefore investing in 4G technologies as method to increase competitiveness via ‘data sales’, while also trying to reduce costs and meet emissions targets (10). Cost and Carbon emissions of 4G base stations can be calculated from literature reviewed earlier, however literature analysing the profitability with migration to 4G services is limited. Paper (21) proposes different pricing models to determine payback time for mobile network operators (MNO) and mobile virtual network operators (MVNO) when investing in 4G LTE. The proposed methodology to calculate both CAPEX and OPEX costs for MNO and MVNO is then used in ‘Real Option Theory’ a method to justify the investment in LTE technology. ‘Real Option Theory’ incorporates financial analysis tools Net Present Value (NPV), present value of future net cash flows (FCF) and weighted average cost of capital (WACC). Although the results conclude that the 4G LTE will produce positive NPV and therefore worthy of investments, the revenue per user and number of user are assumed static at $250 per year and 1,000,000 per year, respectively. These are arbitrary value that do not account for market trends or the revenue generated from network operator offered tariffs and therefore not an accurate estimate of the revenue generated from 4G. Techno-economic analysis in paper (15) shows comparison of UMTS and EDGE technologies and techno-economic analysis in paper (17) makes comparison between 3G LTE and 3G HSPA. In paper (15) and (17) the CAPEX and OPEX costs for deploying each technology across a western company are proposed. The papers also incorporate the user density and market sample in western countries, compared to (21) where it had not. Both papers also consider revenue generated from different data usage profile, accounting for variation in high, medium and low data consumption. However values chosen for average revenue per user in both papers are assumed, which although may be

close, do not provide accurate estimates for the profitability of network operators. Furthermore, the literature reviewed shows no studies have compared the potential profitability of 4G in comparison to 3G with respect to gradual increase in traffic data demand.

### 5.1.1 Literature Review Conclusions

The literature reviewed has shown previous studies that relate to the research topic and highlighted the following areas for further research:

- Throughput gains in 4G and 3G cells with consideration to number of active users present in the cell.
- Power consumption of 3G and 4G base stations with respect to future traffic data demand.
- $CO_2$ emission of 3G and 4G technology with respect to future traffic data demand.
- Mobile network operator profitability considering the migration from 3G to 4G network.

## 5.2 Theoretical Background

This section outlines the theoretical approach used in order to obtain the results and conclusions shown in sections 6.0 and 7.0 respectively.

### 5.2.1 Capacity Gains Migrating from 3G to 4G

Limited Capacity is a concern in all wireless communication networks. With mobile data traffic expected to increase at a compound annual growth rate of 66% percent between 2012 and 2017 (3), LTE is the 4G technologies deployed by network operators in the UK to increase network capacity and meet the expected rise in data traffic. This section explores the theoretical gains expected in the migration from 3G to 4G.

The capacity of cellular system can be defined as the maximum date rate in bits per second that can reliably be transferred from a transmitter to a receiver. In a channel perturbed by white Gaussian noise (AWGN) the *C*, capacity (in bits/s) can be given by the Shannon-Hartely formula (22):

$$C = BWlog_2\left(1 + \frac{S}{N}\right) \qquad \text{[Equation 1]}$$

where *BW, S and N* are the given bandwidth, average received signal power and average received noise power respectively.

Figure 2 shows a simplified model of typical cellular network consisting of several transmission nodes (Base stations/cells). The Signal Interference Noise Ratio (SINR) received by the user from a transmitting cell can be given by:

$$SINR = \frac{S}{W+I} \qquad \text{[Equation 2]}$$

where S, W, I are the desired signal, noise and Interference powers, respectively and the interference powers are a summation over the set of all interfering transmitters (4).

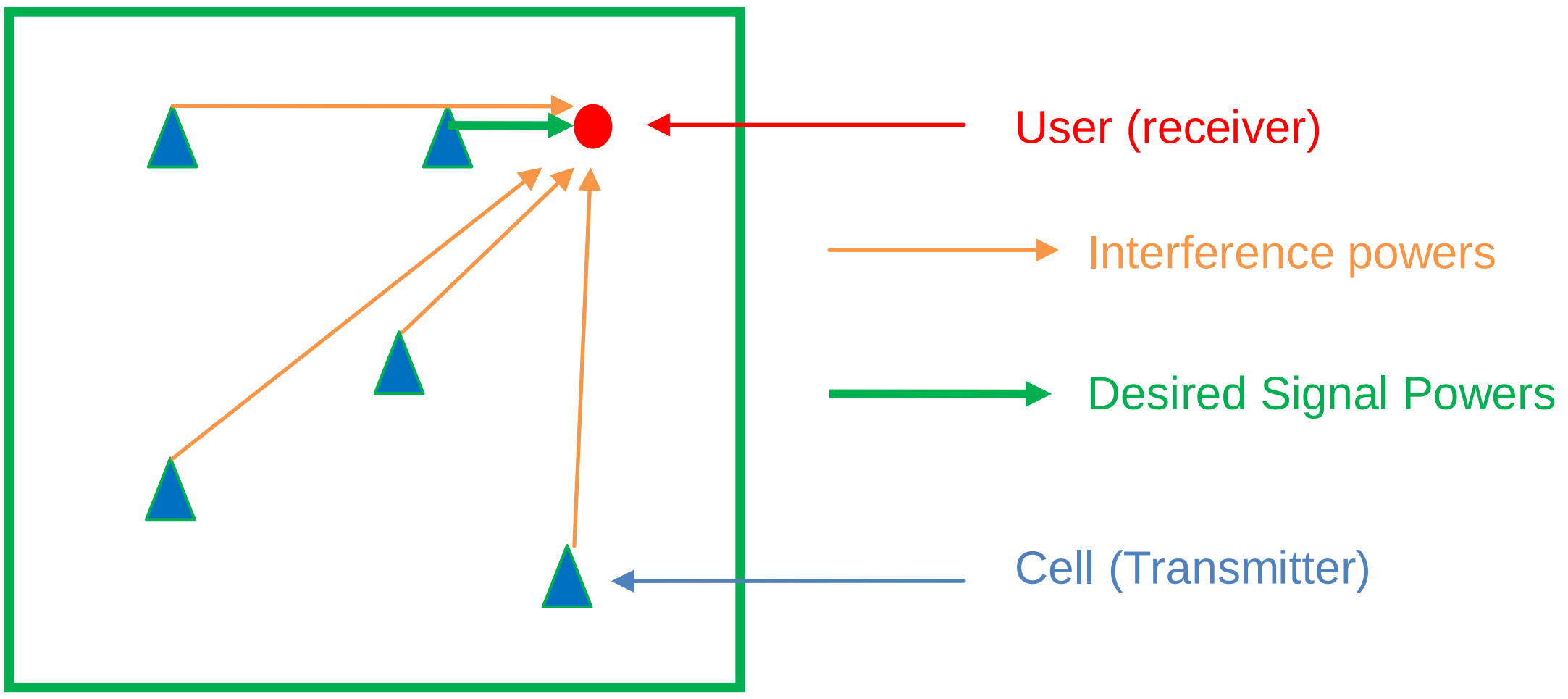

**Figure 2 A simplified model of the SINR experienced by one user in a cellular network (5)**

However in a real system the signal power received by a user from a transmitting cell can be given more accurately by *equation 3* (5) and represents the SINR at a single user location *(x,y)*:

$$\gamma_{x,y,n} = \frac{H_{x,y,k} P_n \lambda\, r_{x,y}^{-\alpha},}{W + \sum_{k=1,k\neq n}^{K} H_{x,y,k} P_k \lambda\, r_{x,y}^{-\alpha},} \qquad [Equation\ 3]$$

Where *H, P, λ,* $r^{-\alpha}$, are the combined Stochastic fading effects (multipath and shadowing), transmitted power, constant and distance loss respectively.

*Equation 3* shows the SINR for a single active user at a location *(x,y)* in a cell *n* and can be used to find the capacity (bits/s/Hz) at that location as shown by *equation 4.*

$$C_{x,y,n} = log_2\left( 1 + \gamma_{x,y,n}\right) \qquad [Equation4]$$

However to calculate the average capacity of the entire cell, it is mainly determined by the SINR distribution over the cell area of ***all*** served users. This can be calculated using *Stochastic Geometry.* Stochastic geometry is beyond the scope of this project; however the key equations and concepts will be outlined to demonstrate how the capacity of cell can be calculated. The resultant capacity findings from applying *stochastic signal processes* (5) can then be used in *section 5.2.4*.

The *stochastic signal processes* are: the spatial variation of location *(x,y)* or equivalently the distance between user and serving cell (*r);* and the temporal variation due to multi-path fading *(H).* Network Capacity is the spatial average of the mean capacity at each location. The formulation can be given by:

$$\bar{C} = \int_0^{+\infty} E[\, log_2\left( 1 + \gamma_{r,n}\right)] f_R(r) dr \qquad [Equation\ 5]$$

where $f_R(r)$ is the spatial Poisson distribution given by:

$$f_R(r) = 2\Lambda\,\mathrm{\pi r}\exp(-\Lambda\,\mathrm{\pi r^2}) \qquad [Equation\ 6]$$

The mean capacity at each location (r), attached to cells *n* is given by:

$$E(C_{r,n} = \exp\left(-\beta r^{\alpha} W\left(2^{\zeta} - 1\right) - \Lambda \pi r^{2} Q(\zeta, \alpha)\right) d\zeta \qquad [\textit{Equation 7}]$$

Where $Q(\zeta, \alpha)$ can be evaluated for $\alpha = 4$, (which is the value for indoor and outdoor urban environments) to give:

$$Q(\zeta, \alpha) = \sqrt{2^{\zeta} - 1}\left[\frac{\pi}{2} - \arctan\left(\frac{1}{\sqrt{2^{\zeta}-1}}\right)\right] \qquad [\textit{Equation 8}]$$

Using the Gauss-Krond quadrature integration technique, the mean capacity integral given by *equation 5* can be solved to give:

$$\bar{C} = \int_0^{+\infty} \frac{1}{1+\sqrt{2^{\zeta}-1}\left[\frac{\pi}{2}-\arctan\left(\frac{1}{\sqrt{2^{\zeta}-1}}\right)\right]} d\zeta = 2.14 \text{Bits/s/Hz} \qquad [\textit{Equation 9}]$$

As the expression above shows, under the conditions proposed in (5), the *spectral efficiency* for a multi-cell network is equal to 2.14 bits/s/Hz. However a real network suffers from inefficiency losses due to coding, modulation, antenna gains and interference from other radio access technologies. As a result the expression above can be modified to incorporate the SINR inefficiency factor μ, where for 3G μ = 8 to 10 and for 4G μ = 1. Using the following expression:

$$\bar{C} = \int_0^{+\infty} \frac{1}{1+\sqrt{\mu 2^{\zeta}-1}\left[\frac{\pi}{2}-\arctan\left(\frac{1}{\sqrt{\mu 2^{\zeta}-1}}\right)\right]} d\zeta \qquad [\textit{Equation10}]$$

Using the μ values stated for 3G and 4G, the following capacities can be found:

3G – **0.82** bits/s/Hz

4G – **2.14** bits/s/Hz

This shows a 2.6 fold increase in the spectral efficiency from 3G to 4G. The maximum downlink bandwidth (BW) available per user in a 3G and 4G cell is given as 5MHz and 20MHz respectively. The throughput achieved per cell can therefore be found using *Equation 13* and is equal to 4Mbps/cell and 43Mbps/cell for 3G and 4G respectively showing 10 fold increase in throughput.

### 5.2.2 Bandwidth Gain

| **Mean Number of Users in Cell** | **BW LTE per user (MHz)** min (20, $\frac{BW}{\#Users}$) | **BW HSPA per user (MHz)** min (5, $\frac{BW}{\#Users}$) | **BW Gain** | **Probability of BW Gain** |
|---|---|---|---|---|
| 1 | 20 | 5 | 4 | 0.07313 |
| 2 | 10 | 5 | 2 | 0.14634 |
| 3 | $6.\dot{6}$ | 5 | $1.\dot{3}$ | 0.19524 |
| 4 | 5 | 5 | 1 | ***0.19537*** |
| 5 | 4 | 4 | 1 | 0.15637 |
| 6 | $3.\dot{3}$ | $3.\dot{3}$ | 1 | 0.10432 |
| 7 | 2.86 | 2.86 | 1 | 0.05964 |

**Table 1 BW Gain when using LTE technology in comparison to HSPA technology**

In both LTE and HSPA technologies 20MHz of BW is available for use, however for HSPA technology only a maximum of 5MHz can be utilised by a single user per cell, whereas a single user in an LTE cell can utilise a maximum of 20MHz of BW. The table above quantifies the BW designated per user, given a certain number of active users in the cell. For LTE the minimum value between 20MHz and $\frac{BW}{\#Users}$ is designated as the 20MHz of spectrum available is allocated equally amongst users. Similarly for HSPA technology the minimum value between 5MHz and $\frac{BW}{\#Users}$ is designated as the 20MHz of spectrum available is allocated equally amongst users.

$$\text{BW GAIN} = \frac{\text{BW LTE per user (MHz)}}{\text{BW HSDPA per user (MHz)}} \qquad \textit{[Equation 11]}$$

The BW Gain column in *Table 1* is calculated using *Equation 11* and indicates the extent BW has been utilised in LTE compared to HSPA. For example when only one user is present in the cell the BW gain is equal to 4, this suggests that the full BW available in LTE has been utilised and also fully utilised in HSPA. In other words, when only one user is active in the cell a 4 fold increase in BW gain can be expected.

$\mathbf{\lambda_N} = 0.0031e^{1.085\log(\overline{\mathbf{R}}_\mathbf{c})}$; $\mathbf{N_u} \sim$ Poisson $(\mathbf{\lambda_N})$ *[Equation 12]*

$\overline{\mathbf{R}}_\mathbf{c} = \mathrm{BW} \cdot S_{eff}$ *[Equation 13]*

In *Table 1* it can be seen that the BW is only utilised effectively for 4G compared to 3G when there are less than 4 users in the network cell (i.e. BW Gain is greater than one). Although this may seem a small number of users, it is entirely dependent on the probability of those users being *active* in a cell. In order to calculate the probability of active users in a cell *Equation 12* was used as suggested in (23), where $\mathbf{N_u}$ is the number of active users and follows a Poisson distribution denoted by parameter $\mathbf{\lambda_N}$ and $\overline{\mathbf{R}}_\mathbf{c}$ denotes the mean desired throughput per cell in bit/s. The estimation for $\mathbf{\lambda_N}$, suggests a strong non-linear exponential dependence on $\overline{\mathbf{R}}_\mathbf{c}$, where $\overline{\mathbf{R}}_\mathbf{c}$ is a function of the *BW* and spectral efficiency $S_{eff}$. The $S_{eff}$ for 3G is calculated to be 0.82 bits/s/Hz as discussed in *Section 5.2.1*.

Using *Equation 13*, the $\overline{\mathbf{R}}_\mathbf{c}$ value for 3G can be calculated as 4 Mbits/s/cell. One can then employ these values to compute the $\mathbf{\lambda_N}$ estimates for 3G HSPA. Parameter $\mathbf{\lambda_N}$ can then used to produce a Poisson distribution showing the probability of a user being active in a cell. The resultant probability distribution function for HSPA can be seen in *figure 3*.

In *section 5.2.1* it was calculated that the throughput per cell for 4G and 3G was 43 Mbits/s/cell and 4Mbits/s/cell respectively. However this assumed that the bandwidth gain was equal to 4 and therefore only one active user present in the cell. However from *figure 3* it can be deduced that when $\overline{\mathbf{R}}_\mathbf{c}$ is equal to 4 Mbits/s/cell, the largest probability of active users occurs when approximately 4 users are active in a cell.

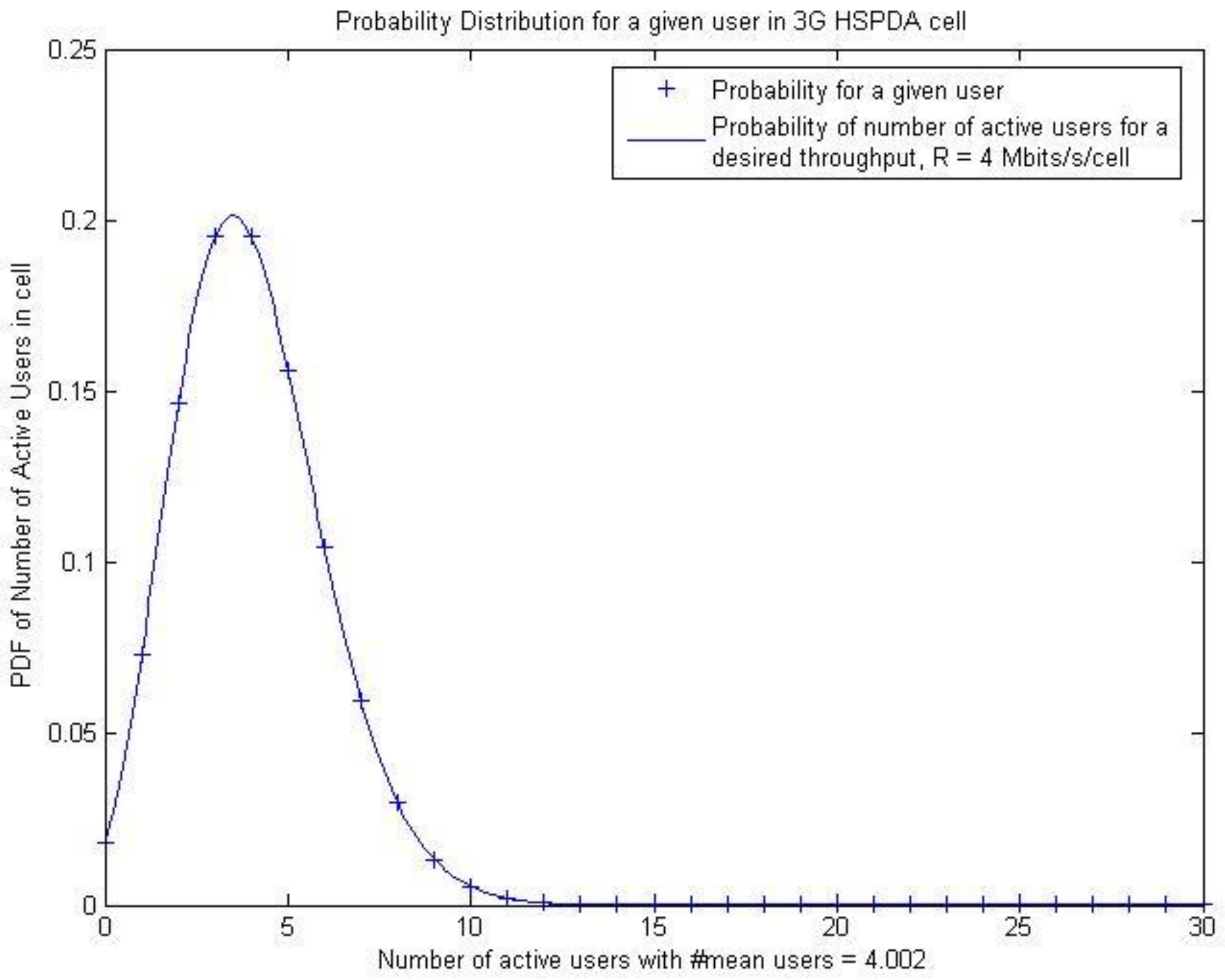

**Figure 3 shows a Poisson probability distribution of the number of active users being present in 3G cell**

The probability distribution shown in *Figure 3* was used to obtain the probability of BW gain. The probability of obtaining a certain BW gain is weighted by the probability of number of active users in a cell (*column 4 & 5, table 1*). From the probabilities obtained for BW gain (*table 1)* it can be observed that the highest probability of occurrence is when BW gain is equal to one, furthermore the probability of BW gain being more than one at any given time in the cell is 0.433 and the probability of BW gain being equal to one is therefore 0.567. In consideration to the BW gain reapplying *equation 13* for when 4 users are present in a cell, the throughput for a 4G cell can be calculated as 10.7 Mbits/s/cell, showing a 2.6 fold increase in throughput. From a probabilistic approach this is throughput more likely to be achieved compared to the suggested 10 fold increase in *section 5.2.1*.

### 5.2.3 Mobile Tariff Analysis

The project aims to examine the 3G and 4G proposed tariffs to see how they can help to increase network operator profit margins. In order to calculate the profit for a network operator, the charge for a consumer to use mobile services provided by network operators has to be calculated. The charge for using mobile services for consumers is dependent on their usage. The desired usage for a consumer can be matched to a mobile tariff, where the mobile tariff has a monthly charge for the consumer for using voice, SMS and internet data services. The Cost of mobile tariffs varies from network operator to network operators, mobile tariff costs also vary depending on the mobile technology used.

Research conducted in (9) showed data services are considered the main driver of power consumption with voice and SMS data contribution considered relatively small in comparison. As suggested in (3) data usage on smart phones is increasing and the project therefore assumes that the 4G tariffs chosen by consumers will be driven by their data usage in the future.[2]

In 2012 only one network operator in the UK had launched a commercial 4G LTE network and hence 4G tariffs on offer to consumers were limited. To obtain a broader insight into the possible 4G tariffs that could be offered by network operators in the UK, tariffs were analysed from different network operators in similar countries to the UK who were currently offering 4G LTE services. The operators chosen for the tariff analysis and their countries of operation can be summarised in *table 2* below.

[2] It should be noted that tariffs are calculated in line with expert estimations that consumers will use more data intensive applications and services such as HDTV, SDTV, mapping services etc and therefore data intensive tariffs for the future will offer more data to consumers than the comparative tariffs of today.

| 3G tariffs analysed | | |
|---|---|---|
| **Network Operator[3]** | **Country** | **Number of tariffs analysed** |
| **T-Mobile** | UK | 16 |
| **Orange** | UK | 10 |
| **Vodafone** | UK | 5 |
| **O2** | UK | 5 |
| **Sprint USA** | USA | 3 |
| **Verizonwireless USA** | USA | 12 |
| **AT&T** | USA | 3 |
| **Vodafone** | Australia | 7 |
| **4G tariffs analysed[4]** | | |
| **EE** | UK | 5 |
| **Verizonwireless** | USA | 11 |
| **Sprint** | USA | 3 |
| **AT&T** | USA | 5 |
| **Optus** | Australia | 6 |
| **Telstra** | Australia | 4 |

**Table 2: The number of 4G & 3G Mobile tariffs analysed from mobile network operators in UK, USA and Australia (tariff details can be found in Appendix A.1 and A.2 for 3G and 4G respectively).**

The collection of tariffs summarised in *table 2* provided a sample large enough to be able to determine the change of cost with respect to tariff minutes and data allowance (3G and 4G tariff details can be found in appendices A.1 & A.2 respectively). In order to ensure a fair comparison across the range of mobile tariffs collected, the criteria for selecting was as follows:

- All tariffs were based on 12 month contracts.
- All Tariffs were based on Sim-Only contracts.
- The data tariffs selected across all countries included Unlimited SMS services.
- The tariff charge for all contracts was converted to GBP currency for fair comparison.
- Australian Contracts which offer minutes in the form of cost (i.e $35 worth of minutes) were represented in the form of pure minutes in the project using a standard conversion.[5]

3 The mobile tariffs obtained were from sources: (55) (56) (57) (58) (59) (60) (61) (62) (63) (64)

4 EE is currently the only UK Company that provides 4G LTE services, while only a limited number of network operators provide 4G LTE services globally, hence the number of tariffs obtained for 4G tariffs was lower in comparison to 3G.

5 The standard charge by Australian operators for a 2 minute national call is approximately $2.36.

- Unlimited minute tariffs were classified as 2000 or more minutes.[6]
- Unlimited data tariffs were classified as 25 GB of data or more. [7]

*Section 5.2.6* will later propose the method for calculating profit per km$^2$. The profit calculation involves a tariff charge $£_{Tariff}$, which is the monthly cost incurred by a subscriber (user) in order to receive the data demanded. *Section 5.2.6* proposes that for given data traffic rate and number of users per Km$^2$, the data consumed by all users over a period of a month can be averaged to produce a monthly tariff which would satisfy the data consumption of all users per km$^2$. To calculate the cost associated with this average monthly tariff, the tariff data collected[8] was used to form a *multiple linear regression model* in Matlab.[9]

```
% Regression Fit
X = [ones(size(Minutes_Data)) log(Minutes_Data) log(Packet_Data) Minutes_Data Packet_Data];
b = regress(Cost_Data,X); % Removes NaN data

% Scatter Plot
scatter3(Minutes_Data,Packet_Data,Cost_Data,'filled'); hold on;
x1fit = min(Minutes_Data):25:max(Minutes_Data);
x2fit = min(Packet_Data):25:max(Packet_Data);
[X1FIT,X2FIT] = meshgrid(x1fit,x2fit);
YFIT = b(1) + b(2)*log(1+X1FIT) + b(3)*log(1+X2FIT) + b(4)*X1FIT + b(5)*X2FIT;
mesh(X1FIT,X2FIT,YFIT)
```

**Code Extract 1 Shows part of the Matlab code used to create the regression model.**

The *regress* function was used in Matlab with regressor variable X equal to *Minutes_Data* and *Packet_Data,* representing the tariff minutes and tariff data respectively. *Code Extract 1* shows how the *regress* function was used in matlab to estimate the regression coefficients *b(1), b(2), b(3), b(4) and b(5).* These are estimated coefficients that show the relationship between the *X variables*

[6] Most network operators offer 'Unlimited' minutes tariff with condition to a *fair usage policy*. This policy caps the minute allowance to 2000 minutes for 'Unlimited' tariffs; therefore 'Unlimited' tariffs were classified as 2000 minutes. (65)
[7] 25 GB was the largest quantifiable data tariff offered by any network operator, beyond this network operators offer truly 'Unlimited' data packages and hence data demand exceeding the 25 GB was considered Unlimited in this project.
[8] Please refer to appendices A.1 & A.2
[9] The programming code used for the regression model can found in Appendices A.3.

*Minutes_Data* and *Packet_data* with the *Y* variable *Cost_Data* which is representative of the tariff cost data collected. *YFIT* shows the estimated *multiple regression equation* used. The equation enabled to populate a mesh grid modelling the relationship between tariff cost, tariff minute and data allowances. *Figure 4* below shows the *regression model* produced in Matlab.

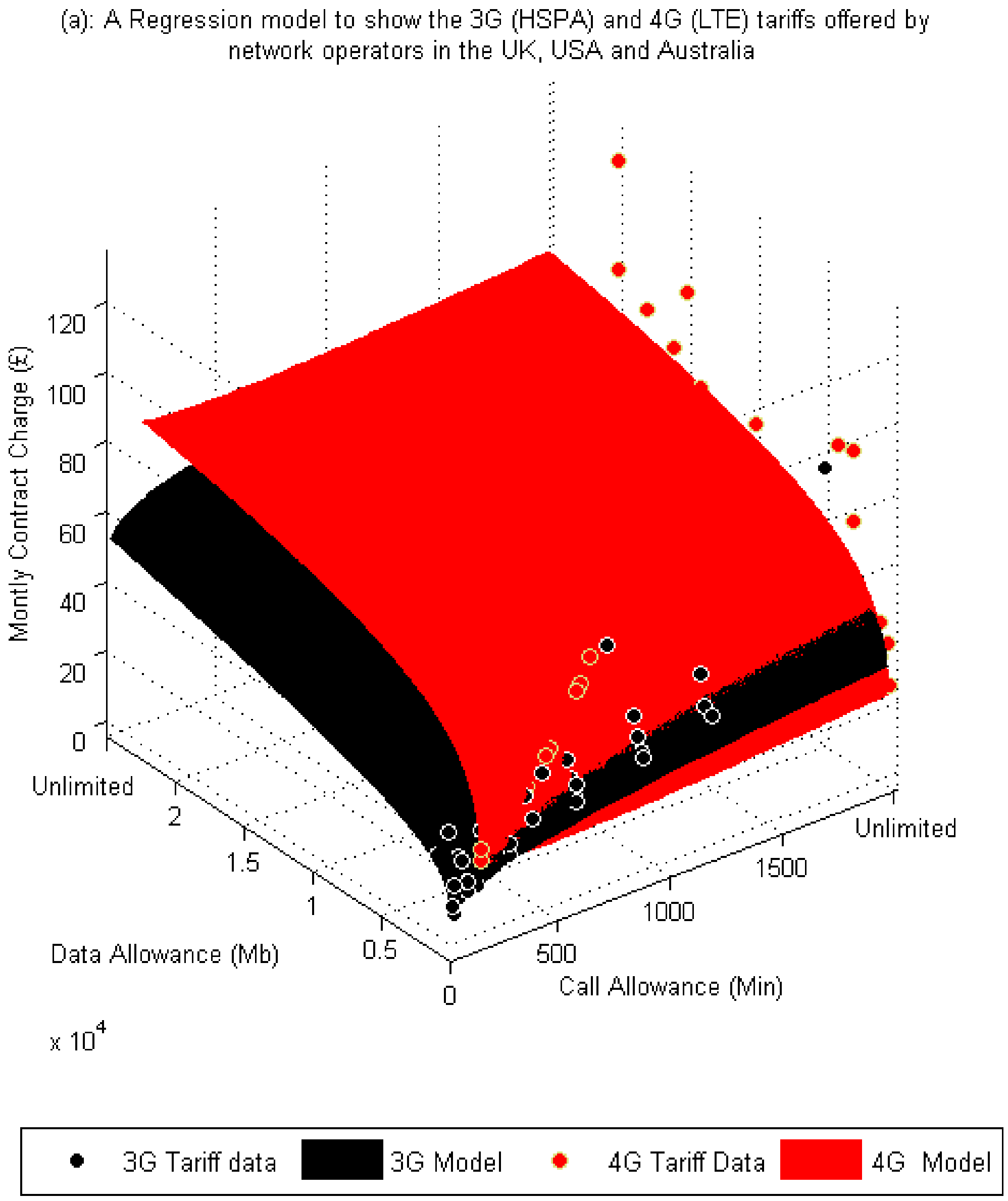

**Figure 4 (a) 4G & 3G regression model**

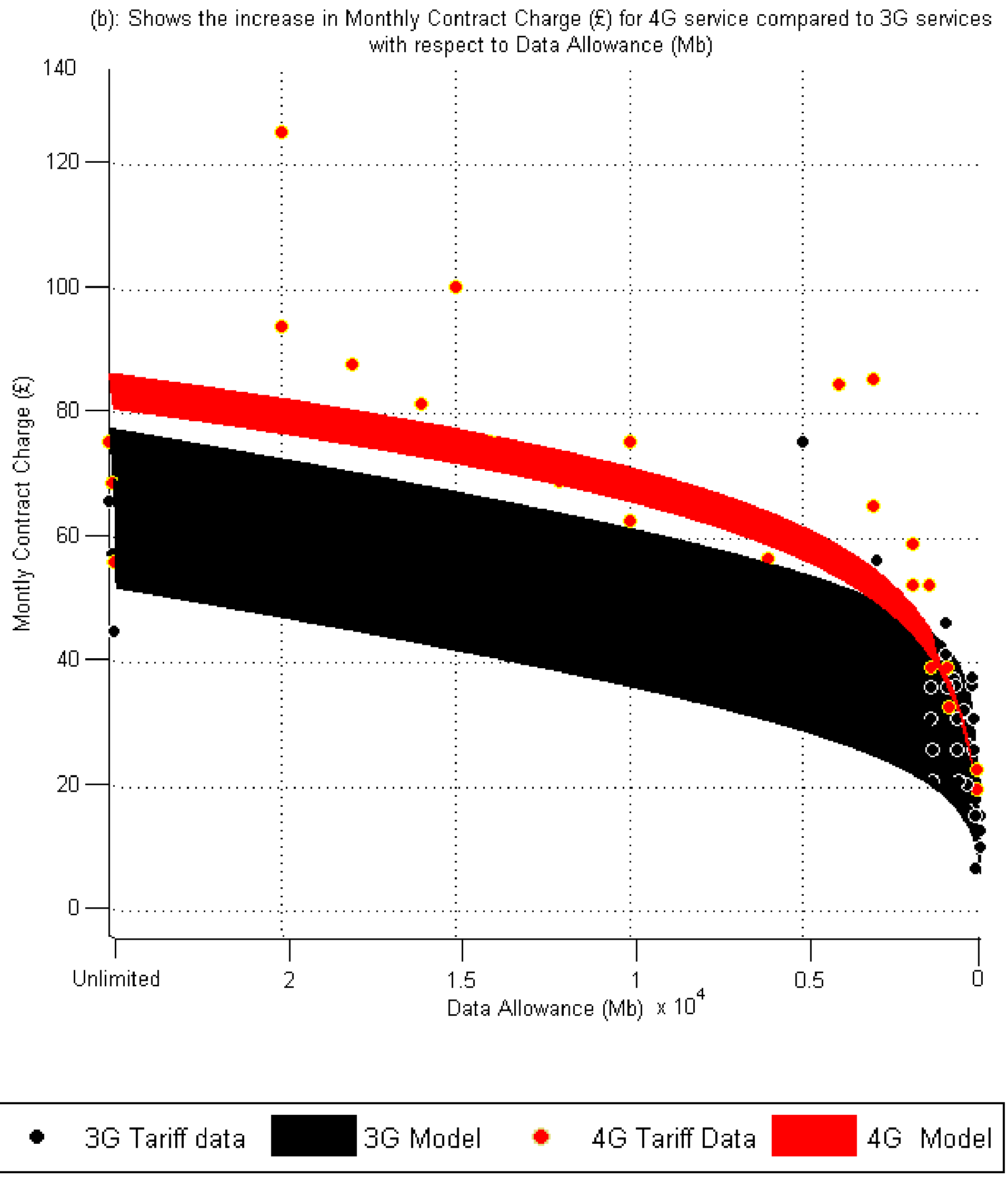

**Figure 4 (b) 4G & 3G regression model showing data allowance against monthly contract charge**

The regression model shown in *figure 4* displays two regression curves, one of 3G tariff data (black) and one of 4G tariff data (red). The characteristic of the regression model show that for small data tariffs (up to 2GB), the cost of 4G tariffs is less than or equal to that of 3G, however for a large data tariff (2GB to Unlimited) the cost for a 4G tariff is significantly higher compared to that of 3G for the same data allowance. For any given data demand by the users, the regression model can be used to show the closest data tariff that meets the data demand and the Monthly contract cost of that tariff. This information can then be used to form the RPS (revenue per subscriber) as proposed in the *Section 5.2.6.*

### 5.2.4 Calculating Power Consumption

The literature review conducted in section 5.1 concluded that outdoor base stations (BS) consume 70% of the total energy consumption of cellular network infrastructure (6). Furthermore report (24) highlights that the power consumption of a macro BS site is dominated by the radio equipment. For this project power consumption was calculated in the interest of network operators (i.e Vodafone). With network operators obligated to provide network services across 99% of UK (25), this coverage requirement (part of the 4G spectrum purchase agreement) means network operators will want to maximise coveragfor this project it was assumed that the network operators will provide mobile services using Macro BS throughout the UK to maximise coverage for 4G LTE and 3G HSPA.

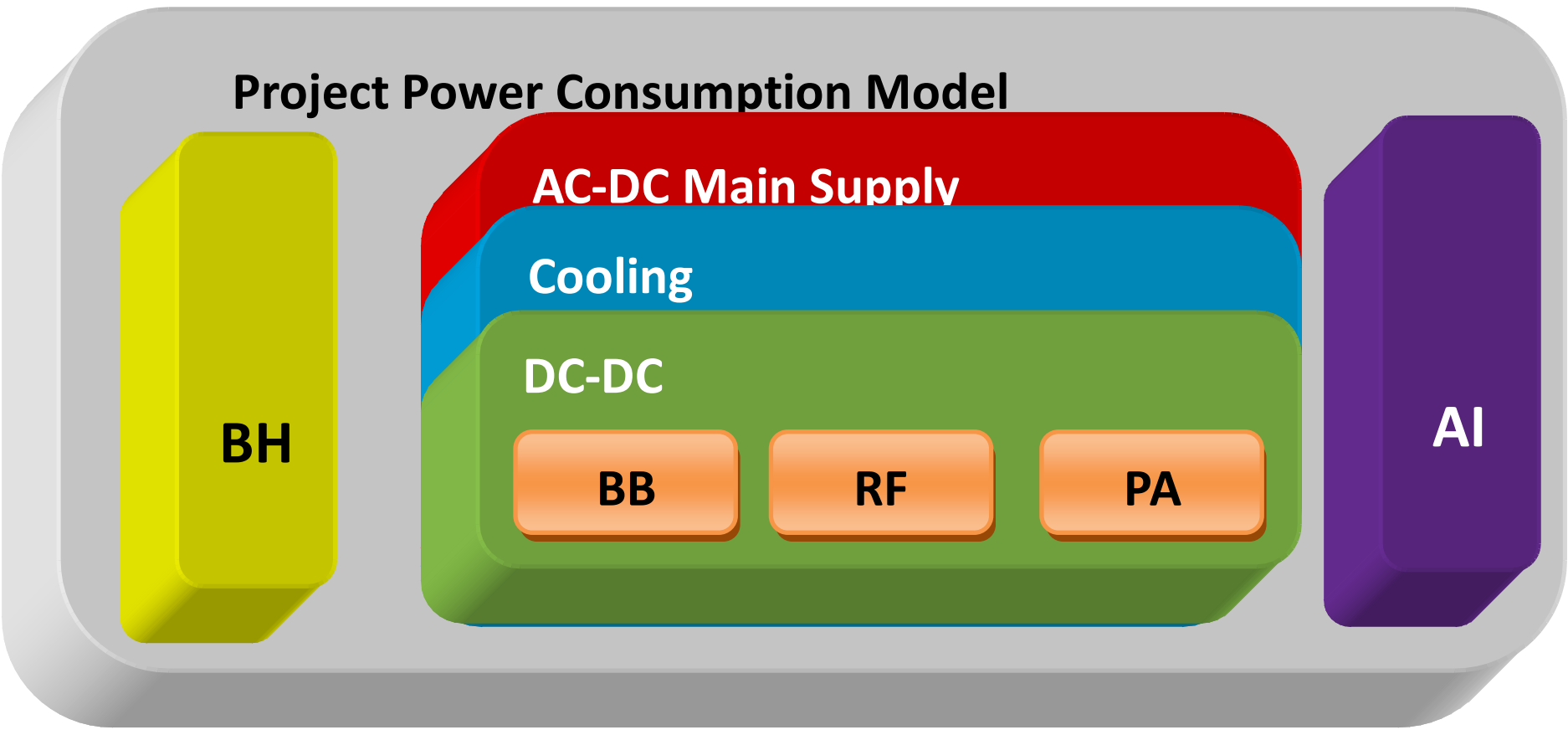

**Figure 5: A simple block diagram showing main components of a BS and connecting Backhaul**

The power consumption model shown in *Figure 6* is similar to that shown in (24) with the addition of the Backhaul (BH). The BH portion compromises of the intermediate link between the core network and base stations. In non technical terms it is the segment of the network that communicates with the global internet. A BS normally consists of multiple transceivers (TRXs) with multiple antennas. The model shows a TRX consisting of a Antenna Interface (AI) Power Amplifier (PA), a Radio frequency (RF- TRX), a Baseband interface (including receiver and transmitter), a DC-DC power supply regulator, cooling system for equipment and AC-DC power supply connecting to power grid. The function of each individual component in the BS can be found in (24)

| Model component | Power consumption per transceiver (W) | Assumptions and Additional Notes |
|---|---|---|
| **Main Supply** | **18.6** | The calculated power consumption after 9% loss due to inefficiencies. |
| **Cooling** | **22.1** | The calculated power consumption after 12% loss due to inefficiencies. |
| **DC-DC** | **13.7** | The calculated power consumption after 8% loss due to inefficiencies. |
| **RF - TRX** | **13.0** | |
| **PA** | **128.2** | The PA efficiency increases with power transmitted. Max transmit power is equal to that of 31.1% of the transceiver power consumption (**39.9 W**). |
| **BB** | **29.5** | Total power consumption including transmitter and receiver section. |
| **BH** | **75** | The value ranges between 50 – 100W per cell, hence the average value has been assumed. |
| | | |
| **Total ($P_{cell}^{OH} + P_{BH}$)** | **300 W** | |

Table 3 The Power consumption associated with different components of the BS

Using the power consumption model shown in *figure 6* and the values obtained from (24) (26), *table 3* can be produced showing the power consumption of individual BS components. The influence of the antenna type on power consumption is considered negligible however the AI does have direct impact on RF and hence accounted for in DC consumption in the PA.

To work out the power consumption of a cell $P_{cell}$ with a number of antennas ($N_a$), (10) (11) the values in *table 1* can be related through the following equation:

$$P_{cell} = N_a \left( \frac{P_T}{\mu_{RH}} \sqrt{L} + P_{cell}^{OH} + P_{BH} \right) \qquad [Equation\ 14]$$

where $P_{cell}$ is the total power consumption of a cell, $P_T$ max transmit power, $\mu_{RH}$ radio head efficiency which increases with traffic load , $L$ is the traffic load $\frac{\bar{R}}{C}$, defined as the ratio between demanded traffic rate $R$ and maximum cell capacity $C_{cell}$, $P_{cell}^{OH}$ is the over head power consumption, this is a constant value and is independent of traffic load, $P_{BH}$ is the power consumption of the Backhaul, $P_{cell}^{OH}$ + $P_{BH}$ constitute as the total value shown in *table 3*.

| Technology | BW | $S_{eff}$ | $C_{cell}$ |
|---|---|---|---|
| Scenario (a): 1 active user in cell | | | |
| **3G** | 20Mhz | 0.82 | 16 |
| **4G** | 20Mhz | 2.14 | 43 |

**Table 4 Table P: Shows the cell capacity for each technology given the number of active users**

*Table 4* shows $C_{cell}$, the maximum achievable capacity per cell in Mbits/s/cell for 4G & 3G technologies and is given by *equation 15,* where *BW* is the bandwidth available for each user.

$$C_{cell} = BW \cdot S_{eff} \qquad [Equation\ 15]$$

In consideration to the project aim, it would be unrealistic to compare 4G and 3G for throughputs of 10.7 and 4 Mbps as the capacity required per km$^2$ would require a large number of cells to deliver the traffic demanded and therefore far exceed the physical deployment capabilities. Therefore in order to obtain an estimate for the power consumption and profitability of 4G and 3G with respect to high traffic data rates (i.e upto 190/km$^2$), the capacity per cell was calculated based on one active user per cell and can be summarised by *Table 4.*

The $P_{cell}$ can also be related to its radius as expressed in (9) however for the project the power consumption of a cell was calculated using *Equation 14.* In order to meet capacity demands network operators deploy more cells in areas of dense populations and fewer cells in areas of lower user densities. The cell area, $A_{cell}$, (*figure 6)* will therefore often vary from location to location based on user density and coverage required. The cell area is based on a homogenous hexagonal deployment with a inter site distance, *ISD*. Different *ISD* were chosen for 3G & 4G depending on the number of cells deployed per km$^2$. The different strategies for 4G and 3G can be summarised by *table 5.* To ensure a fair comparison between 4G and 3G technologies the results obtained in *section 6.1* consider the capacity per km$^2$ to be equal for both technologies. This would then enable a cost comparison between HSPA and LTE in meeting the traffic rate demanded. The deployment strategy required to ensure equal capacity/km$^2$ for 4G & 3G can be seen in *Table 5.*

| ISD (km) | Area (km$^2$) | #Cells /km$^2$ | Deployment Area | User densities/km$^2$ (9) |
|---|---|---|---|---|
| 4G Cell Deployment | | | | |
| **0.50** | 0.216 | 4.62 | Dense-Urban | 3000 |
| 3G Cell Deployment | | | | |
| **0.30** | 0.779 | 12.4 | Dense-Urban | 3000 |

Table 5 Summary of 4G and 3G cell area and related user densities

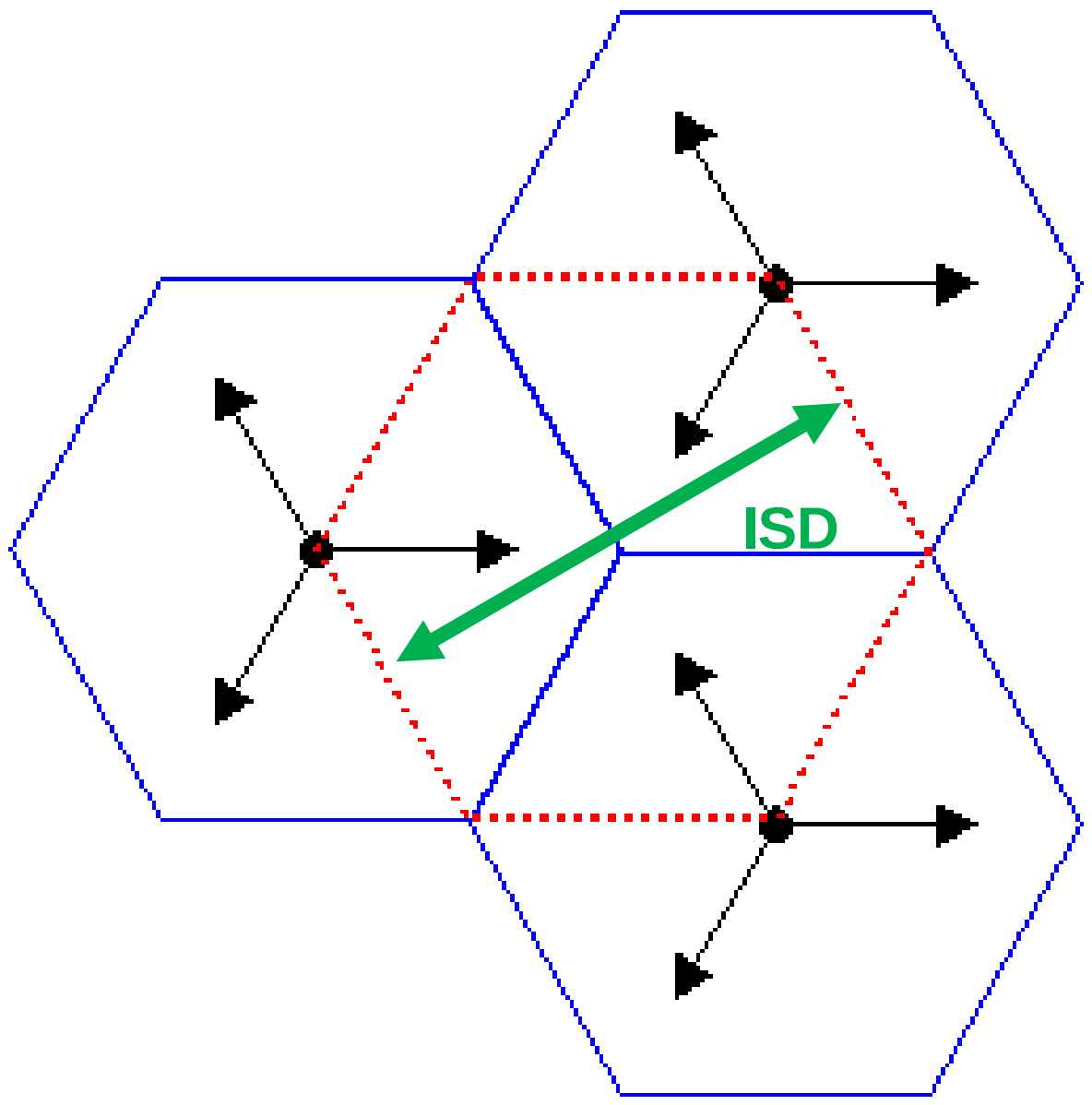

Figure 6 Hexagonal deployment arrangement of cells

To apply the project findings in context to network operators, power consumption must be calculated per km$^2$ and can be achieved by manipulating *equations 14 & 15* to obtain the following:

$$P_{\mathrm{km}^2} = N_a \cdot N_{C/\mathrm{km}^2}\left(\frac{P_T}{\mu_{RH}} \cdot \sqrt{\frac{\mathrm{R}_{\mathrm{km}^2}}{\mathrm{C}_{\mathrm{km}^2}}} + P_{cell}^{OH} + P_{BH}\right) \qquad [\textit{Equation 16}]$$

Where $N_{C/\mathrm{km}^2}$ is the number of cells per $\mathrm{km}^2$, $R_{\mathrm{km}^2}$ is the traffic rate demanded per $\mathrm{km}^2$ and $\mathrm{C}_{\mathrm{km}^2}$ is the Capacity per km$^2$ given by:

$$\mathrm{C}_{\mathrm{km}^2} = N_{\mathrm{c}/\mathrm{km}^2} \cdot \mathrm{C}_{cell} \qquad [\textit{Equation 17}]$$

Using *equation 16* power consumption can be calculated in context to the research project.

### 5.2.5 Calculating Cost

In order to evaluate the profitability of migrating from 3G to 4G for network operators and the consequential costs for subscribers to upgrade to 4G, the cost of deploying a 4G network must first be calculated. Cost of deploying and operating cellular network can be categorised into two sections; CAPEX and OPEX. CAPEX is Capital Expenditure costs, these are usually one-off payments companies must pay in order to acquire or upgrade physical assets. OPEX is Operational Expenditure costs; these are costs that are ongoing for the business. Mobile telecommunications is a very capital intensive industry and so normally CAPEX costs are transformed into OPEX costs via loan and repaid in instalments. The total annual Base station costs per km$^2$, $£_{\mathrm{Total}}$ consisting of $N_{\mathrm{c}/\mathrm{km}^2}$ can be related through *equation 18,* where the CAPEX and OPEX are the respective costs per cell and $i$ is the interest rate over $Y$ years on the loan required to pay for the CAPEX investment.

$$£_{\text{Total}} = N_{\text{c/km}^2} \left( CAPEX \frac{i(1+i)^Y}{(1+i)^Y - 1} + OPEX \right) \quad [\textit{Equation 18}]$$

In order to obtain $£_{\text{Total}}$, the CAPEX and OPEX costs had to initially be calculated using *equations 20 & 21*. The equation used in (11) to calculate CAPEX per cell was modified for this project to include spectrum cost per cell $£_{\text{Spec}}$, considered a significant CAPEX for network operators (27). $£_{\text{Spec}}$ per cell can be obtained using *equation 19*, where $£_{\text{Total}\ Spectrum}$ is that total spectrum purchase cost for the network operator.

$$£_{Spec} = \frac{£_{\text{Total}\ Spectrum}}{Y \cdot N_{cell}} \quad [Equation\ 19]$$

$$CAPEX = £_{\text{cell}} + £_{insertion} + £_{BHi} + £_{Spec} \quad [\textit{Equation 20}]$$

CAPEX per cell shown in *equation 7* comprises of; $£_{\text{cell}}$, the BS equipment costs; $£_{insertion}$, insertion cost of building, installing macro BS equipment combined with core network upgrade costs; $£_{BHi}$, backhaul installation costs; and finally, $£_{Spec}$, cost of spectrum per BS.

$$OPEX = £_{\text{rent}} + £_{BH} + \text{E}_{cell} £_{bill} + £_{\text{M}} + \alpha CAPEX \quad [\textit{Equation 21}]$$

OPEX per cell shown in *equation 21* comprises of; $£_{\text{rent}}$ the site rent; $£_{BH}$ backhaul rent; $\text{E}_{cell}$ annual energy consumption per cell; $£_{bill}$ cost of energy; $£_{\text{M}}$ cost of maintenance per cell; and finally, $\alpha CAPEX$ the costs associated with marketing where $\alpha$ denotes marketing costs as a factor of the cell CAPEX and was calculated based on annual network operator marketing costs having been considered a direct cost (27).

$E_{cell}$, annual energy consumption per cell (kWh) is a function of the OPEX costs as shown in *equation 21* and can be calculated using the following equation:

$$E_{cell} = \frac{(P_{km^2} \cdot N_h \cdot N_d)}{N_{c/km^2}} \qquad [Equation\ 22]$$

| CAPEX and OPEX parameters | | | |
|---|---|---|---|
| *Parameter* | *Symbol* | *Value* | *Source* |
| **Interest rate** | $i$ | 5% | (11) |
| **Years to pay back loan** | $Y$ | 12 | (28) |
| **Number of antennas** | $N_a$ | 3 | (24) |
| **Number of cells in Operator Network**[10] | $N_{cell}$ | 20,000 | (29) |
| **Number of cells per km$^2$** | $N_{c/km^2}$ | 4.6 | Section 5.2.2 |
| **CAPEX parameters** | | | |
| **3 sector Cell equipment costs** | $£_{cell}$ | **£28,000** | (30) |
| **Cell insertion costs** | $£_{insertion}$ | **£100,000** | (30) (31) |
| **Backhaul Installation costs** | $£_{BHi}$ | **£8, 500** | (32) |
| **Spectrum cost per cell** | $£_{Spec}$ | **£3,295** | (33) |
| **OPEX parameters** | | | |
| **Cell Site rental costs** | $£_{rent}$ | **£10,800** | (30) |
| **Backhaul rental costs**[11] | $£_{BH}$ | **£7,500** | (32) |
| **Maintenance costs per cell** | $£_{M}$ | **£3,900** | (30) (32) |
| **Cost of energy (per KWh)** | $£_{bill}$ | **£0.14** | (34) |
| **Factor of costs related to Marketing.** | $\alpha$ | **0.0233** | (35) |
| **Number of hours** | $N_h$ | **24** | |
| **Number of days** | $N_d$ | **365** | |

Table 6 Summary of parameter values

The parameters shown in *equations 19, 20 & 21* can be summarised in *table 6*. The values shown in *table 6* were extracted from various sources including equipment vendors, reviewed literature, marketing reports, energy suppliers and network operator reports.

### 5.2.6 Calculating Profit

Profitability for network operators was calculated per km$^2$ based on dense-urban areas (cities). Dense-urban user density was chosen as 90.1% of the UK population lives in cities and hence the most profitable deployment area for network operators (36). The Subscriber (user) density and number of cells per km$^2$ is shown in *table 5.*

[10] All figures that require information based on network operators are based on Vodafone UK throughout the project.

[11] For wireless backhaul, we assume a point-to-multipoint (PMP) that uses microwave spectrum and provides 300 mbps capacity. (32)

In order to work out the profit per km$^2$ the revenue per subscriber (RPS) must be determined. The RPS can vary depending on the subscriber tariff. This in turn depends on the traffic demand per km$^2$ and $S_u$, the subscriber uptake[12] per km$^2$. This can be shown by *equations 23 & 24:*

$S_u$ = % uptake · user density/ km$^2$ [*Equation 23*]

$R_s = \frac{\bar{R}_{\mathrm{km}^2}}{S_u}$ [*Equation 24]*

Where $R_s$, is the data demand per subscriber in Mbps/user, for demanded traffic rate of $\bar{R}_{\mathrm{km}^2}$. Given $R_s$ the subscriber data usage $D_u$, over a period of a month[13] was calculated using *equation 25*, where $D_u$ is GB/month/subscriber/km$^2$ and $K_f$ is a constant equal to $\frac{1}{8192}$, used to convert Mbits into GB. The closest tariff giving an equivalent usage to $D_u$ was then chosen from the 4G and 3G tariff regression curves formed in *section 5.2.3* and the cost of user data consumption could then be determined.

$D_u = R_s \cdot (3600 \cdot 12 \cdot 30) \cdot K_f$ [*Equation 25*]

This tariff used is assumed the average tariff for all subscribers per km$^2$, for this then quantifies the demanded data rate to satisfy all subscribers. The charge associated with the chosen tariff $£_{Tariff}$, then forms the RPS for the network operator and is calculated using the methodology proposed in *section 5.2.3*. In order to obtain the annual revenue for all subscribers/km$^2$, the following equation was used:

$$£_R = £_{Tariff} \cdot S_u \cdot 12\, months \qquad [Equation\ 26]$$

[12] The subscriber uptake will be determined by applying percentage uptake increments.

[13] In equation 12 the numbers of hours in a day have been classified as 12, this is based on the assumption that no user is active for more than a maximum of 12 hours.

Where $£_R$ is the annual revenue generated from all subscribers per km$^2$. Using *equation 26* and *equation 27* an estimate for $\mathrm{Pr}_{\mathbf{km^2}}$ the pre-tax and pre-retail profit per km$^2$, could then be generated for network providers. The $\mathrm{Pr}_{\mathbf{km^2}}$ can be estimated using the following equation.

$$\mathrm{Pr}_{\mathbf{km^2}} = £_R - £_{\mathrm{Total}} \qquad [Equation\ 27]$$

The $\mathrm{Pr}_{\mathbf{km^2}}$ was calculated for a demanded traffic rate between 5 and 190 Mbps/km$^2$ for both 3G & 4G. The percentage uptake increments in 4G were chosen based on the predicted uptake of 4G globally and corresponding to the actual percentage uptake in countries of similar consumer lifestyle (3) (37). The assumption made is that 4G & 3G subscription percentage combined equals the total number of subscriptions i.e. 3% uptake in 4G subscription results in the remaining 97% being 3G subscriptions[14]. The results obtained are analysed in *Section 6.*

### 5.2.7 Calculating $CO_2$ Emissions

The UK government is legally obliged to reducing green house gas emissions by 80% by 2050, under the Climate Change act of 2008. As a major contributor to green house gases, reducing the levels of $CO_2$ emitted into the atmosphere is a key area of focus for the UK and world governing authorities (38). As a result businesses and consumers are encouraged to reduce energy consumption in order to lower $CO_2$ emissions. With rising electricity costs and increasing demand for more data, mobile operators must satisfy data demand while reducing their carbon footprint at low costs in order to still remain profitable. This section proposes how $CO_2$ emissions can be calculated per cell, per km$^2$ and for a cellular network for 3G & 4G.

[14] This was based on the assumption that the introduction of 4G services will result in 2G users migrating to 3G, making the 2G service redundant, although 2G services will still exist.

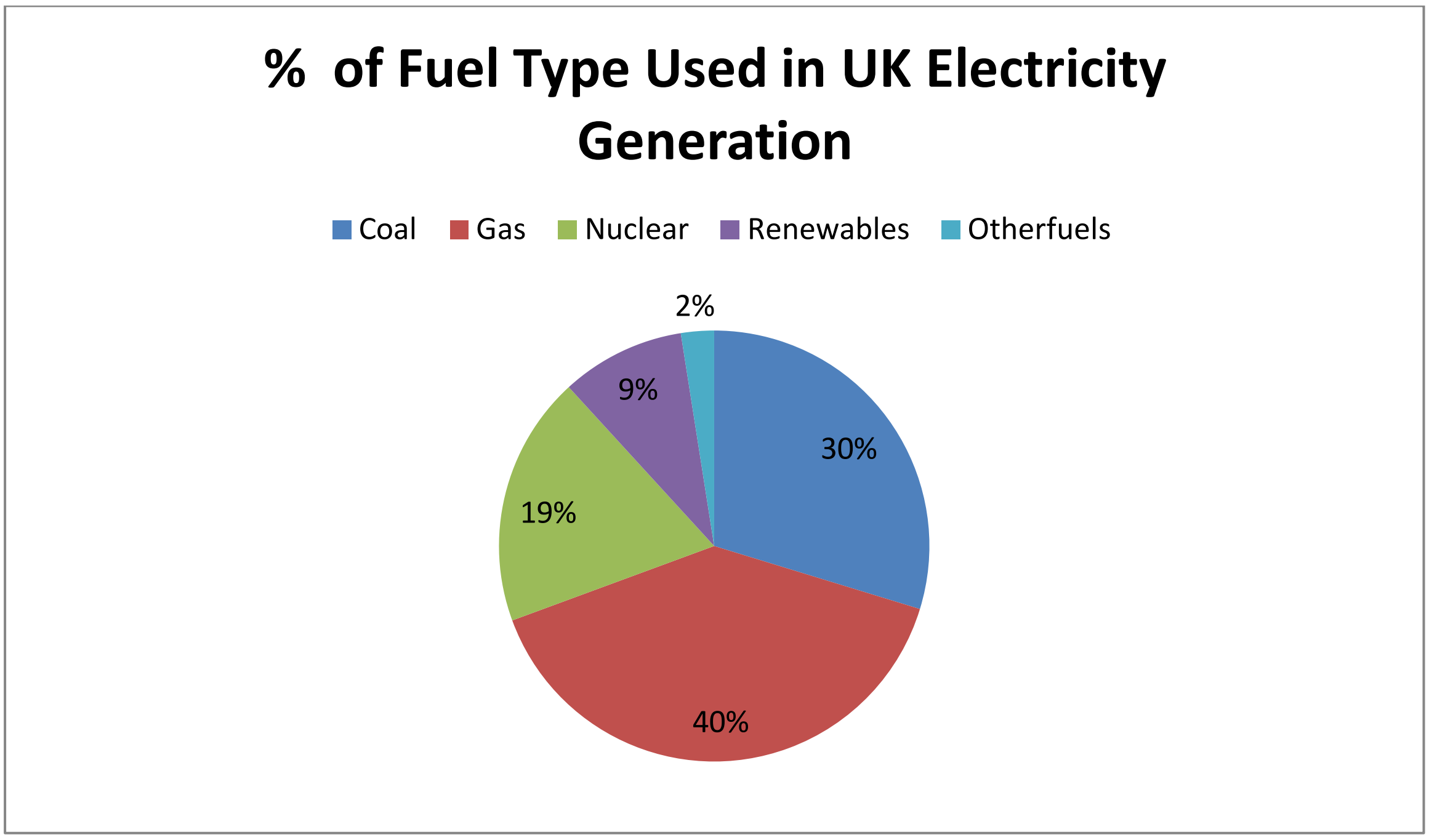

Figure 7 Shows the percentage of different fuels used for electricity generation in the UK

| Types of Fuel ($F_T$) | % employed in the UK ($\rho_{UK}$) | Grams of $CO_2$ produced per kWh ($G_{UK}$) |
|---|---|---|
| Coal | 30 % | 960 |
| Gas | 40 % | 443 |
| Nuclear | 19 % | 66 |
| Renewable | 9 % | 11 |
| Other fuels | 2 % | 25 |

Table 7 Shows the $CO_2$ produced from different fuel types in the UK[15]

In the UK electricity is generated using a variety of fuels. The type of fuels, the percentage used to generate electricity, and the associated $CO_2$ can be summarised by *figure 7 & table 7* (39) (40). *Section 5.2.5* earlier proposed how energy consumption can be calculated per km$^2$ for a given traffic rate. Given the energy consumption per km$^2$ and the information presented in *table 7* an approximation for the quantity of $CO_2$ produced per km$^2$ for each fuel type $F_T$ can be obtained by:

$$F_T = E_{cell} \cdot N_{c/km^2} \cdot G_{UK} \cdot \rho_{UK} \qquad [Equation\ 28]$$

[15] *Table 7* defines 'Renewables' as energy sources: Wind (on-shore and off-shore), Hydroelectric (reservoir and run-of-river) and solar thermal; where the average $CO_2$ production (in grams) of the three sources is assumed. Other fuel is largely defined as Bio gas and Biomass where the average $CO_2$ production (in grams) of the two sources is assumed.

The total $CO_2$ emitted per km$^2$ ($CO_{km2}$), can then be given by the summation for $CO_2$ emitted for each fuel type as shown by *equation 29*:

$$CO_{km2} = \sum F_T \qquad [Equation\ 29]$$

The results obtained for 3G and 4G can then be compared to establish if the migration to 4G will result in lower levels of $CO_2$ emissions. This information can also be used by network operators to estimate if the companies $CO_2$ emissions can be further reduced from the migration to 4G.

# 6.0 Results & Analysis

This section applies the theory discussed previously to obtain results showing: the power consumption of 3G and 4G; related OPEX costs and saving; profitability of 4G; and finally, carbon emission analysis.

## 6.1 Power Consumption Analysis

By applying the theory outlined in section *5.2.4* the power consumption per km$^2$ with increasing traffic demand was calculated for 4G & 3G and is shown in *tables 8 and 9* respectively.

| 4G Power Consumption with 4.6 cells/km² | | | | | | |
|---|---|---|---|---|---|---|
| $P_T$ | $\mu_{RH}$ | $R_{km^2}$ | $C_{km^2}$ | $P_{cell}^{OH} + P_{BH}$ | $P_{cell}$ (W) | $P_{km^2}$ (kW) |
| 39.80 | 0.31 | 0.00 | 198.66 | 300.00 | 900.00 | 4.14 |
| 39.80 | 0.31 | 5.00 | 198.66 | 300.00 | 961.10 | 4.42 |
| 39.80 | 0.31 | 10.00 | 198.66 | 300.00 | 986.41 | 4.54 |
| 39.80 | 0.31 | 20.00 | 198.66 | 300.00 | 1022.21 | 4.70 |
| 39.80 | 0.31 | 30.00 | 198.66 | 300.00 | 1049.67 | 4.83 |
| 39.80 | 0.31 | 40.00 | 198.66 | 300.00 | 1072.83 | 4.94 |
| 39.80 | 0.31 | 50.00 | 198.66 | 300.00 | 1093.23 | 5.03 |
| 39.80 | 0.31 | 60.00 | 198.66 | 300.00 | 1111.67 | 5.11 |
| 39.80 | 0.31 | 70.00 | 198.66 | 300.00 | 1128.63 | 5.19 |
| 39.80 | 0.31 | 80.00 | 198.66 | 300.00 | 1144.42 | 5.26 |
| 39.80 | 0.31 | 90.00 | 198.66 | 300.00 | 1159.24 | 5.33 |
| 39.80 | 0.31 | 100.00 | 198.66 | 300.00 | 1173.27 | 5.40 |
| 39.80 | 0.31 | 110.00 | 198.66 | 300.00 | 1186.60 | 5.46 |
| 39.80 | 0.31 | 120.00 | 198.66 | 300.00 | 1199.35 | 5.52 |
| 39.80 | 0.31 | 130.00 | 198.66 | 300.00 | 1211.57 | 5.57 |
| 39.80 | 0.31 | 140.00 | 198.66 | 300.00 | 1223.33 | 5.63 |
| 39.80 | 0.31 | 150.00 | 198.66 | 300.00 | 1234.68 | 5.68 |
| 39.80 | 0.31 | 160.00 | 198.66 | 300.00 | 1245.66 | 5.73 |
| 39.80 | 0.31 | 170.00 | 198.66 | 300.00 | 1256.30 | 5.78 |
| 39.80 | 0.31 | 180.00 | 198.66 | 300.00 | 1266.63 | 5.83 |
| 39.80 | 0.31 | 190.00 | 198.66 | 300.00 | 1276.67 | 5.87 |

Table 8 4G power consumption with increasing traffic demand

| 3G Power Consumption with 12.4 cells/km² | | | | | | |
|---|---|---|---|---|---|---|
| $P_T$ | $\mu_{RH}$ | $R_{km^2}$ | $C_{km^2}$ | $P_{cell}^{OH} + P_{BH}$ | $P_{cell}$ (W) | $P_{km^2}$ (kW) |
| **39.80** | 0.31 | 0.00 | 198.40 | 300.00 | 900.00 | 11.16 |
| **39.80** | 0.31 | 5.00 | 198.40 | 300.00 | 961.10 | 11.92 |
| **39.80** | 0.31 | 10.00 | 198.40 | 300.00 | 986.41 | 12.23 |
| **39.80** | 0.31 | 20.00 | 198.40 | 300.00 | 1022.21 | 12.68 |
| **39.80** | 0.31 | 30.00 | 198.40 | 300.00 | 1049.67 | 13.02 |
| **39.80** | 0.31 | 40.00 | 198.40 | 300.00 | 1072.83 | 13.30 |
| **39.80** | 0.31 | 50.00 | 198.40 | 300.00 | 1093.23 | 13.56 |
| **39.80** | 0.31 | 60.00 | 198.40 | 300.00 | 1111.67 | 13.78 |
| **39.80** | 0.31 | 70.00 | 198.40 | 300.00 | 1128.63 | 14.00 |
| **39.80** | 0.31 | 80.00 | 198.40 | 300.00 | 1144.42 | 14.19 |
| **39.80** | 0.31 | 90.00 | 198.40 | 300.00 | 1159.24 | 14.37 |
| **39.80** | 0.31 | 100.00 | 198.40 | 300.00 | 1173.27 | 14.55 |
| **39.80** | 0.31 | 110.00 | 198.40 | 300.00 | 1186.60 | 14.71 |
| **39.80** | 0.31 | 120.00 | 198.40 | 300.00 | 1199.35 | 14.87 |
| **39.80** | 0.31 | 130.00 | 198.40 | 300.00 | 1211.57 | 15.02 |
| **39.80** | 0.31 | 140.00 | 198.40 | 300.00 | 1223.33 | 15.17 |
| **39.80** | 0.31 | 150.00 | 198.40 | 300.00 | 1234.68 | 15.31 |
| **39.80** | 0.31 | 160.00 | 198.40 | 300.00 | 1245.66 | 15.45 |
| **39.80** | 0.31 | 170.00 | 198.40 | 300.00 | 1256.30 | 15.58 |
| **39.80** | 0.31 | 180.00 | 198.40 | 300.00 | 1266.63 | 15.71 |
| **39.80** | 0.31 | 190.00 | 198.40 | 300.00 | 1276.67 | 15.83 |

Table 9 3G power consumption with increasing traffic demand

Comparison of the above tables shows that due to the lower spectral efficiency of 3G technology the maximum capacity (throughput) achieved per cell is far less than that of 4G. In order to therefore achieve the same capacity the number of BS deployed per $km^2$ must be increased. The following is a comparison of 4G & 3G power consumption with respect to traffic rate demanded.

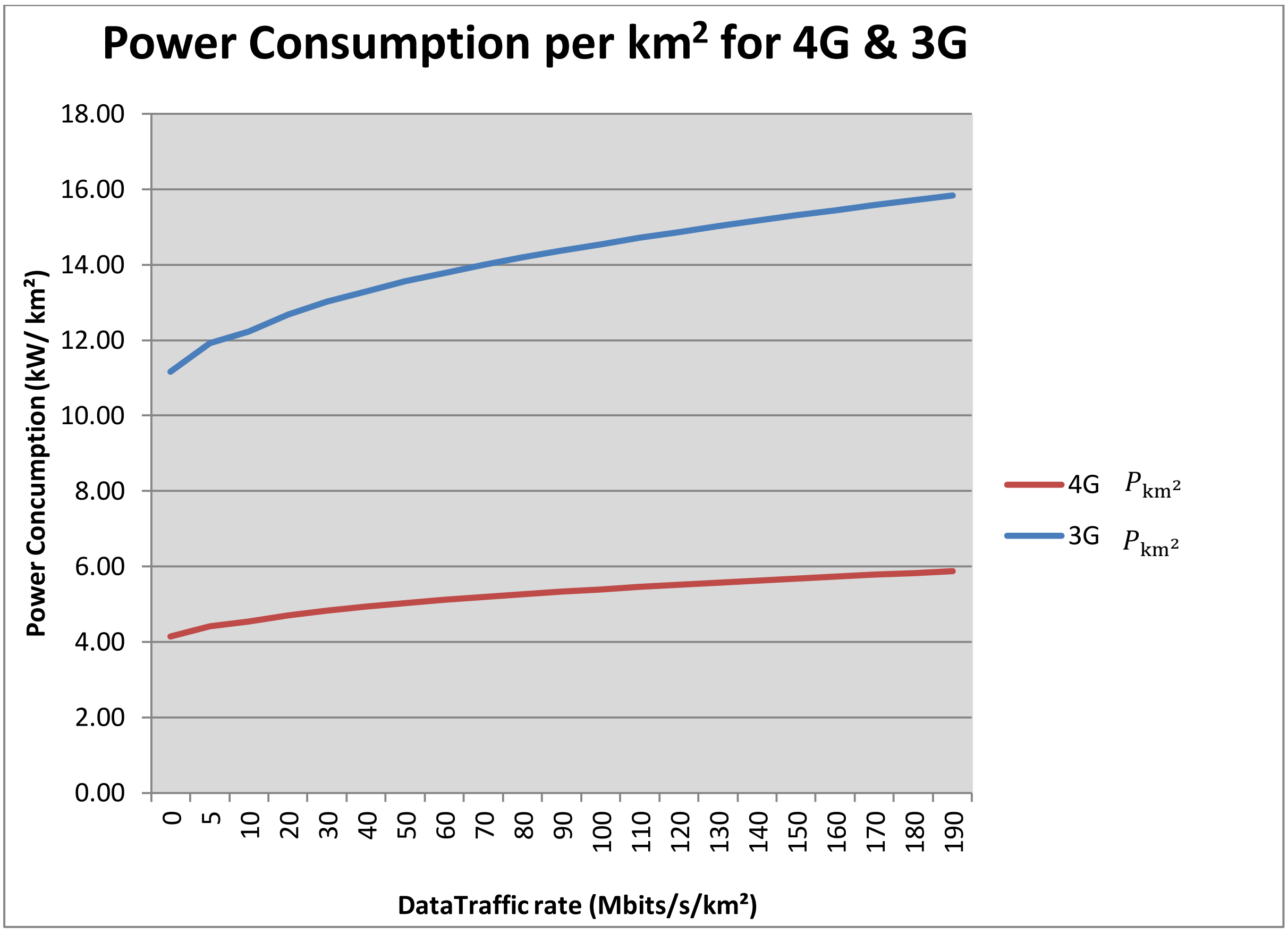

Figure 8 shows 4G & 3G power consumption per $km^2$

The comparison of 4G and 3G reveals that 4G power consumption per $km^2$ is almost 63% lower than 3G power consumption. This suggests 4G can deliver the same data throughput per $km^2$ as 3G consuming far less power. It can also be noted that the power increases exponentially, this is due to the power amplifier efficiency increasing with respective to data traffic, as a result with data demand expected to increase in the future and data traffic demand expected to rise far beyond 190 mbps/$km^2$ the power consumption will further improve. Achieving greater spectral efficiency per cell in 4G has meant fewer 4G cells are required to deliver the same throughput per $km^2$ compared to 3G. As a result consumers can now achieve higher throughput speeds using 4G compared to 3G but network operators can benefit from lower power consumption per $km^2$.

## 6.2 OPEX Analysis

The OPEX costs for operating a network are considered the main cost for network providers for any given technology. A detailed comparison based on the theory presented earlier revealed that the total OPEX cost for operating a 4G network per $km^2$ is approximately 34% less than that of an equivalent 3G network. *For a complete breakdown of OPEX costs for 4G & 3G please refer to appendix B.1 & B.2 respectively.*

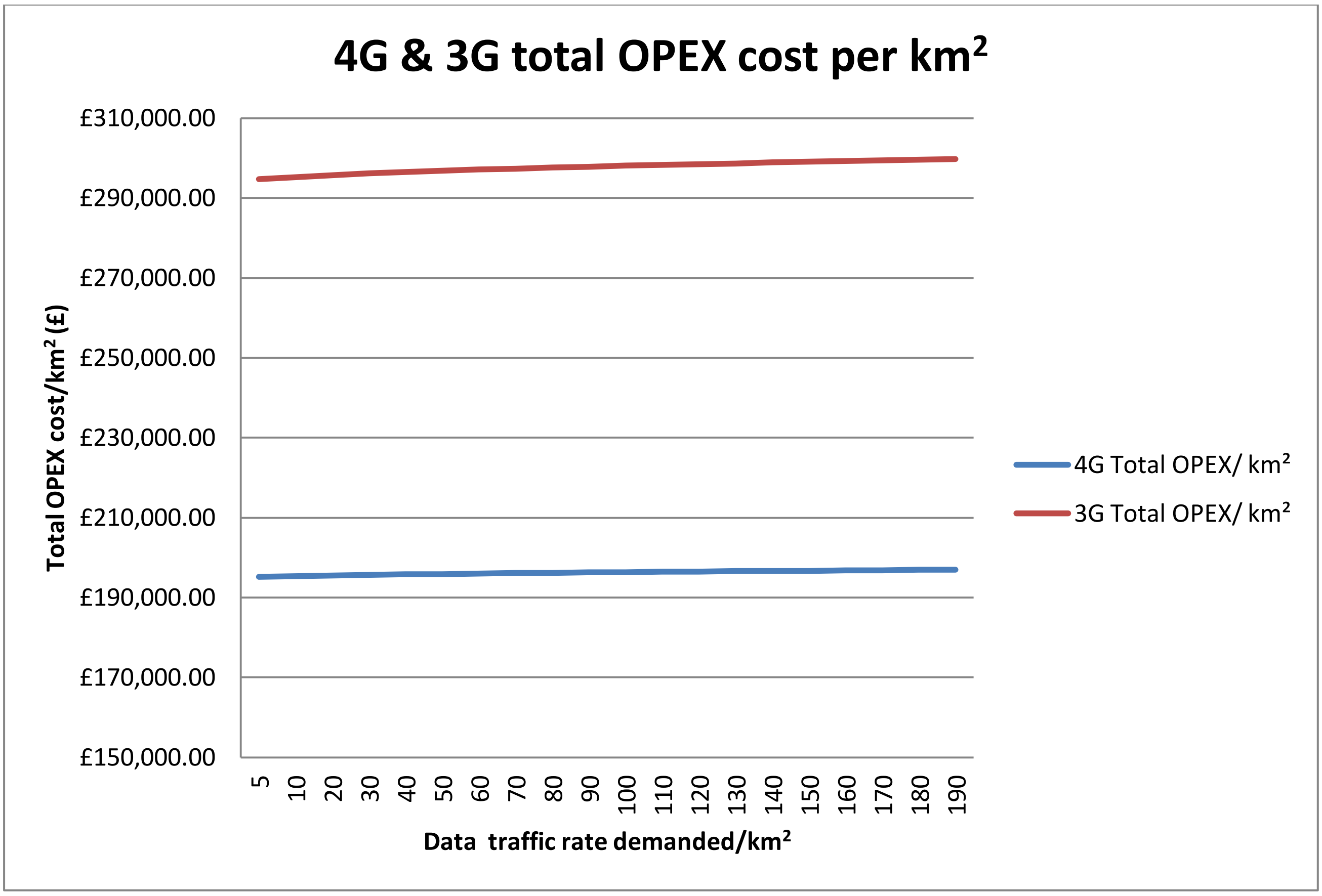

**Figure 9 4G & 3G OEPX costs per km$^2$**

Analysis of the above OPEX costs shows the overall trend that OPEX costs increase for both 4G and 3G with respect to traffic demand. Furthermore, it can be noticed that the change in OPEX costs when operating at high traffic loads and low traffic loads is very small. In 4G the OPEX cost for a demanded traffic rate of 190 Mbps/km$^2$ and 5Mbps/km$^2$ is £196,956 and £195,152. This shows a change of £1,804 between the maximum and minimum offered data rate. Similarly, applying the same analysis to 3G OPEX costs, the change in cost for the maximum and minimum demanded traffic rate is £4, 942. Although a larger change in cost compared to 4G, the change represents approximately 1.5% of the total annual OPEX cost, which in context is insignificant. One can therefore view the OPEX cost per km$^2$ relatively *static* with increase in demanded traffic rate. However it should be noted that depending on the size of the entire network, the change in OPEX costs with increasing traffic could bare greater significance.

A more prominent observation shows a 34% reduction in the 4G OPEX costs in comparison to 3G. From a network operator's perspective, it enables providing higher throughput speeds at lower costs and deploying fewer BS' by migrating from 3G to 4G.

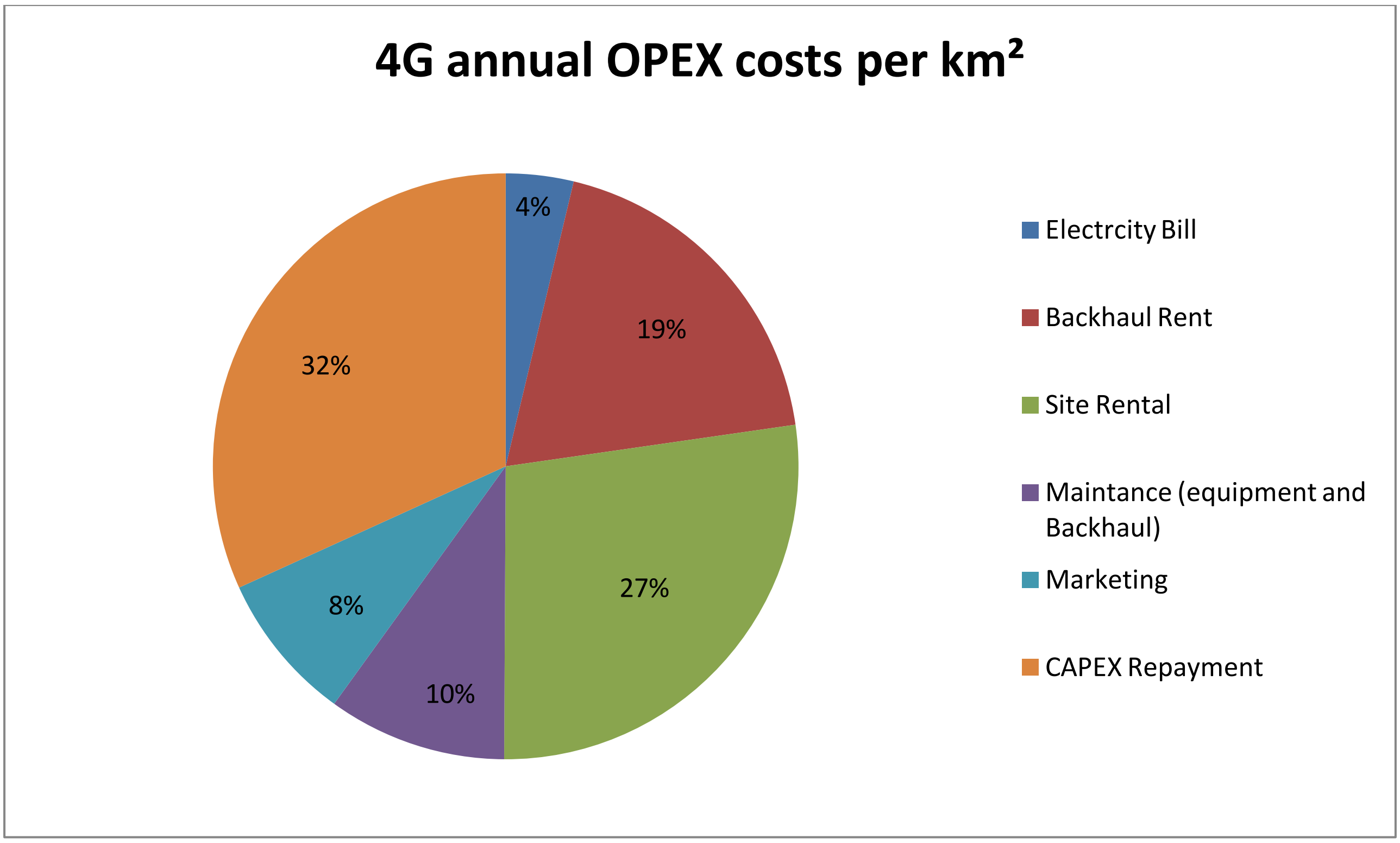

Figure 10 Breakdown of individual 4G OPEX costs element

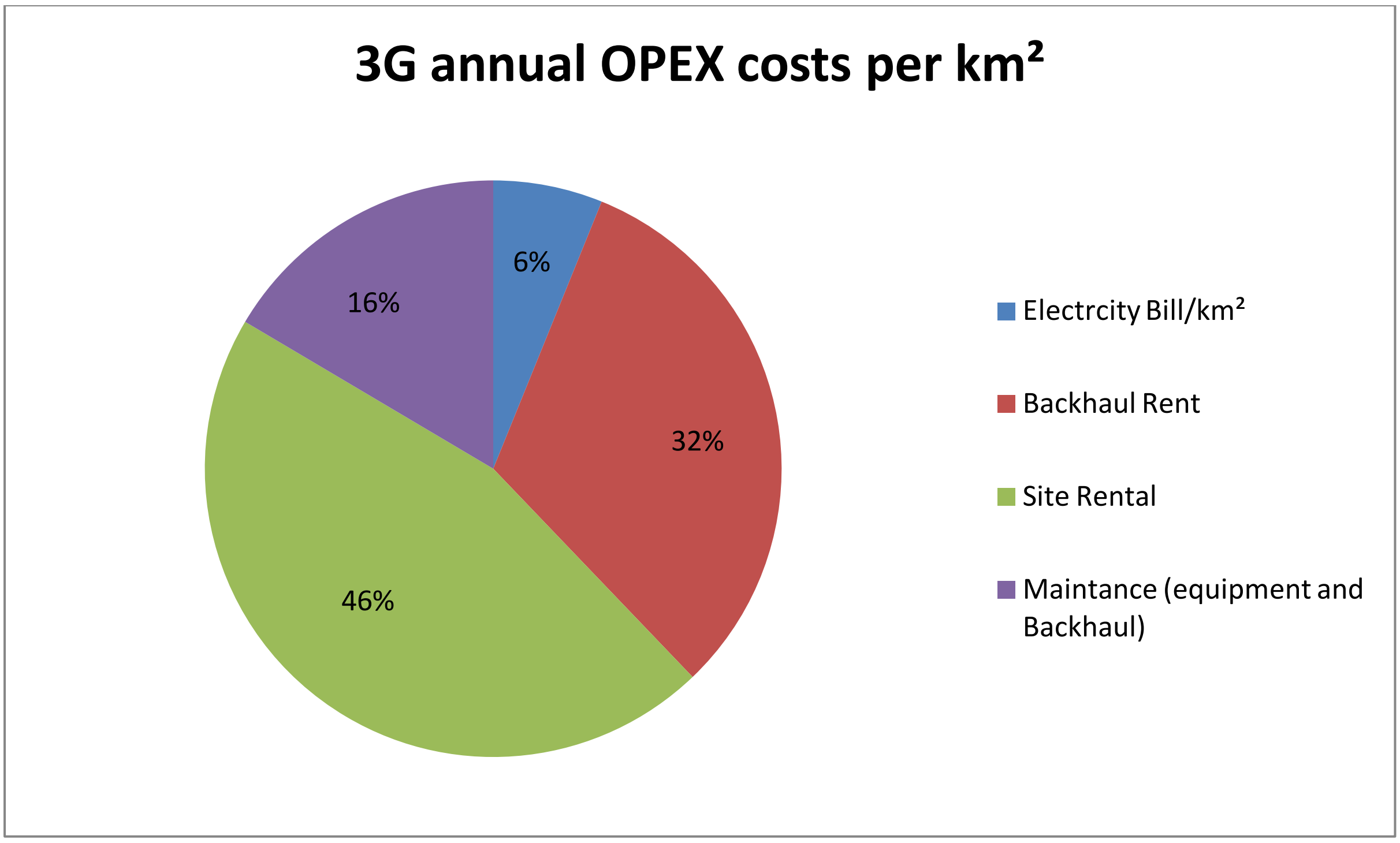

**Figure 11 Breakdown of individual 3G OPEX cost elements**

From *figures 10 & 11* a comparison can be made between the individual OPEX costs elements for 3G & 4G. A comparison between the two technologies reveals that the electricity bill is lower percentage of the total OPEX cost in 4G in comparison to 3G. However comparing the 3G OPEX cost breakdown with the findings proposed in (10), one can see discrepancy of 10% in the electricity bill. This change in cost is a result of the project proposing more recent cost evaluations of Backhaul and site rental compared to that presented in (10). It can also be noted that the 3G breakdown of OPEX does not incorporate CAPEX costs, as literature from Vodafone showed that the CAPEX costs associated with 3G have been repaid over last 12 years (28). The CAPEX repayment however forms 32% of 4G OPEX costs, the largest contributor to the OPEX costs. Although not considered for this project, research into cell site sharing suggest CAPEX and OPEX cost savings of up to 40%, these savings are a result of network operators sharing the cost of active network elements and initial site build costs (41). This is also something that has gained momentum in 2013, network operators Vodafone and O2 having been granted approval in cell site sharing to expand the potential 4G network (42). However further research would have to be conducted into the actual OPEX savings from cell site sharing.

## 6.3 Profit Analysis

*Section 5.2.6* outlined the theoretical approach to determine the profit for a network operator in a dense urban environment consisting of 3000 users/km$^2$. This section analyses the results obtained from the application of that theory. The results shows profitability per km$^2$ as 4G percentage uptake increases and the resultant 3G percentage decrease. The percentage increments for 3G & 4G uptake are shown in *Table 10.* The tables showing profitability with percentage uptake can be found in Appendix C. *Tables 11 & 12* below exemplify the structure of the profitability tables found in Appendix C.

| 4G % uptake increments | Resulting 3G subscription % |
|---|---|
| 3% | 97% |
| 6% | 94% |
| 9% | 91% |
| 20% | 80% |
| 40% | 60% |
| 90% | 10% |

Table 10 shows the percentage uptake increments for 4G

| 4G Profitability - 3% uptake in subscription demand | | | | | | | | | | |
|---|---|---|---|---|---|---|---|---|---|---|
| % Uptake | Rate of Traffic demand Mbps/km² | Subscribers uptake per km² | Mbps/ subscriber | Mb/month/subscriber | GB/month/subscriber | Equivlent tarrif (GB) per month | Tarrif cost per month | (Charge per subscriber x number of subscribers) x 12 months | Cost for cells per km² | Profit/Loss (Before Retail Expenditure & Tax) |
| 3% | 5 | 90 | 0.0556 | 72000.00 | 70.31 | Unlimited | £82.00 | £88,560.00 | £195,152.00 | -£106,592 |
| 3% | 10 | 90 | 0.1111 | 144000.00 | 140.63 | Unlimited | £82.00 | £88,560.00 | £195,296.00 | -£106,736 |
| 3% | 20 | 90 | 0.2222 | 288000.00 | 281.25 | Unlimited | £82.00 | £88,560.00 | £195,501.00 | -£106,941 |
| 3% | 30 | 90 | 0.3333 | 432000.00 | 421.88 | Unlimited | £82.00 | £88,560.00 | £195,658.00 | -£107,098 |
| 3% | 40 | 90 | 0.4444 | 576000.00 | 562.50 | Unlimited | £82.00 | £88,560.00 | £195,790.00 | -£107,230 |
| 3% | 50 | 90 | 0.5556 | 720000.00 | 703.13 | Unlimited | £82.00 | £88,560.00 | £195,907.00 | -£107,347 |
| 3% | 60 | 90 | 0.6667 | 864000.00 | 843.75 | Unlimited | £82.00 | £88,560.00 | £196,012.00 | -£107,452 |
| 3% | 70 | 90 | 0.7778 | 1008000.00 | 984.38 | Unlimited | £82.00 | £88,560.00 | £196,109.00 | -£107,549 |
| 3% | 80 | 90 | 0.8889 | 1152000.00 | 1125.00 | Unlimited | £82.00 | £88,560.00 | £196,200.00 | -£107,640 |
| 3% | 90 | 90 | 1.0000 | 1296000.00 | 1265.63 | Unlimited | £82.00 | £88,560.00 | £196,284.00 | -£107,724 |
| 3% | 100 | 90 | 1.1111 | 1440000.00 | 1406.25 | Unlimited | £82.00 | £88,560.00 | £196,365.00 | -£107,805 |
| 3% | 110 | 90 | 1.2222 | 1584000.00 | 1546.88 | Unlimited | £82.00 | £88,560.00 | £196,441.00 | -£107,881 |
| 3% | 120 | 90 | 1.3333 | 1728000.00 | 1687.50 | Unlimited | £82.00 | £88,560.00 | £196,514.00 | -£107,954 |
| 3% | 130 | 90 | 1.4444 | 1872000.00 | 1828.13 | Unlimited | £82.00 | £88,560.00 | £196,584.00 | -£108,024 |
| 3% | 140 | 90 | 1.5556 | 2016000.00 | 1968.75 | Unlimited | £82.00 | £88,560.00 | £196,651.00 | -£108,091 |
| 3% | 150 | 90 | 1.6667 | 2160000.00 | 2109.38 | Unlimited | £82.00 | £88,560.00 | £196,716.00 | -£108,156 |
| 3% | 160 | 90 | 1.7778 | 2304000.00 | 2250.00 | Unlimited | £82.00 | £88,560.00 | £196,779.00 | -£108,219 |
| 3% | 170 | 90 | 1.8889 | 2448000.00 | 2390.63 | Unlimited | £82.00 | £88,560.00 | £196,839.00 | -£108,279 |
| 3% | 180 | 90 | 2.0000 | 2592000.00 | 2531.25 | Unlimited | £82.00 | £88,560.00 | £196,898.00 | -£108,338 |
| 3% | 190 | 90 | 2.1111 | 2736000.00 | 2671.88 | Unlimited | £82.00 | £88,560.00 | £196,956.00 | -£108,396 |

**Table 11 shows the profit/loss per km2 with 3% 4G uptake.**

| 3G Profitability - 97% uptake in subscription demand | | | | | | | | | | | Total Profitability with 3% 4G uptake |
|---|---|---|---|---|---|---|---|---|---|---|---|
| Uptake of % | Rate of Traffic demand Mbps/km² | Subscribers uptake per km² | Mbps/subscriber | Mb/month/subscriber | GB/month/subscriber | Equivlent tarrif (GB) | Tarrif cost per month | (Charge per subscriber x number of subscribers) x 12 months | Cost for cells per km² | Profit/Loss (Before Retail Expenditure & Tax) | Total Profitability = (4G profit + 3G Profit) |
| 97% | 5 | 2910 | 0.0017 | 2226.80 | 2.17 | 2.50 | £23.00 | £803,160.00 | £294,773.00 | £508,387 | £401,795 |
| 97% | 10 | 2910 | 0.0034 | 4453.61 | 4.35 | 4.50 | £28.50 | £995,220.00 | £295,169.00 | £700,051 | £593,315 |
| 97% | 20 | 2910 | 0.0069 | 8907.22 | 8.70 | 9.00 | £34.00 | £1,187,280.00 | £295,730.00 | £891,550 | £784,609 |
| 97% | 30 | 2910 | 0.0103 | 13360.82 | 13.05 | 13.00 | £39.50 | £1,379,340.00 | £296,160.00 | £1,083,180 | £976,082 |
| 97% | 40 | 2910 | 0.0137 | 17814.43 | 17.40 | 18.00 | £44.00 | £1,536,480.00 | £296,523.00 | £1,239,957 | £1,132,727 |
| 97% | 50 | 2910 | 0.0172 | 22268.04 | 21.75 | 22.00 | £49.00 | £1,711,080.00 | £296,824.00 | £1,414,256 | £1,306,909 |
| 97% | 60 | 2910 | 0.0206 | 26721.65 | 26.10 | Unlimited | £52.00 | £1,815,840.00 | £297,131.00 | £1,518,709 | £1,411,257 |
| 97% | 70 | 2910 | 0.0241 | 31175.26 | 30.44 | Unlimited | £52.00 | £1,815,840.00 | £297,397.00 | £1,518,443 | £1,410,894 |
| 97% | 80 | 2910 | 0.0275 | 35628.87 | 34.79 | Unlimited | £52.00 | £1,815,840.00 | £297,644.00 | £1,518,196 | £1,410,556 |
| 97% | 90 | 2910 | 0.0309 | 40082.47 | 39.14 | Unlimited | £52.00 | £1,815,840.00 | £297,876.00 | £1,517,964 | £1,410,240 |
| 97% | 100 | 2910 | 0.0344 | 44536.08 | 43.49 | Unlimited | £52.00 | £1,815,840.00 | £298,096.00 | £1,517,744 | £1,409,939 |
| 97% | 110 | 2910 | 0.0378 | 48989.69 | 47.84 | Unlimited | £52.00 | £1,815,840.00 | £298,305.00 | £1,517,535 | £1,409,654 |
| 97% | 120 | 2910 | 0.0412 | 53443.30 | 52.19 | Unlimited | £52.00 | £1,815,840.00 | £298,504.00 | £1,517,336 | £1,409,382 |
| 97% | 130 | 2910 | 0.0447 | 57896.91 | 56.54 | Unlimited | £52.00 | £1,815,840.00 | £298,696.00 | £1,517,144 | £1,409,120 |
| 97% | 140 | 2910 | 0.0481 | 62350.52 | 60.89 | Unlimited | £52.00 | £1,815,840.00 | £298,880.00 | £1,516,960 | £1,408,869 |
| 97% | 150 | 2910 | 0.0515 | 66804.12 | 65.24 | Unlimited | £52.00 | £1,815,840.00 | £299,058.00 | £1,516,782 | £1,408,626 |
| 97% | 160 | 2910 | 0.0550 | 71257.73 | 69.59 | Unlimited | £52.00 | £1,815,840.00 | £299,229.00 | £1,516,611 | £1,408,392 |
| 97% | 170 | 2910 | 0.0584 | 75711.34 | 73.94 | Unlimited | £52.00 | £1,815,840.00 | £299,396.00 | £1,516,444 | £1,408,165 |
| 97% | 180 | 2910 | 0.0619 | 80164.95 | 78.29 | Unlimited | £52.00 | £1,815,840.00 | £299,558.00 | £1,516,282 | £1,407,944 |
| 97% | 190 | 2910 | 0.0653 | 84618.56 | 82.64 | Unlimited | £52.00 | £1,815,840.00 | £299,715.00 | £1,516,125 | £1,407,729 |

**Table 12 shows the profit/loss per km2 with resultant 97% 3G subscriptions and total profitability for network operator (Column 11).**

*Table 11 & 12* show the percentage uptake in 4G services and the remaining 97% 3G subscriptions. The tables show the number of subscribers per km$^2$ 4G & 3G, given 3000 users/km$^2$ in total. As the traffic rate increases the proportion of Mbps demanded per subscriber is then shown by the column *Mbps/subscriber*. The demanded traffic rate over the period of a month is then shown as an *Equivalent tariff (GB)* and the equivalent cost of that tariff (obtained using the 3G & 4G regression model) is shown in the column *Tariff cost per month*. The *Total Profitability* for the network operator is then shown in the last column of *table 12.*

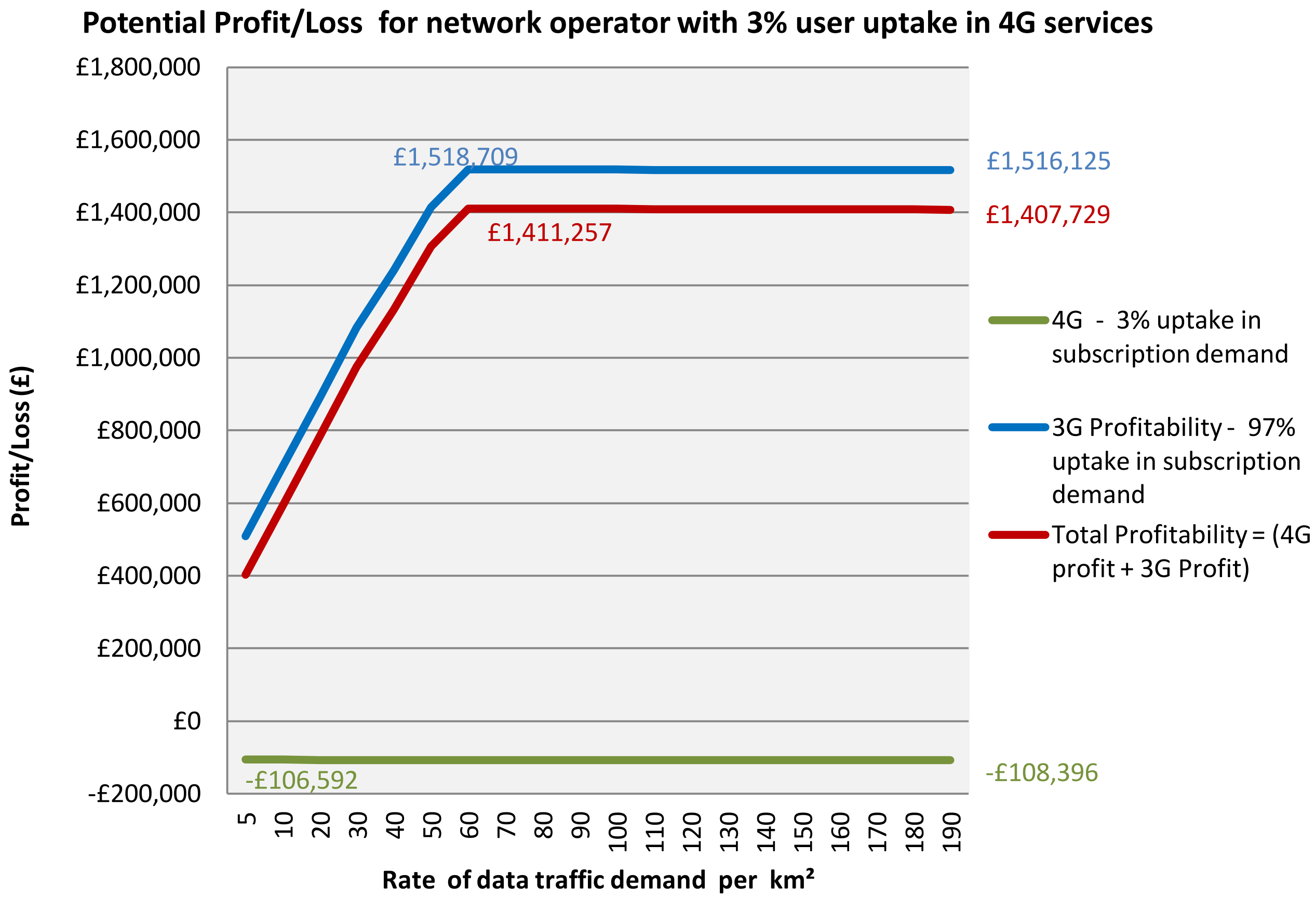

**Figure 12 shows the total profitability for network operators per km$^2$**

The figure above shows the operational profit expected from 4G & 3G services. Relating *figure 12* with *tables 11 & 12,* for 4G services, subscribers would require unlimited data tariff to satisfy their data usage. For 3G services, it is observed from *table 12* that beyond 60Mbps of demanded traffic rate per km$^2$, 3G users would also require unlimited data tariffs to satisfy data demand. The overall trend shows that the network operator will be profitable. However with 3% uptake in 4G services, network operators will incur a loss, although remaining profitable overall due to the revenue generated from 3G subscription. Applying the findings to the previous research, in (3) Cisco estimate that the in 2013 the global uptake in 4G would be 3%, similarly to that estimated in (37) for countries with similar consumer lifestyles to that of the UK. This would therefore suggest that operation 4G services in 2013 with the tariffs proposed, network operators would expect to incur a loss.

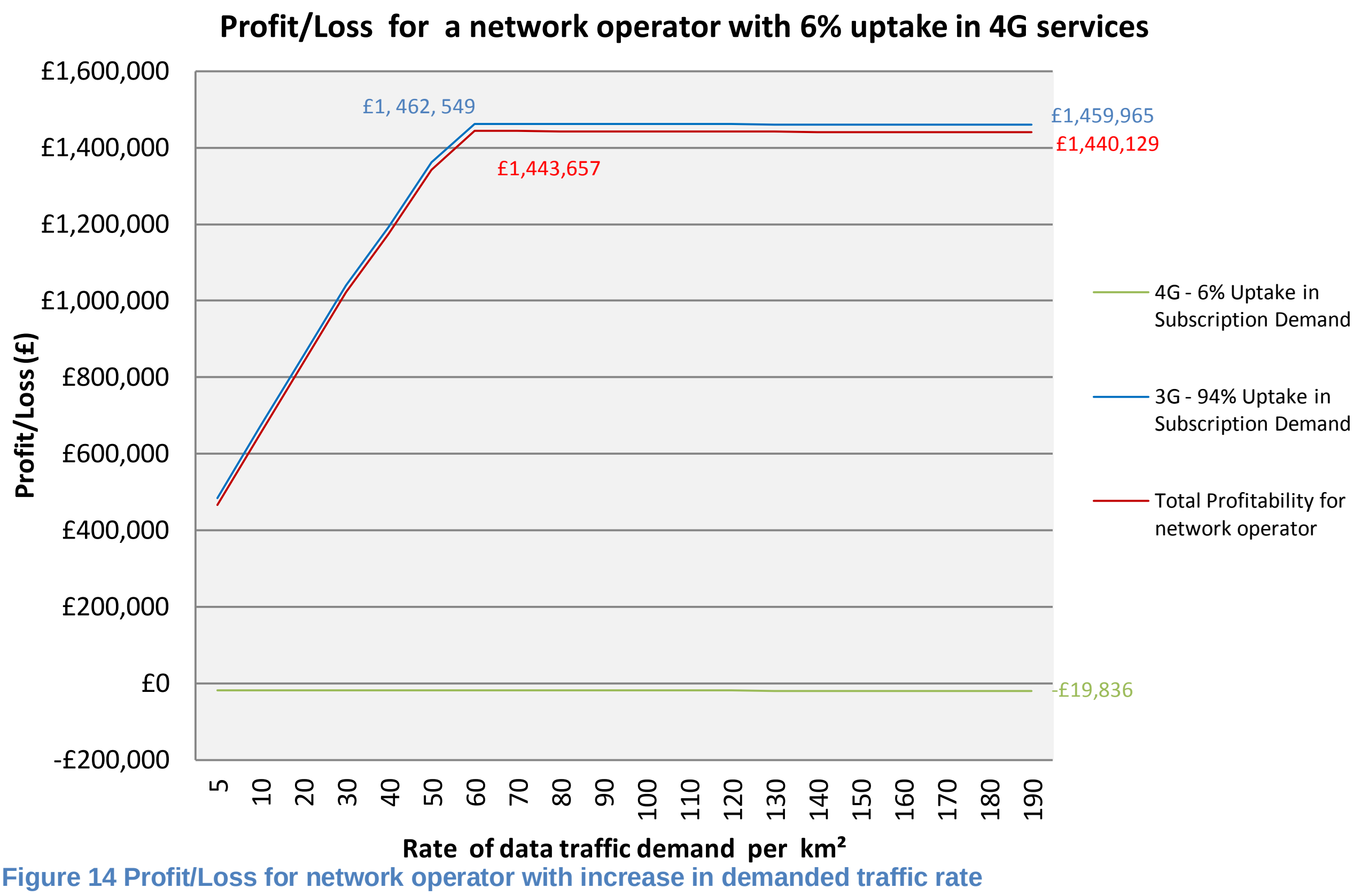

Figure 14 Profit/Loss for network operator with increase in demanded traffic rate

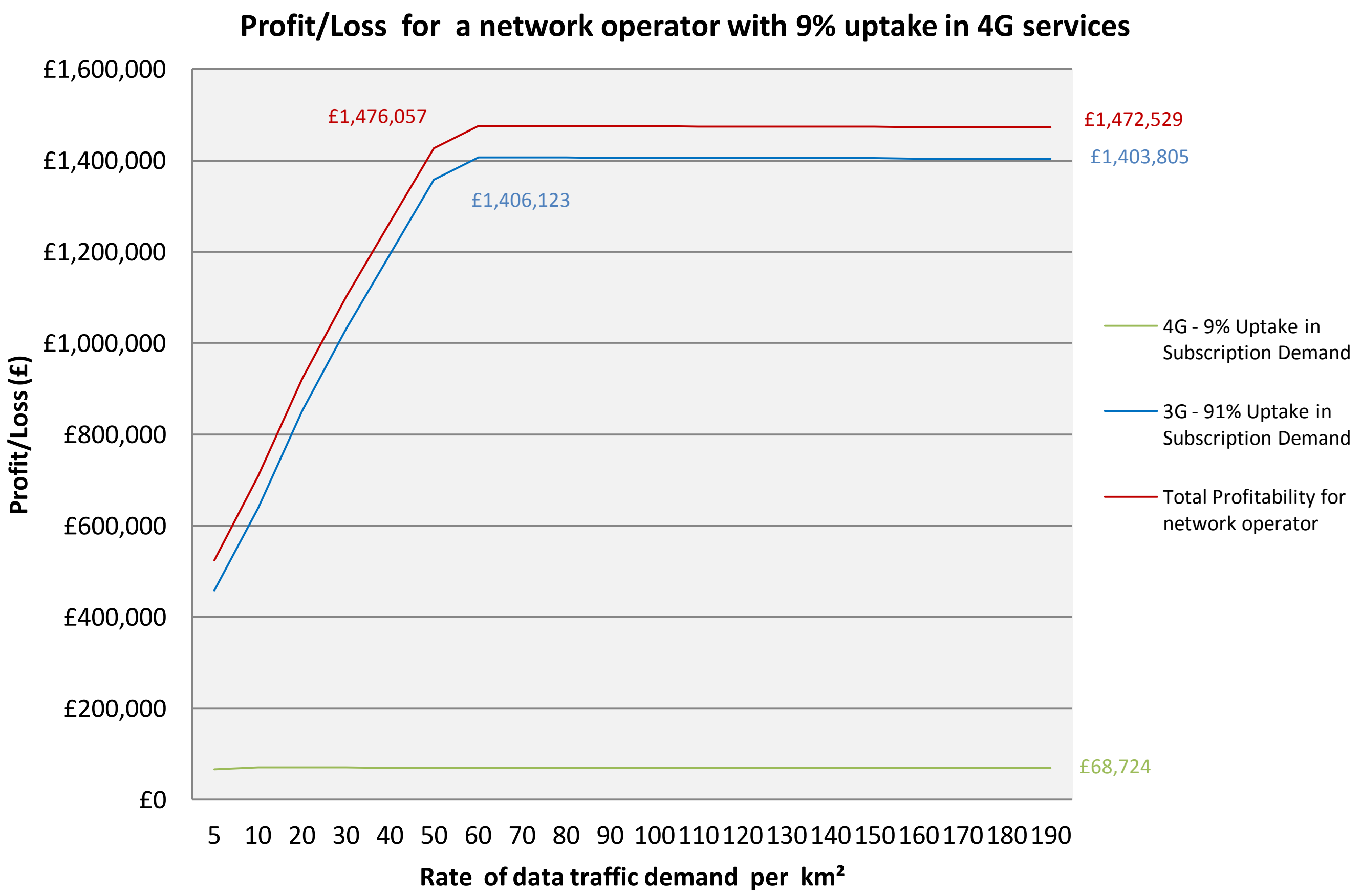

Figure 13 Profit/Loss for network operator with increase in demanded traffic rate

Analysis of the graphs above shows that the network operators will continue to remain profitable overall with the operation of both 3G & 4G service when meeting traffic demands. However with 6% uptake in 4G, results shows that network operators would continue to incur a loss. The loss is due to not generating enough revenue, as 180 subscribers per km$^2$ although on the maximum tariff (unlimited data) is not sufficient to cover network operational costs for 4G. *Figure 14* shows that network operators can expect to see a profit return on 4G as subscription uptake approaches 9% of users/km$^2$. The majority of the users will still however require the unlimited tariff to satisfy the data demand[16], with the exception of when a low data traffic rate of 5 Mbps/km$^2$ is demanded. It should be noted that beyond 60 Mbps/km$^2$, profitability becomes saturated as the high data demanded from users would only be satisfied by unlimited tariffs. Although profitability no longer increases, the results show the profitability is sustainable.

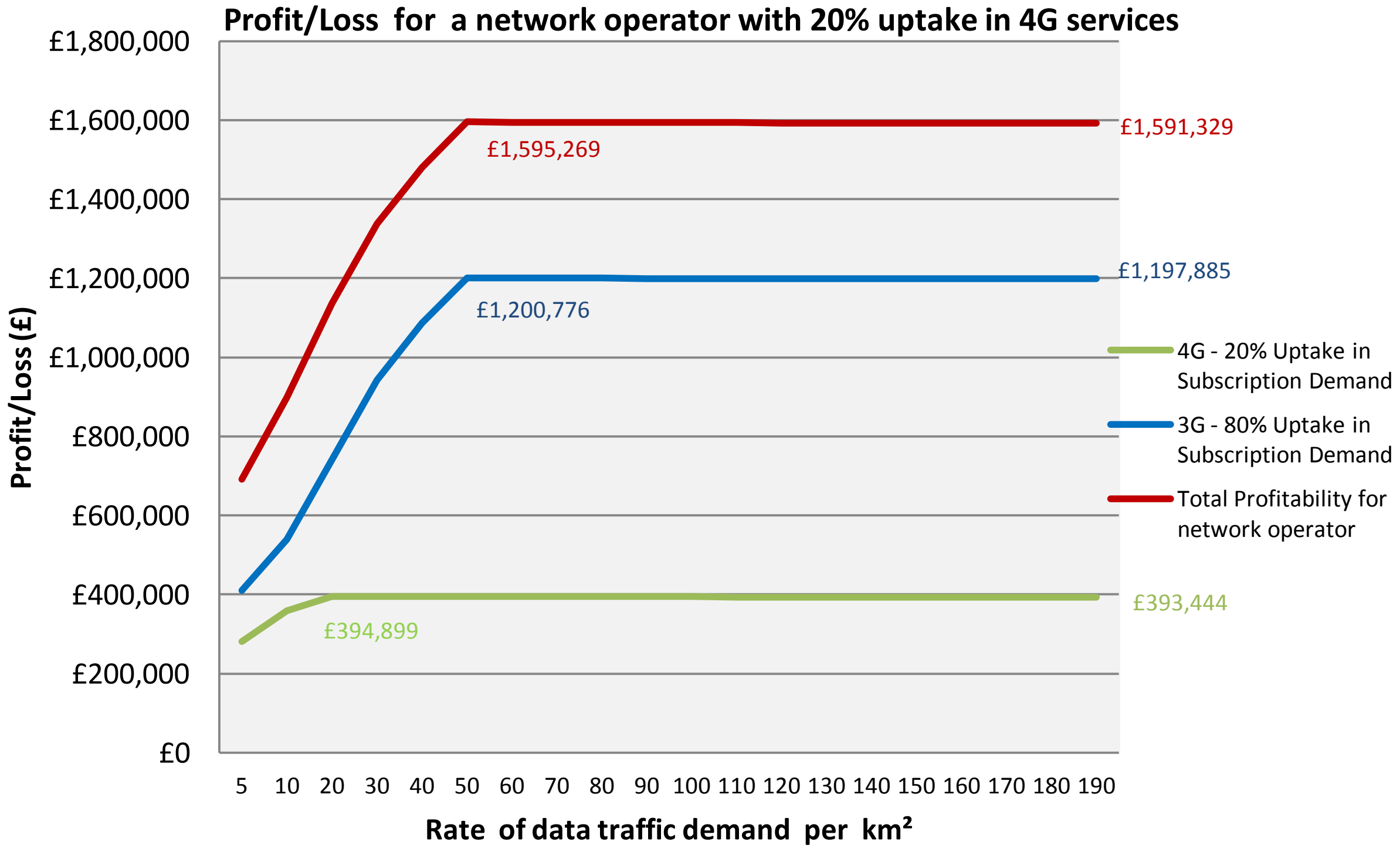

Figure 15 Profit/Loss for network operator with increase in demanded traffic rate

[16] Please refer to Appendix C.2 for more detail on tariff choice

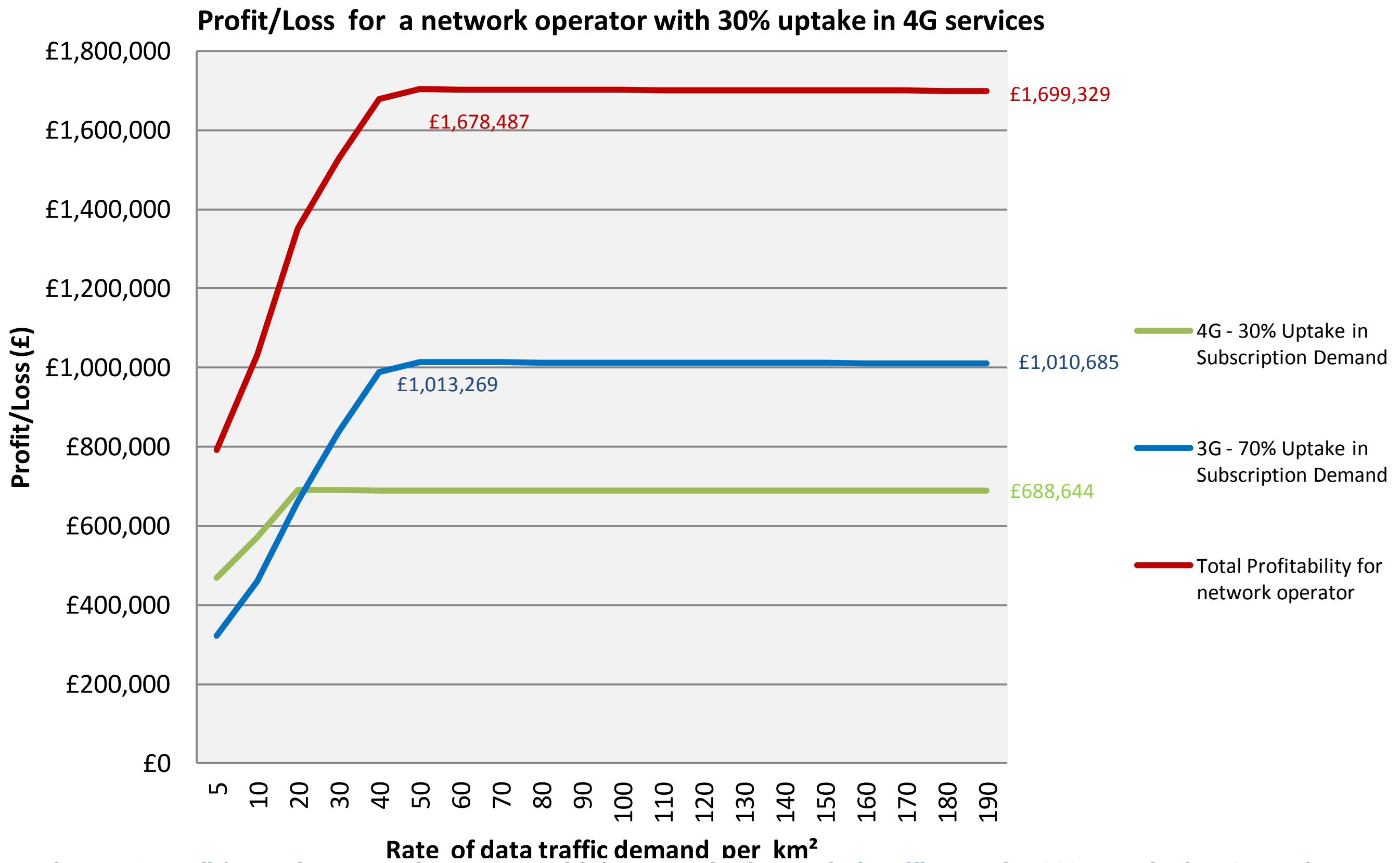

Figure 16 Profit/Loss for network operator with increase in demanded traffic rate for 30% uptake in 4G services

Analysis of *Figure 15* shows that at 20% uptake profit peaks are expected to occur at 20Mbps/km$^2$ for 4G services. It can be noted that 3G profits far exceed 4G profits at 20% uptake; 3G services still remain 67% more profitable than 4G. *Figure 16* shows however that with the current tariff strategy employed by network operators, 4G will become more profitable than 3G, when an uptake by 30% of the users occurs. It can be seen that for low traffic data rates below 20 Mbps/km$^2$ 4G is more profitable than 3G. However due to the current tariff strategy employed beyond 20 Mbps/km$^2$ users would require unlimited tariff to satisfy the high data demand rate, as a result 4G profits would peak at 20Mbps/km$^2$ and become saturated beyond this demand rate.

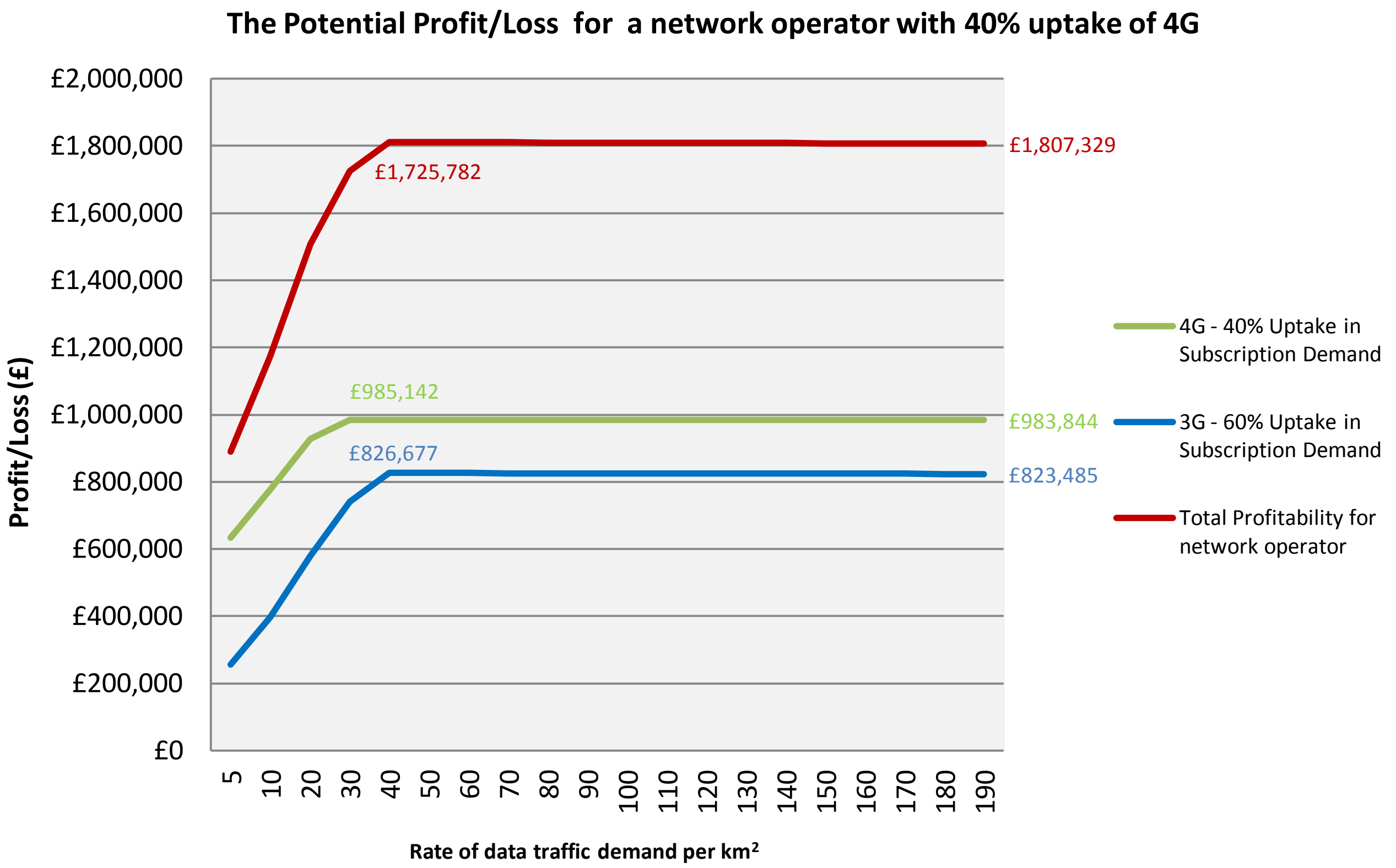

Figure 18 Profit/Loss for network operator with increase in demanded traffic rate for 40% uptake in 4G services

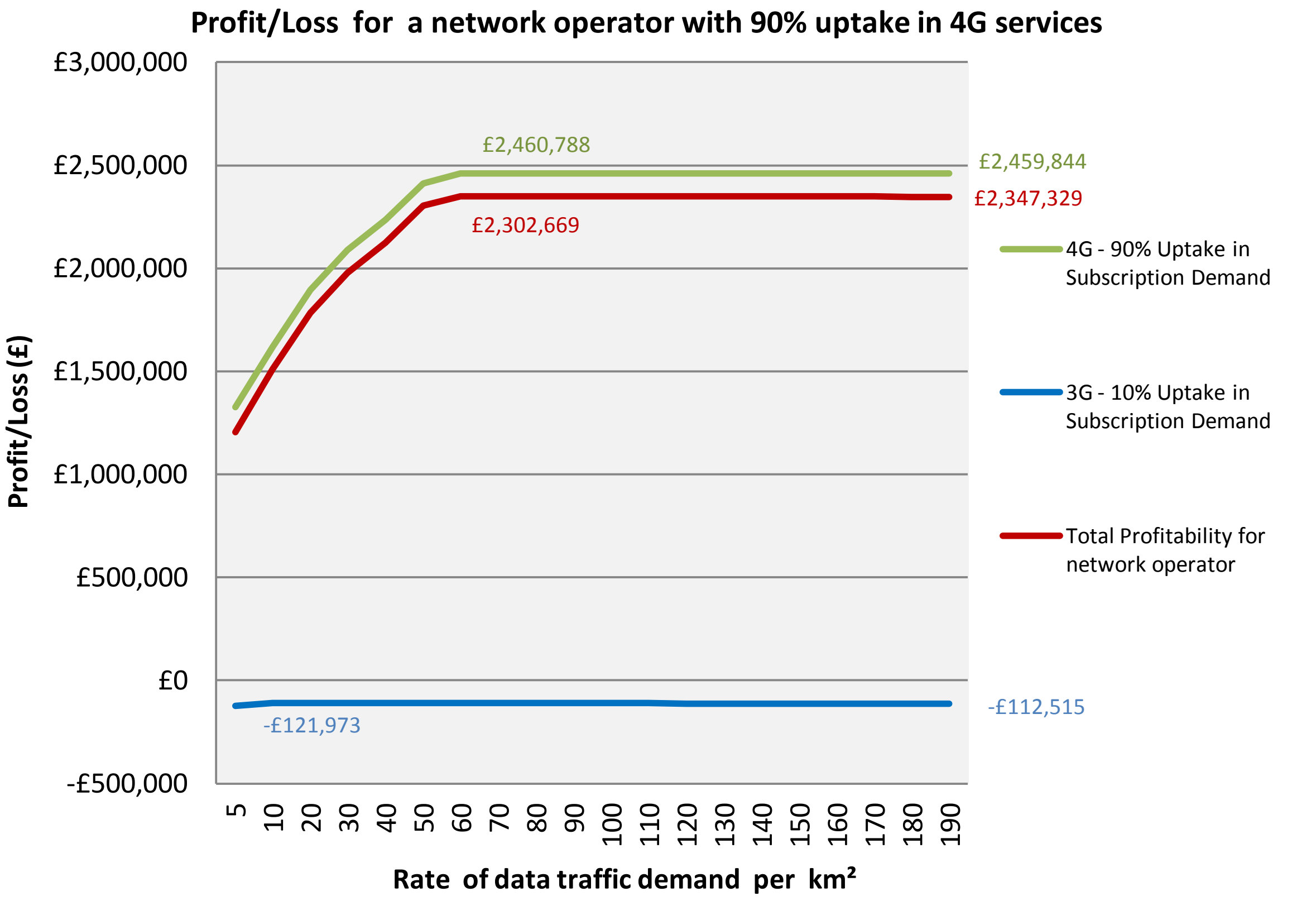

Figure 17 Profit/Loss for network operator with increase in demanded traffic rate for 90% uptake in 4G services

*Figures 17 & 18* show the total profitability for a network operator with 40 and 90 percent uptake in 4G services respectively. The results show that 4G will become more profitable than 3G for the first time when user uptake in 4G services approaches 40%. However with traffic demand beyond 30Mbps/km$^2$, only 'unlimited' tariffs would suffice in satisfying users given the current tariff strategy. *Figure 18* shows the profitability for a network operator at 90% uptake in 4G services. This is to model the profitability of network operators per km$^2$ at a maximum 4G user uptake. Based on the current tariff strategy, results shows that a peak profit in the range of £2.4 million is anticipated, however the results also show at that at such a high percentage uptake of 4G, profitability of 3G will cease to exist.

### 6.3.1 User Trends

The previous section discussed the profitability of operating a 4G network. The profit analysis conducted assumed that the mobile tariffs offered by network operators and therefore used in the analysis would be accepted and paid by subscribers. This section explores the research justifying the assumption made.

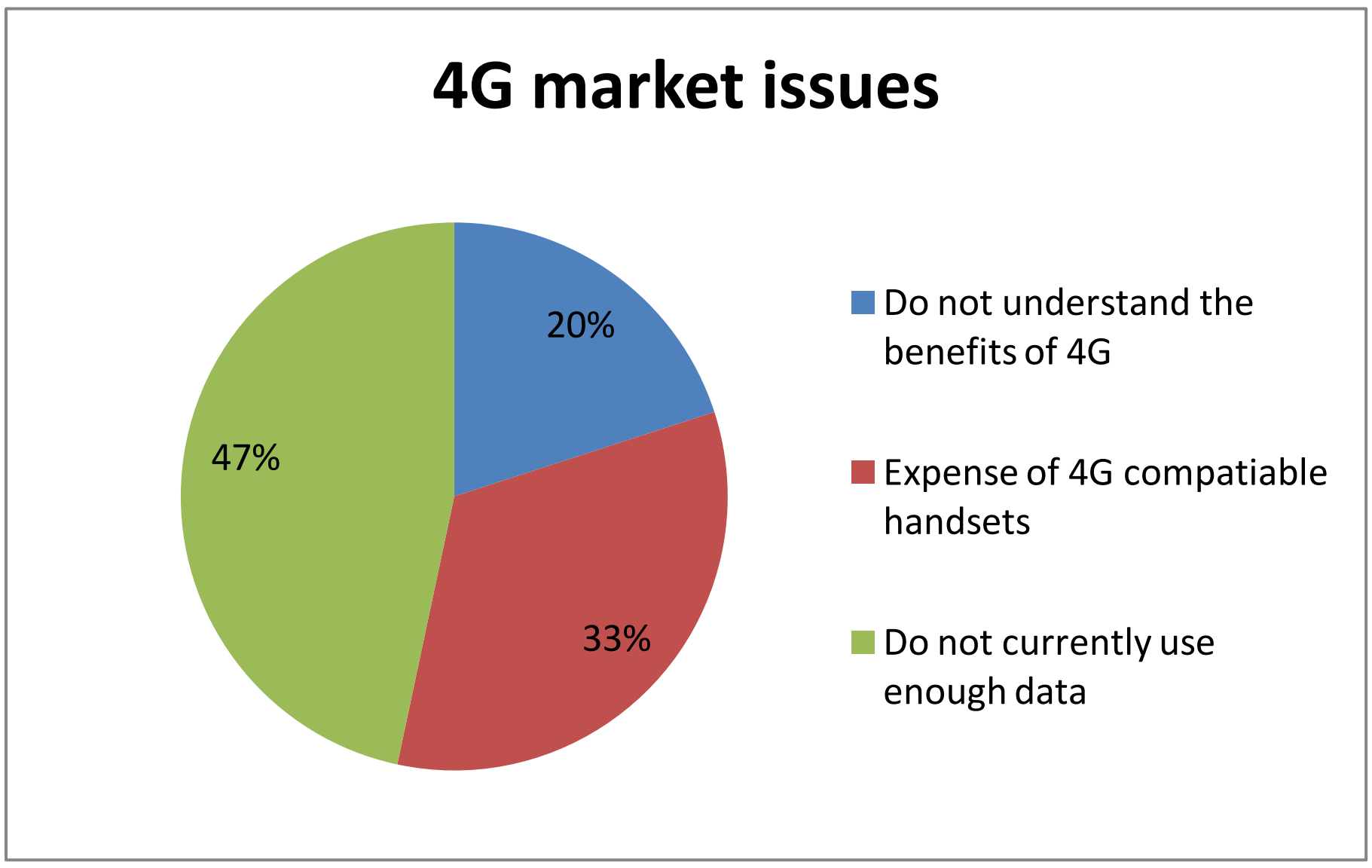

Figure 19 Mintel Oxygen exclusive 4G market issues survey

The figure above shows the response of mobile users to a public survey conducted regarding 4G services (42). From the results of the survey it can be seen that a majority of users said they would not upgrade to 4G as currently they do not use enough data. However this suggests that should user data usage increase, with regards to the research conducted by Cisco (3) suggests that a majority of users would consider the uptake of 4G services. A fifth of users surveyed said that the benefits of 4G were not understood; as a consequence these users would not consider switching to the 4G service. A different survey in the same research also found that users who understood the benefits of 4G would only consider upgrading to the service should 'unlimited' tariffs be offered. It should be noted that cost of tariffs was not viewed as a primary concern to users when considering 4G services. Research conducted in (42) also suggested that eventually a "vast majority" would switch to 4G services in the UK.

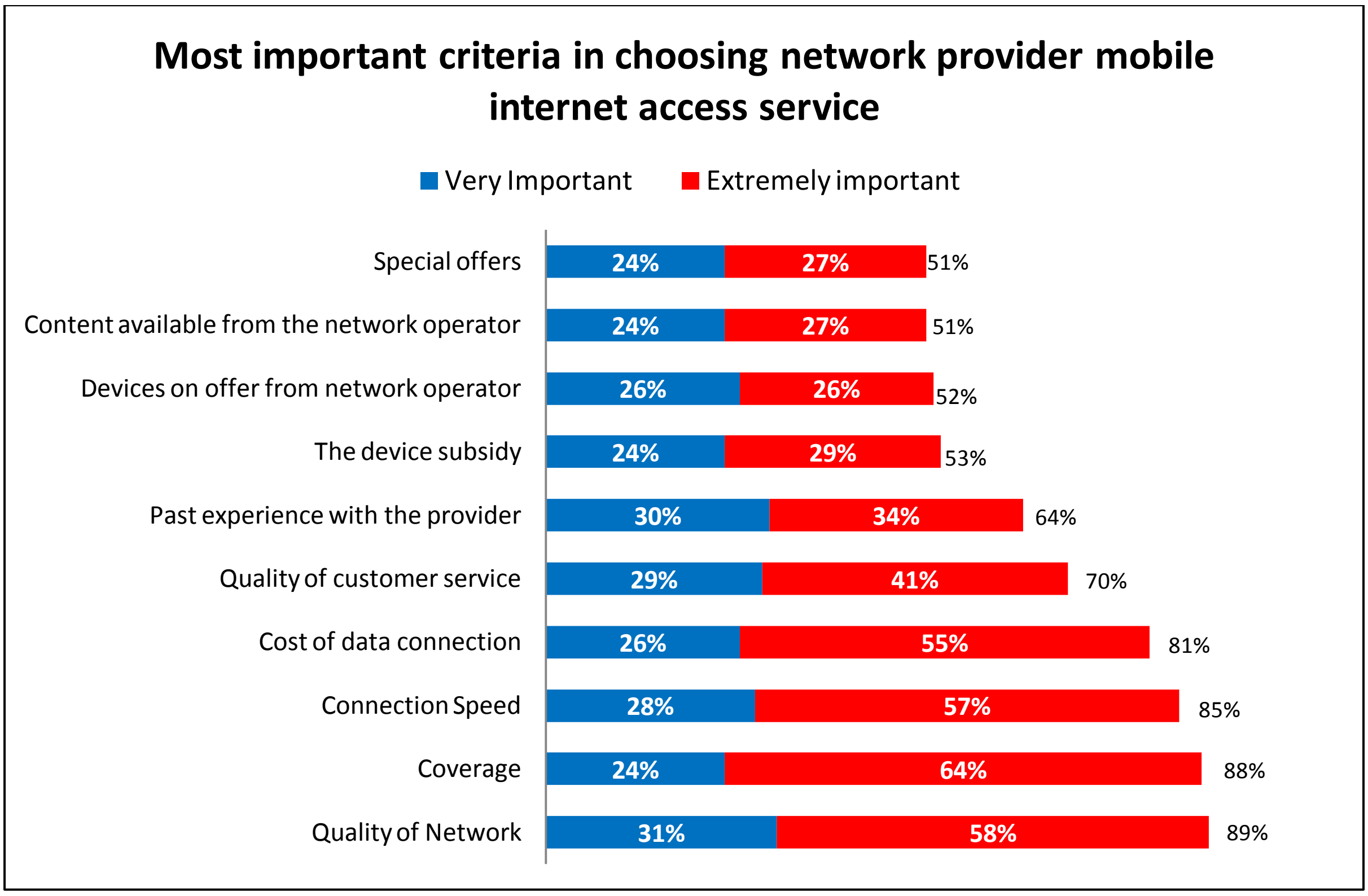

Figure 20 Shows factors affecting internet access service

*Figure 20* shows the views of mobile users as part of the market research carried out in (43). The research clearly shows the three most important factors to mobile users are quality of network, coverage and connection speed. Given that the quality of service in LTE is greater than HSPA and will meet any current and future needs (44) and the coverage of LTE is expected to be 99% of the UK by 2017 (25), it can be seen that the most important criteria in choosing network operator internet access will be met by LTE. The research also indicates that although *cost of data connection* (tariff costs) is the fourth most important criterion for users, other more important criteria for which 4G is a more proficient technology exists.

### *6.3.2 Summary of Profit Analysis & User trends*

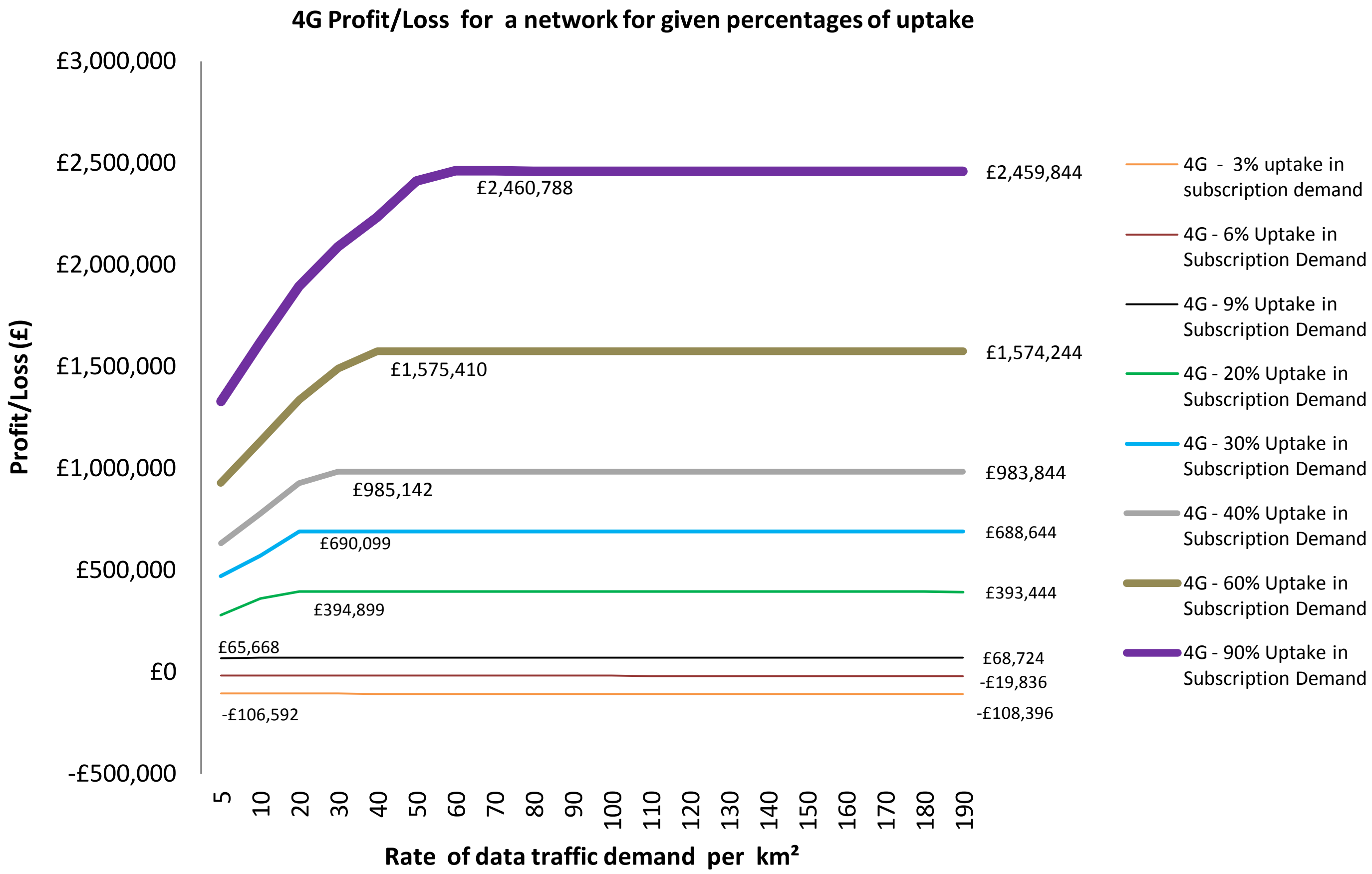

Figure 21 projected 4G profitability with increase in user percentage uptake

The profit analysis and user trends analysis has shown the potential profitability of 4G networks and key issues regarding the uptake of 4G services. The key findings of the profit and user trends analysis are as follows:

- As user uptake increases the profitability of 4G services will rise (*figure 21)*. Profitability however will occur when 4G uptake approaches 9% and beyond.
- Current offers of ‘unlimited’ tariffs will result in saturated profits as future high data usage can only be satisfied with current ‘unlimited’ data tariffs. With high data demands only satisfied with ‘unlimited’ tariffs and OPEX costs increasing with demanded traffic rate, network operators will see profits ‘shrink’ at high traffic data rates as revenue per subscriber will be limited by offering ‘unlimited’ data tariffs.
- User trends shows demand for 4G services exists. The most important criterions for user when choosing an internet access service are network quality, coverage and connection speed. Although cost of mobile internet access is considered the fourth most important criterion.
- The vast majority of users will eventually uptake 4G services; however current mobile users with understanding of 4G will only consider upgrading to 4G services which offer ‘unlimited’ data as part of the tariff.

## 6.4 Carbon Emission Analysis

From the results obtained with reference to energy consumption, it can be concluded that the energy consumption per BS for 4G LTE technology is less than that of 3G HSPA technology. With consideration to literature reviewed and with mobile operators committing to reducing their carbon footprint (18), an estimation of the $CO_2$ emissions of 4G and 3G technologies was produced in order to determine the effect of each technology on the environment.

| Mobile Technology | Annual Energy Consumption per km$^2$ (kWh) |
|---|---|
| **4G** | 51,445 |
| **3G** | 138,677 |

Table 13: The maximum annual energy consumption for 4G & 3G cells per km$^2$ for a demanded maximum traffic rate of 190 Mbps/km$^2$

*Table 14* shows the different power generation plants and their utilisation percentage in producing electricity for the UK. Using the values in *Table 13* and applying the percentages of fuels employed in the UK with their associated $CO_2$ production, one can obtain the annual $CO_2$ emission for the maximum demanded traffic rate of 190 Mbps/km$^2$ for 4G and 3G (*Table 14* column 4 & 5)[17].

| Types of Fuel | % employed in the UK | Grams of $CO_2$ produced per kWh | | 4G $CO_2$ emission (tonnes) | 3G $CO_2$ emission (tonnes) |
|---|---|---|---|---|---|
| Coal | 30 % | 960 | | **14.82** | **39.94** |
| Gas | 40 % | 443 | | **9.11** | **24.65** |
| Nuclear | 19 % | 66 | | **0.65** | **1.74** |
| Renewable | 9% | 11 | | **0.05** | **0.14** |
| Other fuels | 2 % | 25 | | **0.03** | **0.07** |
| **Annual $CO_2$ emission per km$^2$ (tonnes)** | | | | **24.66** | **66.54** |

Table 14: The quantity of CO2 produced using various energy sources when operating to meet maximum traffic demand (39) (40)

From *table 14* it can be noted that the annual $CO_2$ emission per km$^2$ for 4G and 3G is 24.66 and 66.54 tonnes, respectively. This shows that the migration from 3G to 4G can produce a 63% reduction in $CO_2$ emissions with a potential 41.88 tonnes of $CO_2$ prevented from being released into the atmosphere. Given the deployment ratio of 3G BS to 4G BS per km$^2$, the information in *table 14* can be applied in context to network operators for a given number of BS's in the UK. *Table 15* summarises the expected $CO_2$ emissions for the network operator Vodafone UK.

[17]*Table 2* defines 'renewables' as energy sources: Wind (on-shore and off-shore), Hydroelectric (reservoir and run-of-river) and solar thermal; where the average $CO_2$ production (in grams) of the three sources is assumed. Other fuel is largely defined as Bio gas and Biomass where the average $CO_2$ production (in grams) of the two sources is assumed.

| Vodafone UK | 4G $CO_2$ emission (tonnes) | 3G $CO_2$ emission (tonnes) |
|---|---|---|
| **Total $CO_2$ emission for Vodafone BS[18]** | **39 135** | **105 619** |

Table 15 Total annual carbon emission from Vodafone BS's using 4G & 3G technology

In the year 2010/2011 Vodafone UK emitted 159,000 tonnes of $CO_2$ as shown by *figure 22*. Based on the information published by Vodafone UK (8) the $CO_2$ emission from a direct result of energy consumption was 144, 402 tonnes. This includes the energy consumption from Vodafone BS's and all other aspects of the business such ICT equipment, offices and retail outlets. From the information calculated in *Table 15* the $CO_2$ emissions from BS's can be reduced by 66, 484 tonnes by Vodafone UK by migrating to 4G from their current 3G network. This can then further reduce their overall $CO_2$ emissions from 159, 000 tonnes to 92 516 tonnes, a potential maximum carbon emission saving of 42%. The migration from 3G to 4G can therefore have a huge implication on the carbon emission targets of network operators.

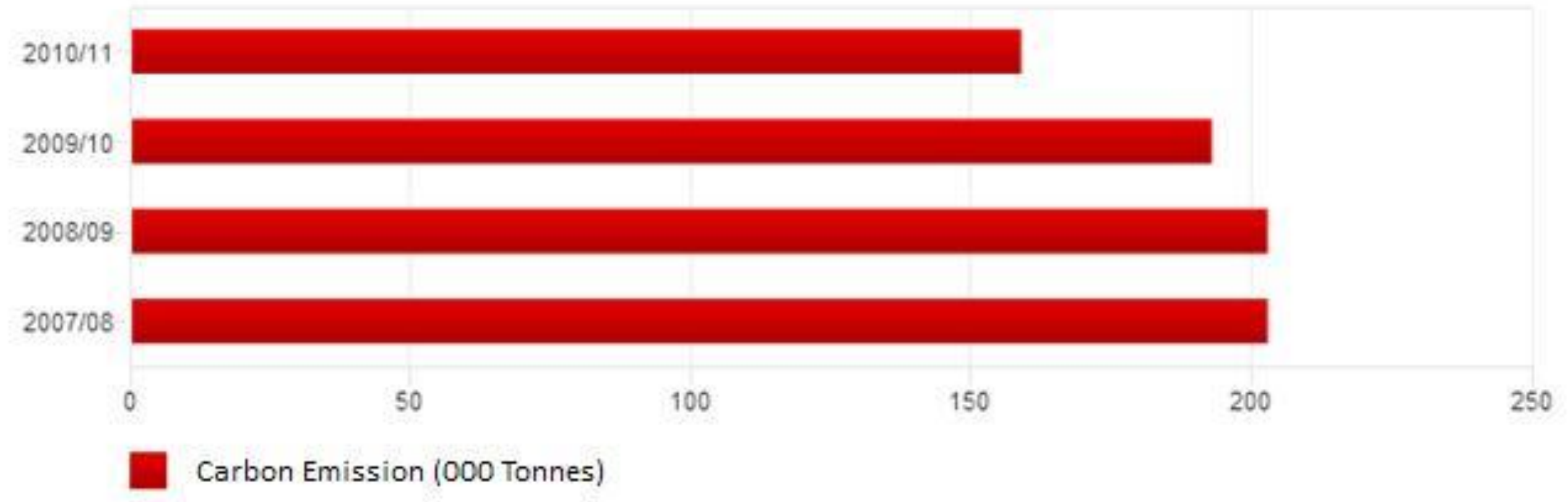

Figure 22 Vodafone UK total $CO_2$ emission from energy consumption. (45)

Although this study relates $CO_2$ emission in context to Vodafone, the findings can be applied to all network operators in the industry. One can conclude that from an environmental perspective the switch to 4G LTE network services will be beneficial for all stakeholders involved. The government will benefit as the decrease in emissions from the mobile telecommunication industry will help contribute to the overall reduction in $CO_2$ production by the UK. Operators will benefit as the growing pressure from consumers for 'green tariffs' which either use more renewable energy sources, or produce less $CO_2$ can be introduced with a switch to 4G and will therefore be seen as a positive for consumers. Growing emphasis on corporate responsibility and sustainability being a key issue, lower levels of $CO_2$ emission will aid in increasing a network provider's profile in 'going green', while certain experts also suggest it can form a competitive advantage. (46) Operators can also take advantage of the global 'carbon credit' trading schemes (47). Carbon credits place a monitoring price on polluting activities; as a result the businesses can earn money for their reduction in carbon emissions. By reducing carbon emissions below industry standards businesses can gain carbon credits which can then be sold to other businesses or governments. One carbon credit is equal to one tonne of carbon and is currently valued at approximately £14 per tonne. (48)

[18] Estimates for Vodafone $CO_2$ emissions based on Vodafone's current network of 20, 000 BS's

## 7.0 Conclusions

The project has demonstrated how greater spectral efficiency in 4G has resulted in higher capacity gains in comparison to 3G. As a result the number of cells required per $km^2$ to deliver the traffic rate demanded can be reduced. Power consumption analysis revealed that a network operator using 4G services can reduce their power consumption by 63% compared to 3G. While OPEX analysis showed that cost per $km^2$ for network operators can be reduced by 34% with the switch to 4G services. Carbon emission analysis of 3G & 4G services revealed, 41.88 tonnes of $CO_2$ production could be prevented, resulting in direct benefits for government, users and network operators.

The project was able to demonstrate the profitability of 4G, showing network operators to be profitable beyond 9% uptake in 4G services. Given the current tariffs available, high data demands however would result in saturated profits. Research into user trends found that eventually the vast majority in the UK would upgrade to 4G services given that, 4G is able fulfil the most important internet access criterions for users. An increase of uptake in 4G will require educating consumers in the benefits of 4G as 20% of users are still unaware of 4G benefits.

In conclusion, one can determine that with consumers wanting unlimited data and network operators wanting to maximise profits a trade-off must be made between user satisfaction and network operator profitability. The results show that the network operators will be profitable with the deployment of 4G with profits exceeding that of 3G. Network operators have the opportunity to implement two tariff strategies; offer unlimited data packages to consumers with focus to increase user uptake; or alternatively have tiered data packages (where unlimited data tariffs is not an option) and therefore a strategy to maximise profits. However research suggests that such data packages will deter users from upgrading to 4G services and could further result in environmental benefits not being realised. Network operators should therefore offer ‘unlimited’ data packages as tariffs and consider the deployment of 4G as a technology that enhances the user experience rather than a technology that could enhance network operator profitability.

## 8.0 Recommendation for Further Work

The research project focused on the profitability and environmental impact of 3rd generation HSPA and 4th generation LTE technologies in dense urban areas.

Research into technological, economic and environmental impact of telecommunication technologies still remains understudied and further research is required to fully understand the subject matter.

A continuation of this study is essential and should advance on the research into profitability with consideration to increase in system capacity. Femtocells offer network operators a method of increasing indoor coverage and data throughput via installed wireless data access points inside homes. Femtocells are advantageous in overcoming signal loss ratio from outdoor to indoor, providing voice and data coverage in homes where mobile signals are not received due to the physics of EM waves. Furthermore, where a large number of users are requesting data services, it is infeasible for macro cells to deliver due to limited transmission power and spectrum availability; femtocells however can deliver data services simultaneously over the entire bandwidth, providing massive spatial spectral efficiency improvements. Further research into the profitability of 4G Macro & femto cells can therefore prove beneficial to both consumers and network operators. Useful papers are *Femtocell Networks: A survey* (49) and *A study of femtocells for LTE* (50).

Developments in reducing carbon emission with implementation of sleep mode mechanisms could prove beneficial. Sleep Mode is when no active users are present in the cell and its sectors, and the cell enters sleep mode. Although findings show that power consumption saving in macro BS's are negligible, power consumption in small cells (micro) BS's can be reduced by 48%, 84% and 96% for heavy, medium and light traffic loads respectively. If active participation from network operators is possible, the number of small cells operated by network providers could provide research grounds to investigate if implementation of Sleep Mode Mechanism could be environmentally beneficial enough for capital investment. Useful papers for further research *Green Cellular Network: Deployment Solutions* (51)

Additional research in profitability analysis of Urban, suburban and rural areas could also be beneficial. Network operators could collaborate and provide the commercial sensitive data required to calculate profitability and carbon emissions more accurately relating to BS's across the UK.

## 9.0 Project Costing

The project undertaken was purely theoretical and therefore cost involved in designing and build were not applicable. The cost associated with the projected can be summarised by the following table.

| Cost element | Cost per element | Total time used (hours) | Total cost of element |
|---|---|---|---|
| Meetings with Dr Guo | £50.00 | 14 | £700 |
| Computational analysis | £15.00 | 200 | £3,000 |
| Project report | £15.00 | 80 | £1,200 |
| Total Project Costs: | | | **£4,900** |

The total cost estimated for this project is £4,900. Although the project was conducted for academic purposes, the cost can be justified by its benefit to industry, consumers and government alike. The project has demonstrated factors relating to profitability and therefore of interest to the telecommunications industry; factors relating carbon emissions and therefore of interest to government; and finally, factors relating to tariff strategy to increase user satisfaction and therefore of interest to consumers. The cost is therefore justified as the project will benefit all key stakeholders of the telecommunication industry.

# 9.0 Appendices

## Appendix A.1 & A.2 – 3G & 4G Mobile Tariffs[19]

**Appendix A.1: 3G Tariffs offered by Network providers**

| Minutes | Data Allowance (GB) | Total Cost (£) | Notes (Operator) |
|---|---|---|---|
| 50 | 0.25 | 7 | T-Mobile UK 2012 |
| 100 | 0.25 | 15.5 | T-Mobile UK 2012 |
| 300 | 0.25 | 21 | T-Mobile UK 2012 |
| 50 | 1.5 | 21 | T-Mobile UK 2012 |
| 100 | 0.75 | 21 | T-Mobile UK 2012 |
| 100 | 1.5 | 26 | T-Mobile UK 2012 |
| 300 | 0.75 | 26 | T-Mobile UK 2012 |
| 600 | 0.25 | 26 | T-Mobile UK 2012 |
| 900 | 0.25 | 31 | T-Mobile UK 2012 |
| 600 | 0.75 | 31 | T-Mobile UK 2012 |
| 300 | 1.5 | 31 | T-Mobile UK 2012 |
| 900 | 0.75 | 36 | T-Mobile UK 2012 |
| 1200 | 0.25 | 36 | T-Mobile UK 2012 |
| 600 | 1.5 | 36 | T-Mobile UK 2012 |
| 50 | 0.25 | 15.5 | Orange UK 2012 |
| 100 | 0.5 | 20.5 | Orange UK 2012 |
| 200 | 0.75 | 26 | Orange UK 2012 |
| 400 | 0.75 | 31 | Orange UK 2012 |
| 600 | 1 | 36 | Orange UK 2012 |
| 900 | 1 | 41 | Orange UK 2012 |
| 1200 | 1 | 46 | Orange UK 2012 |
| 30 | 0.1 | 10.5 | Orange UK 2012 |
| 100 | 0.1 | 15.5 | Orange UK 2012 |
| 400 | 0.25 | 26 | Orange UK 2012 |
| 600 | 0.25 | 31 | Orange UK 2012 |
| 100 | 0.1 | 13 | Vodafone UK 2012 |
| 300 | 0.25 | 18 | Vodafone UK 2012 |
| 600 | 0.5 | 23 | Vodafone UK 2012 |
| Unlimited | 1 | 20.5 | Vodafone UK 2012 |
| Unlimited | 2 | 26 | Vodafone UK 2012 |
| 100 | 0.5 | 16.5 | O2 UK 2012 |
| 300 | 0.5 | 21.5 | O2 UK 2012 |
| 600 | 0.5 | 27 | O2 UK 2012 |
| 900 | 0.5 | 32 | O2 UK 2012 |
| 1200 | 0.75 | 37 | O2 UK 2012 |

[19] The data shown in appendices A1 & A2 was used to form the Multiple Linear Model regression used in the project. It should be noted that Unlimited minutes were modelled as 2000 minutes and Unlimited data as 25GB for reasons explained in *Section 6.*

| | | | |
|---|---|---|---|
| 153 | 0.2 | 13.08 | Vodafone Aus 2012 |
| 254 | 0.5 | 19.61 | Vodafone Aus 2012 |
| 254 | 0.5 | 19.61 | Vodafone Aus 2012 |
| 424 | 0.75 | 22.88 | Vodafone Aus 2012 |
| 424 | 0.75 | 22.88 | Vodafone Aus 2012 |
| 593 | 1 | 29.42 | Vodafone Aus 2012 |
| Unlimited | 2 | 42.50 | Vodafone Aus 2012 |
| 450 | 25 | 49.91 | Sprint USA 2012 |
| 900 | Unlimited | 62.39 | Sprint USA 2012 |
| Unlimited | Unlimited | 68.63 | Sprint USA 2012 |
| Unlimited | 0.3 | 24.96 | Verizonwireless USA 2012 |
| Unlimited | 1 | 31.20 | Verizonwireless USA 2012 |
| Unlimited | 2 | 37.43 | Verizonwireless USA 2012 |
| Unlimited | 4 | 43.67 | Verizonwireless USA 2012 |
| Unlimited | 6 | 49.91 | Verizonwireless USA 2012 |
| Unlimited | 8 | 56.15 | Verizonwireless USA 2012 |
| Unlimited | 10 | 62.39 | Verizonwireless USA 2012 |
| Unlimited | 12 | 68.63 | Verizonwireless USA 2012 |
| Unlimited | 14 | 74.87 | Verizonwireless USA 2012 |
| Unlimited | 16 | 81.11 | Verizonwireless USA 2012 |
| Unlimited | 18 | 87.35 | Verizonwireless USA 2012 |
| Unlimited | 20 | 93.59 | Verizonwireless USA 2012 |
| 450 | 0.3 | 37.43 | AT&T Mobile share 2012 |
| 900 | 3 | 56.15 | AT&T Mobile share 2012 |
| Unlimited | 5 | 74.87 | AT&T Mobile share 2012 |

**Appendix A.2: 4G Tariffs offered by Network providers**

| Minutes | Data Allowance (GB) | Total Cost (£) | Notes (Operator) |
|---|---|---|---|
| Unlimited | 0.5 | 36 | EE UK 2012 |
| Unlimited | 1 | 41 | EE UK 2012 |
| Unlimited | 3 | 46 | EE UK 2012 |
| Unlimited | 5 | 51 | EE UK 2012 |
| Unlimited | 8 | 56 | EE UK 2012 |
| *N/A Launch of Vodafone and 02 4G services in the UK expected in late 2013* | | | Vodafone UK 2012 |
| | | | Vodafone UK 2012 |
| | | | Vodafone UK 2012 |
| | | | Vodafone UK 2012 |
| | | | Vodafone UK 2012 |
| | | | O2 UK 2012 |
| | | | O2 UK 2012 |
| | | | O2 UK 2012 |
| | | | O2 UK 2012 |
| | | | O2 UK 2012 |
| 169 | 0.2 | 19.61 | Optus Aus 2012 |
| 169 | 0.2 | 22.88 | Optus Aus 2012 |
| 424 | 1 | 32.69 | Optus Aus 2012 |
| 551 | 1.5 | 39.23 | Optus Aus 2012 |
| 720 | 2 | 52.30 | Optus Aus 2012 |
| Unlimited | 3 | 64.73 | Optus Aus 2012 |
| Unlimited | 4 | 84.34 | Optus Aus 2012 |
| 508 | 1 | 39.23 | Telstra Aus 2012 |
| 678 | 1.5 | 52.30 | Telstra Aus 2012 |
| 763 | 2 | 58.84 | Telstra Aus 2012 |
| Unlimited | 3 | 85.00 | Telstra Aus 2012 |
| Unlimited | 0.3 | 24.96 | Verizonwireless USA 2012 |
| Unlimited | 1 | 31.20 | Verizonwireless USA 2012 |
| Unlimited | 2 | 37.43 | Verizonwireless USA 2012 |
| Unlimited | 4 | 43.67 | Verizonwireless USA 2012 |
| Unlimited | 6 | 49.91 | Verizonwireless USA 2012 |
| Unlimited | 8 | 56.15 | Verizonwireless USA 2012 |
| Unlimited | 10 | 62.39 | Verizonwireless USA 2012 |
| Unlimited | 12 | 68.63 | Verizonwireless USA 2012 |
| Unlimited | 14 | 74.87 | Verizonwireless USA 2012 |
| Unlimited | 16 | 81.11 | Verizonwireless USA 2012 |
| Unlimited | 18 | 87.35 | Verizonwireless USA 2012 |
| Unlimited | 20 | 93.59 | Verizonwireless USA 2012 |
| Unlimited | 1 | 24.956 | AT&T Mobile share 2012 |
| Unlimited | 4 | 43.673 | AT&T Mobile share 2012 |
| Unlimited | 6 | 56.151 | AT&T Mobile share 2012 |
| Unlimited | 10 | 74.868 | AT&T Mobile share 2012 |
| Unlimited | 15 | 99.824 | AT&T Mobile share 2012 |
| Unlimited | 20 | 124.78 | AT&T Mobile share 2012 |
| 450 | 25 | 56.15 | Sprint USA 2012 |
| 900 | Unlimited | 68.63 | Sprint USA 2012 |
| Unlimited | Unlimited | 81.11 | Sprint USA 2012 |

## Appendix A.3– Regression Model Matlab Code

```
Tariff_Data = xlsread('Tariff Data analysis.xls','A1:C62');
Minutes_Data = Tariff_Data(:,1);
Packet_Data = Tariff_Data(:,2)*1000;
Cost_Data = Tariff_Data(:,3);

% 3G Regression Fit

X = [ones(size(Minutes_Data)) log(Minutes_Data) log(Packet_Data) Minutes_Data
Packet_Data];
b = regress(Cost_Data,X); % Removes NaN data

% 3G Scatter Plot
```

```
scatter3(Minutes_Data,Packet_Data,Cost_Data,'filled'); hold on;
x1fit = min(Minutes_Data):25:max(Minutes_Data);
x2fit = min(Packet_Data):25:max(Packet_Data);
[X1FIT,X2FIT] = meshgrid(x1fit,x2fit);
YFIT = b(1) + b(2)*log(1+X1FIT) + b(3)*log(1+X2FIT) + b(4)*X1FIT + b(5)*X2FIT;
mesh(X1FIT,X2FIT,YFIT)
xlabel('Call Allowance (Min)')
ylabel('Data Allowance (Mb)')
zlabel('Montly Contract (£)')

Tariff_Data = xlsread('Tariff Data analysis.xls','A67:C113');
Minutes_Data = Tariff_Data(:,1);
Packet_Data = Tariff_Data(:,2)*1000;
Cost_Data = Tariff_Data(:,3);

% 4G Regression Fit

X = [ones(size(Minutes_Data)) log(Minutes_Data) log(Packet_Data) Minutes_Data Packet_Data];
b = regress(Cost_Data,X); % Removes NaN data

% 4G Scatter Plot

scatter3(Minutes_Data,Packet_Data,Cost_Data,'filled'); hold on;
x1fit = min(Minutes_Data):25:max(Minutes_Data);
x2fit = min(Packet_Data):25:max(Packet_Data);
[X1FIT,X2FIT] = meshgrid(x1fit,x2fit);
YFIT = b(1) + b(2)*log(1+X1FIT) + b(3)*log(1+X2FIT) + b(4)*X1FIT + b(5)*X2FIT;
mesh(X1FIT,X2FIT,YFIT)
xlabel('Call Allowance (Min)')
ylabel('Data Allowance (Mb)')
zlabel('Montly Contract (£)')
```

# Appendix B

## B.1- 4G OPEX costs per km²

4G OPEX per km²

| Cost elements per cell | Traffic rate demanded Mbits/km² | | | | | | | | | | | | | | | | | | | |
|---|---|---|---|---|---|---|---|---|---|---|---|---|---|---|---|---|---|---|---|---|
| | 5 | 10 | 20 | 30 | 40 | 50 | 60 | 70 | 80 | 90 | 100 | 110 | 120 | 130 | 140 | 150 | 160 | 170 | 180 | 190 |
| Electrcity Bill/km² | £5,496 | £5,640 | £5,845 | £6,002 | £6,134 | £6,251 | £6,357 | £6,454 | £6,544 | £6,629 | £6,709 | £6,785 | £6,858 | £6,928 | £6,995 | £7,060 | £7,123 | £7,184 | £7,243 | £51,445 |
| Electrcity Bill | £1,195 | £1,226 | £1,271 | £1,305 | £1,334 | £1,359 | £1,382 | £1,403 | £1,423 | £1,441 | £1,458 | £1,475 | £1,491 | £1,506 | £1,521 | £1,535 | £1,548 | £1,562 | £1,574 | £1,587 |
| Backhaul Rent | £7,500 | £7,500 | £7,500 | £7,500 | £7,500 | £7,500 | £7,500 | £7,500 | £7,500 | £7,500 | £7,500 | £7,500 | £7,500 | £7,500 | £7,500 | £7,500 | £7,500 | £7,500 | £7,500 | £7,500 |
| Site Rental | £10,800 | £10,800 | £10,800 | £10,800 | £10,800 | £10,800 | £10,800 | £10,800 | £10,800 | £10,800 | £10,800 | £10,800 | £10,800 | £10,800 | £10,800 | £10,800 | £10,800 | £10,800 | £10,800 | £10,800 |
| Maintance | £3,900 | £3,900 | £3,900 | £3,900 | £3,900 | £3,900 | £3,900 | £3,900 | £3,900 | £3,900 | £3,900 | £3,900 | £3,900 | £3,900 | £3,900 | £3,900 | £3,900 | £3,900 | £3,900 | £3,900 |
| Marketing Cost | £3,257 | £3,257 | £3,257 | £3,257 | £3,257 | £3,257 | £3,257 | £3,257 | £3,257 | £3,257 | £3,257 | £3,257 | £3,257 | £3,257 | £3,257 | £3,257 | £3,257 | £3,257 | £3,257 | £3,257 |
| CAPEX per cell | £139,795 | £139,795 | £139,795 | £139,795 | £139,795 | £139,795 | £139,795 | £139,795 | £139,795 | £139,795 | £139,795 | £139,795 | £139,795 | £139,795 | £139,795 | £139,795 | £139,795 | £139,795 | £139,795 | £139,795 |
| CAPEX repayment per cell over 12 years | £12,553 | £12,553 | £12,553 | £12,553 | £12,553 | £12,553 | £12,553 | £12,553 | £12,553 | £12,553 | £12,553 | £12,553 | £12,553 | £12,553 | £12,553 | £12,553 | £12,553 | £12,553 | £12,553 | £12,553 |
| **4G Total OPEX/ km²** | £195,152.00 | £195,296.72 | £195,501.39 | £195,658.44 | £195,790.84 | £195,907.48 | £196,012.94 | £196,109.92 | £196,200.18 | £196,284.96 | £196,365.14 | £196,441.41 | £196,514.28 | £196,584.17 | £196,651.42 | £196,716.32 | £196,779.08 | £196,839.91 | £196,898.97 | £196,956.42 |

The tables seen above are the 4G OPEX costs relating to *section 6.2* in the report.

## B.2- 3G OPEX costs per km²

| 3G OPEX per km² | | | | | | | | | |
|---|---|---|---|---|---|---|---|---|---|
| Cost elements/cell | Traffic rate demanded Mbits/km² | | | | | | | | |
| | 5 | 10 | 20 | 30 | 40 | 50 | 60 | 70 | 80 |
| Electrcity Bill/km² | £14,814 | £15,204 | £15,756 | £16,179 | £16,536 | £16,851 | £17,135 | £17,396 | £17,640 |
| Electrcity Bill | £1,195 | £1,226 | £1,271 | £1,305 | £1,334 | £1,359 | £1,382 | £1,403 | £1,423 |
| Backhaul Rent | £7,500 | £7,500 | £7,500 | £7,500 | £7,500 | £7,500 | £7,500 | £7,500 | £7,500 |
| Site Rental | £10,800 | £10,800 | £10,800 | £10,800 | £10,800 | £10,800 | £10,800 | £10,800 | £10,800 |
| Maintance | £3,900 | £3,900 | £3,900 | £3,900 | £3,900 | £3,900 | £3,900 | £3,900 | £3,900 |
| 3G Total OPEX/ km² | £294,773.16 | £295,169.58 | £295,730.20 | £296,160.38 | £296,523.03 | £296,842.54 | £297,131.40 | £297,397.03 | £297,644.27 |

| 3G OPEX per km² | | | | | | | | | | |
|---|---|---|---|---|---|---|---|---|---|---|
| Traffic rate demanded Mbits/km131 | | | | | | | | | | |
| 90 | 100 | 110 | 120 | 130 | 140 | 150 | 160 | 170 | 180 | 190 |
| £17,868 | £18,084 | £18,290 | £18,486 | £18,675 | £18,856 | £19,031 | £19,200 | £19,364 | £19,523 | £138,677 |
| £1,441 | £1,458 | £1,475 | £1,491 | £1,506 | £1,521 | £1,535 | £1,548 | £1,562 | £1,574 | £1,587 |
| £7,500 | £7,500 | £7,500 | £7,500 | £7,500 | £7,500 | £7,500 | £7,500 | £7,500 | £7,500 | £7,500 |
| £10,800 | £10,800 | £10,800 | £10,800 | £10,800 | £10,800 | £10,800 | £10,800 | £10,800 | £10,800 | £10,800 |
| £3,900 | £3,900 | £3,900 | £3,900 | £3,900 | £3,900 | £3,900 | £3,900 | £3,900 | £3,900 | £3,900 |
| £297,876.49 | £298,096.13 | £298,305.03 | £298,504.63 | £298,696.08 | £298,880.29 | £299,058.04 | £299,229.95 | £299,396.57 | £299,558.35 | £299,715.70 |

# Appendix C

Appendix C contains the profitability breakdown of all the profitability graphs shown in *section 6.3.*

| 3G - 94% Uptake in Subscription Demand | | | | | | | | | | | |
|---|---|---|---|---|---|---|---|---|---|---|---|
| Uptake of % | Rate of Traffic demand Mbps/km² | Subscribers uptake per km² | Mbps/subscriber | Mb/month/subscriber | GB/month/subscriber | Equivlent tarrif (GB) | Tarrif cost per month (to the closes .5 tarriff) | (Charge per subscriber x number of subscribers) x 12 months | Cost for cells per km² | Profit/Loss (Before Retail Expenditure & Tax) | Total Profitability for network operator |
| 94% | 5 | 2820 | 0.0018 | 2297.87 | 2.24 | 2.50 | £23.00 | £778,320.00 | £294,773.00 | £483,547 | £465,515 |
| 94% | 10 | 2820 | 0.0035 | 4595.74 | 4.49 | 4.50 | £28.50 | £964,440.00 | £295,169.00 | £669,271 | £651,095 |
| 94% | 20 | 2820 | 0.0071 | 9191.49 | 8.98 | 9.00 | £34.00 | £1,150,560.00 | £295,730.00 | £854,830 | £836,449 |
| 94% | 30 | 2820 | 0.0106 | 13787.23 | 13.46 | 13.50 | £39.50 | £1,336,680.00 | £296,160.00 | £1,040,520 | £1,021,982 |
| 94% | 40 | 2820 | 0.0142 | 18382.98 | 17.95 | 18.00 | £44.00 | £1,488,960.00 | £296,523.00 | £1,192,437 | £1,173,767 |
| 94% | 50 | 2820 | 0.0177 | 22978.72 | 22.44 | 22.50 | £49.00 | £1,658,160.00 | £296,824.00 | £1,361,336 | £1,342,549 |
| 94% | 60 | 2820 | 0.0213 | 27574.47 | 26.93 | Unlimited | £52.00 | £1,759,680.00 | £297,131.00 | £1,462,549 | £1,443,657 |
| 94% | 70 | 2820 | 0.0248 | 32170.21 | 31.42 | Unlimited | £52.00 | £1,759,680.00 | £297,397.00 | £1,462,283 | £1,443,294 |
| 94% | 80 | 2820 | 0.0284 | 36765.96 | 35.90 | Unlimited | £52.00 | £1,759,680.00 | £297,644.00 | £1,462,036 | £1,442,956 |
| 94% | 90 | 2820 | 0.0319 | 41361.70 | 40.39 | Unlimited | £52.00 | £1,759,680.00 | £297,876.00 | £1,461,804 | £1,442,640 |
| 94% | 100 | 2820 | 0.0355 | 45957.45 | 44.88 | Unlimited | £52.00 | £1,759,680.00 | £298,096.00 | £1,461,584 | £1,442,339 |
| 94% | 110 | 2820 | 0.0390 | 50553.19 | 49.37 | Unlimited | £52.00 | £1,759,680.00 | £298,305.00 | £1,461,375 | £1,442,054 |
| 94% | 120 | 2820 | 0.0426 | 55148.94 | 53.86 | Unlimited | £52.00 | £1,759,680.00 | £298,504.00 | £1,461,176 | £1,441,782 |
| 94% | 130 | 2820 | 0.0461 | 59744.68 | 58.34 | Unlimited | £52.00 | £1,759,680.00 | £298,696.00 | £1,460,984 | £1,441,520 |
| 94% | 140 | 2820 | 0.0496 | 64340.43 | 62.83 | Unlimited | £52.00 | £1,759,680.00 | £298,880.00 | £1,460,800 | £1,441,269 |
| 94% | 150 | 2820 | 0.0532 | 68936.17 | 67.32 | Unlimited | £52.00 | £1,759,680.00 | £299,058.00 | £1,460,622 | £1,441,026 |
| 94% | 160 | 2820 | 0.0567 | 73531.91 | 71.81 | Unlimited | £52.00 | £1,759,680.00 | £299,229.00 | £1,460,451 | £1,440,792 |
| 94% | 170 | 2820 | 0.0603 | 78127.66 | 76.30 | Unlimited | £52.00 | £1,759,680.00 | £299,396.00 | £1,460,284 | £1,440,565 |
| 94% | 180 | 2820 | 0.0638 | 82723.40 | 80.78 | Unlimited | £52.00 | £1,759,680.00 | £299,558.00 | £1,460,122 | £1,440,344 |
| 94% | 190 | 2820 | 0.0674 | 87319.15 | 85.27 | Unlimited | £52.00 | £1,759,680.00 | £299,715.00 | £1,459,965 | £1,440,129 |

| Uptake of % | Rate of Traffic demand Mbps/km² | Subscribers uptake per km² | Mbps/subscriber | Mb/month/subscriber | GB/month/subscriber | Equivlent tarrif (GB) | Tarrif cost per month | (Charge per subscriber x number of subscribers) x 12 months | Cost for cells per km² | Profit/Loss (Before Retail Expenditure & Tax) |
|---|---|---|---|---|---|---|---|---|---|---|
| 4G - 6% Uptake in Subscription Demand | | | | | | | | | | |
| 6% | 5 | 180 | 0.0278 | 36000.00 | 35.16 | Unlimited | £82.00 | £177,120.00 | £195,152.00 | -£18,032 |
| 6% | 10 | 180 | 0.0556 | 72000.00 | 70.31 | Unlimited | £82.00 | £177,120.00 | £195,296.00 | -£18,176 |
| 6% | 20 | 180 | 0.1111 | 144000.00 | 140.63 | Unlimited | £82.00 | £177,120.00 | £195,501.00 | -£18,381 |
| 6% | 30 | 180 | 0.1667 | 216000.00 | 210.94 | Unlimited | £82.00 | £177,120.00 | £195,658.00 | -£18,538 |
| 6% | 40 | 180 | 0.2222 | 288000.00 | 281.25 | Unlimited | £82.00 | £177,120.00 | £195,790.00 | -£18,670 |
| 6% | 50 | 180 | 0.2778 | 360000.00 | 351.56 | Unlimited | £82.00 | £177,120.00 | £195,907.00 | -£18,787 |
| 6% | 60 | 180 | 0.3333 | 432000.00 | 421.88 | Unlimited | £82.00 | £177,120.00 | £196,012.00 | -£18,892 |
| 6% | 70 | 180 | 0.3889 | 504000.00 | 492.19 | Unlimited | £82.00 | £177,120.00 | £196,109.00 | -£18,989 |
| 6% | 80 | 180 | 0.4444 | 576000.00 | 562.50 | Unlimited | £82.00 | £177,120.00 | £196,200.00 | -£19,080 |
| 6% | 90 | 180 | 0.5000 | 648000.00 | 632.81 | Unlimited | £82.00 | £177,120.00 | £196,284.00 | -£19,164 |
| 6% | 100 | 180 | 0.5556 | 720000.00 | 703.13 | Unlimited | £82.00 | £177,120.00 | £196,365.00 | -£19,245 |
| 6% | 110 | 180 | 0.6111 | 792000.00 | 773.44 | Unlimited | £82.00 | £177,120.00 | £196,441.00 | -£19,321 |
| 6% | 120 | 180 | 0.6667 | 864000.00 | 843.75 | Unlimited | £82.00 | £177,120.00 | £196,514.00 | -£19,394 |
| 6% | 130 | 180 | 0.7222 | 936000.00 | 914.06 | Unlimited | £82.00 | £177,120.00 | £196,584.00 | -£19,464 |
| 6% | 140 | 180 | 0.7778 | 1008000.00 | 984.38 | Unlimited | £82.00 | £177,120.00 | £196,651.00 | -£19,531 |
| 6% | 150 | 180 | 0.8333 | 1080000.00 | 1054.69 | Unlimited | £82.00 | £177,120.00 | £196,716.00 | -£19,596 |
| 6% | 160 | 180 | 0.8889 | 1152000.00 | 1125.00 | Unlimited | £82.00 | £177,120.00 | £196,779.00 | -£19,659 |
| 6% | 170 | 180 | 0.9444 | 1224000.00 | 1195.31 | Unlimited | £82.00 | £177,120.00 | £196,839.00 | -£19,719 |
| 6% | 180 | 180 | 1.0000 | 1296000.00 | 1265.63 | Unlimited | £82.00 | £177,120.00 | £196,898.00 | -£19,778 |
| 6% | 190 | 180 | 1.0556 | 1368000.00 | 1335.94 | Unlimited | £82.00 | £177,120.00 | £196,956.00 | -£19,836 |

**Table 16Profitability table showing profit calculation for 6% and 94% uptake in 4G and 3G subscriptions respectively**

| 4G - 9% Uptake in Subscription Demand | | | | | | | | | | |
|---|---|---|---|---|---|---|---|---|---|---|
| Uptake of % | Rate of Traffic demand Mbps/km² | Subscribers uptake per km² | Mbps/subscriber | Mb/month/subscriber | GB/month/subscriber | Equivlent tarrif (GB) | Tarrif cost per month | (Charge per subscriber x number of subscribers) x 12 months | Cost for cells per km² | Profit/Loss (Before Retail Expenditure & Tax) |
| 9% | 5 | 270 | 0.0185 | 24000.00 | 23.44 | 23.50 | £80.50 | £260,820.00 | £195,152.00 | £65,668 |
| 9% | 10 | 270 | 0.0370 | 48000.00 | 46.88 | Unlimited | £82.00 | £265,680.00 | £195,296.00 | £70,384 |
| 9% | 20 | 270 | 0.0741 | 96000.00 | 93.75 | Unlimited | £82.00 | £265,680.00 | £195,501.00 | £70,179 |
| 9% | 30 | 270 | 0.1111 | 144000.00 | 140.63 | Unlimited | £82.00 | £265,680.00 | £195,658.00 | £70,022 |
| 9% | 40 | 270 | 0.1481 | 192000.00 | 187.50 | Unlimited | £82.00 | £265,680.00 | £195,790.00 | £69,890 |
| 9% | 50 | 270 | 0.1852 | 240000.00 | 234.38 | Unlimited | £82.00 | £265,680.00 | £195,907.00 | £69,773 |
| 9% | 60 | 270 | 0.2222 | 288000.00 | 281.25 | Unlimited | £82.00 | £265,680.00 | £196,012.00 | £69,668 |
| 9% | 70 | 270 | 0.2593 | 336000.00 | 328.13 | Unlimited | £82.00 | £265,680.00 | £196,109.00 | £69,571 |
| 9% | 80 | 270 | 0.2963 | 384000.00 | 375.00 | Unlimited | £82.00 | £265,680.00 | £196,200.00 | £69,480 |
| 9% | 90 | 270 | 0.3333 | 432000.00 | 421.88 | Unlimited | £82.00 | £265,680.00 | £196,284.00 | £69,396 |
| 9% | 100 | 270 | 0.3704 | 480000.00 | 468.75 | Unlimited | £82.00 | £265,680.00 | £196,365.00 | £69,315 |
| 9% | 110 | 270 | 0.4074 | 528000.00 | 515.63 | Unlimited | £82.00 | £265,680.00 | £196,441.00 | £69,239 |
| 9% | 120 | 270 | 0.4444 | 576000.00 | 562.50 | Unlimited | £82.00 | £265,680.00 | £196,514.00 | £69,166 |
| 9% | 130 | 270 | 0.4815 | 624000.00 | 609.38 | Unlimited | £82.00 | £265,680.00 | £196,584.00 | £69,096 |
| 9% | 140 | 270 | 0.5185 | 672000.00 | 656.25 | Unlimited | £82.00 | £265,680.00 | £196,651.00 | £69,029 |
| 9% | 150 | 270 | 0.5556 | 720000.00 | 703.13 | Unlimited | £82.00 | £265,680.00 | £196,716.00 | £68,964 |
| 9% | 160 | 270 | 0.5926 | 768000.00 | 750.00 | Unlimited | £82.00 | £265,680.00 | £196,779.00 | £68,901 |
| 9% | 170 | 270 | 0.6296 | 816000.00 | 796.88 | Unlimited | £82.00 | £265,680.00 | £196,839.00 | £68,841 |
| 9% | 180 | 270 | 0.6667 | 864000.00 | 843.75 | Unlimited | £82.00 | £265,680.00 | £196,898.00 | £68,782 |
| 9% | 190 | 270 | 0.7037 | 912000.00 | 890.63 | Unlimited | £82.00 | £265,680.00 | £196,956.00 | £68,724 |

| 3G - 91% Uptake in Subscription Demand | | | | | | | | | | | |
|---|---|---|---|---|---|---|---|---|---|---|---|
| Uptake of % | Rate of Traffic demand Mbps/km² | Subscribers uptake per km² | Mbps/subscriber | Mb/month/subscriber | GB/month/subscriber | Equivlent tarrif (GB) | Tarrif cost per month (to the closes .5 tarriff) | (Charge per subscriber x number of subscribers) x 12 months | Cost for cells per km² | Profit/Loss (Before Retail Expenditure & Tax) | Total Profitability for network operator |
| 91% | 5 | 2730 | 0.0018 | 2373.63 | 2.32 | 2.50 | £23.00 | £753,480.00 | £294,773.00 | £458,707 | £524,375 |
| 91% | 10 | 2730 | 0.0037 | 4747.25 | 4.64 | 5.00 | £28.50 | £933,660.00 | £295,169.00 | £638,491 | £708,875 |
| 91% | 20 | 2730 | 0.0073 | 9494.51 | 9.27 | 9.50 | £35.00 | £1,146,600.00 | £295,730.00 | £850,870 | £921,049 |
| 91% | 30 | 2730 | 0.0110 | 14241.76 | 13.91 | 14.00 | £40.50 | £1,326,780.00 | £296,160.00 | £1,030,620 | £1,100,642 |
| 91% | 40 | 2730 | 0.0147 | 18989.01 | 18.54 | 18.50 | £45.50 | £1,490,580.00 | £296,523.00 | £1,194,057 | £1,263,947 |
| 91% | 50 | 2730 | 0.0183 | 23736.26 | 23.18 | 23.50 | £50.50 | £1,654,380.00 | £296,824.00 | £1,357,556 | £1,427,329 |
| 91% | 60 | 2730 | 0.0220 | 28483.52 | 27.82 | Unlimited | £52.00 | £1,703,520.00 | £297,131.00 | £1,406,389 | £1,476,057 |
| 91% | 70 | 2730 | 0.0256 | 33230.77 | 32.45 | Unlimited | £52.00 | £1,703,520.00 | £297,397.00 | £1,406,123 | £1,475,694 |
| 91% | 80 | 2730 | 0.0293 | 37978.02 | 37.09 | Unlimited | £52.00 | £1,703,520.00 | £297,644.00 | £1,405,876 | £1,475,356 |
| 91% | 90 | 2730 | 0.0330 | 42725.27 | 41.72 | Unlimited | £52.00 | £1,703,520.00 | £297,876.00 | £1,405,644 | £1,475,040 |
| 91% | 100 | 2730 | 0.0366 | 47472.53 | 46.36 | Unlimited | £52.00 | £1,703,520.00 | £298,096.00 | £1,405,424 | £1,474,739 |
| 91% | 110 | 2730 | 0.0403 | 52219.78 | 51.00 | Unlimited | £52.00 | £1,703,520.00 | £298,305.00 | £1,405,215 | £1,474,454 |
| 91% | 120 | 2730 | 0.0440 | 56967.03 | 55.63 | Unlimited | £52.00 | £1,703,520.00 | £298,504.00 | £1,405,016 | £1,474,182 |
| 91% | 130 | 2730 | 0.0476 | 61714.29 | 60.27 | Unlimited | £52.00 | £1,703,520.00 | £298,696.00 | £1,404,824 | £1,473,920 |
| 91% | 140 | 2730 | 0.0513 | 66461.54 | 64.90 | Unlimited | £52.00 | £1,703,520.00 | £298,880.00 | £1,404,640 | £1,473,669 |
| 91% | 150 | 2730 | 0.0549 | 71208.79 | 69.54 | Unlimited | £52.00 | £1,703,520.00 | £299,058.00 | £1,404,462 | £1,473,426 |
| 91% | 160 | 2730 | 0.0586 | 75956.04 | 74.18 | Unlimited | £52.00 | £1,703,520.00 | £299,229.00 | £1,404,291 | £1,473,192 |
| 91% | 170 | 2730 | 0.0623 | 80703.30 | 78.81 | Unlimited | £52.00 | £1,703,520.00 | £299,396.00 | £1,404,124 | £1,472,965 |
| 91% | 180 | 2730 | 0.0659 | 85450.55 | 83.45 | Unlimited | £52.00 | £1,703,520.00 | £299,558.00 | £1,403,962 | £1,472,744 |
| 91% | 190 | 2730 | 0.0696 | 90197.80 | 88.08 | Unlimited | £52.00 | £1,703,520.00 | £299,715.00 | £1,403,805 | £1,472,529 |

**Table 17 Profitability table showing profit calculation for 9% and 91% uptake in 4G and 3G subscriptions respectively**

**3G - 80% Uptake in Subscription Demand**

| Uptake of % | Rate of Traffic demand Mbps/km² | Subscribers uptake per km² | Mbps/subscriber | Mb/month/subscriber | GB/month/subscriber | Equivlent tariff (GB) | Tarrif cost per month (to the closes .5 tarriff) | (Charge per subscriber x number of subscribers) x 12 months | Cost for cells per km² | Profit/Loss (Before Retail Expenditure & Tax) | Total Profitability for network operator |
|---|---|---|---|---|---|---|---|---|---|---|---|
| 80% | 5 | 2400 | 0.0021 | 2700.00 | 2.64 | 3.00 | £24.50 | £705,600.00 | £294,773.00 | £410,827 | £690,875 |
| 80% | 10 | 2400 | 0.0042 | 5400.00 | 5.27 | 5.50 | £29.00 | £835,200.00 | £295,169.00 | £540,031 | £899,135 |
| 80% | 20 | 2400 | 0.0083 | 10800.00 | 10.55 | 10.50 | £36.00 | £1,036,800.00 | £295,730.00 | £741,070 | £1,135,969 |
| 80% | 30 | 2400 | 0.0125 | 16200.00 | 15.82 | 16.00 | £43.00 | £1,238,400.00 | £296,160.00 | £942,240 | £1,336,982 |
| 80% | 40 | 2400 | 0.0167 | 21600.00 | 21.09 | 21.00 | £48.00 | £1,382,400.00 | £296,523.00 | £1,085,877 | £1,480,487 |
| 80% | 50 | 2400 | 0.0208 | 27000.00 | 26.37 | Unlimited | £52.00 | £1,497,600.00 | £296,824.00 | £1,200,776 | £1,595,269 |
| 80% | 60 | 2400 | 0.0250 | 32400.00 | 31.64 | Unlimited | £52.00 | £1,497,600.00 | £297,131.00 | £1,200,469 | £1,594,857 |
| 80% | 70 | 2400 | 0.0292 | 37800.00 | 36.91 | Unlimited | £52.00 | £1,497,600.00 | £297,397.00 | £1,200,203 | £1,594,494 |
| 80% | 80 | 2400 | 0.0333 | 43200.00 | 42.19 | Unlimited | £52.00 | £1,497,600.00 | £297,644.00 | £1,199,956 | £1,594,156 |
| 80% | 90 | 2400 | 0.0375 | 48600.00 | 47.46 | Unlimited | £52.00 | £1,497,600.00 | £297,876.00 | £1,199,724 | £1,593,840 |
| 80% | 100 | 2400 | 0.0417 | 54000.00 | 52.73 | Unlimited | £52.00 | £1,497,600.00 | £298,096.00 | £1,199,504 | £1,593,539 |
| 80% | 110 | 2400 | 0.0458 | 59400.00 | 58.01 | Unlimited | £52.00 | £1,497,600.00 | £298,305.00 | £1,199,295 | £1,593,254 |
| 80% | 120 | 2400 | 0.0500 | 64800.00 | 63.28 | Unlimited | £52.00 | £1,497,600.00 | £298,504.00 | £1,199,096 | £1,592,982 |
| 80% | 130 | 2400 | 0.0542 | 70200.00 | 68.55 | Unlimited | £52.00 | £1,497,600.00 | £298,696.00 | £1,198,904 | £1,592,720 |
| 80% | 140 | 2400 | 0.0583 | 75600.00 | 73.83 | Unlimited | £52.00 | £1,497,600.00 | £298,880.00 | £1,198,720 | £1,592,469 |
| 80% | 150 | 2400 | 0.0625 | 81000.00 | 79.10 | Unlimited | £52.00 | £1,497,600.00 | £299,058.00 | £1,198,542 | £1,592,226 |
| 80% | 160 | 2400 | 0.0667 | 86400.00 | 84.38 | Unlimited | £52.00 | £1,497,600.00 | £299,229.00 | £1,198,371 | £1,591,992 |
| 80% | 170 | 2400 | 0.0708 | 91800.00 | 89.65 | Unlimited | £52.00 | £1,497,600.00 | £299,396.00 | £1,198,204 | £1,591,765 |
| 80% | 180 | 2400 | 0.0750 | 97200.00 | 94.92 | Unlimited | £52.00 | £1,497,600.00 | £299,558.00 | £1,198,042 | £1,591,544 |
| 80% | 190 | 2400 | 0.0792 | 102600.00 | 100.20 | Unlimited | £52.00 | £1,497,600.00 | £299,715.00 | £1,197,885 | £1,591,329 |

**4G - 20% Uptake in Subscription Demand**

| Uptake of % | Rate of Traffic demand Mbps/km² | Subscribers uptake per km² | Mbps/subscriber | Mb/month/subscriber | GB/month/subscriber | Equivlent tariff (GB) | Tarrif cost per month | (Charge per subscriber x number of subscribers) x 12 months | Cost for cells per km² | Profit/Loss (Before Retail Expenditure & Tax) |
|---|---|---|---|---|---|---|---|---|---|---|
| 20% | 5 | 600 | 0.0083 | 10800.00 | 10.55 | 10.50 | £66.00 | £475,200.00 | £195,152.00 | £280,048 |
| 20% | 10 | 600 | 0.0167 | 21600.00 | 21.09 | 21.00 | £77.00 | £554,400.00 | £195,296.00 | £359,104 |
| 20% | 20 | 600 | 0.0333 | 43200.00 | 42.19 | Unlimited | £82.00 | £590,400.00 | £195,501.00 | £394,899 |
| 20% | 30 | 600 | 0.0500 | 64800.00 | 63.28 | Unlimited | £82.00 | £590,400.00 | £195,658.00 | £394,742 |
| 20% | 40 | 600 | 0.0667 | 86400.00 | 84.38 | Unlimited | £82.00 | £590,400.00 | £195,790.00 | £394,610 |
| 20% | 50 | 600 | 0.0833 | 108000.00 | 105.47 | Unlimited | £82.00 | £590,400.00 | £195,907.00 | £394,493 |
| 20% | 60 | 600 | 0.1000 | 129600.00 | 126.56 | Unlimited | £82.00 | £590,400.00 | £196,012.00 | £394,388 |
| 20% | 70 | 600 | 0.1167 | 151200.00 | 147.66 | Unlimited | £82.00 | £590,400.00 | £196,109.00 | £394,291 |
| 20% | 80 | 600 | 0.1333 | 172800.00 | 168.75 | Unlimited | £82.00 | £590,400.00 | £196,200.00 | £394,200 |
| 20% | 90 | 600 | 0.1500 | 194400.00 | 189.84 | Unlimited | £82.00 | £590,400.00 | £196,284.00 | £394,116 |
| 20% | 100 | 600 | 0.1667 | 216000.00 | 210.94 | Unlimited | £82.00 | £590,400.00 | £196,365.00 | £394,035 |
| 20% | 110 | 600 | 0.1833 | 237600.00 | 232.03 | Unlimited | £82.00 | £590,400.00 | £196,441.00 | £393,959 |
| 20% | 120 | 600 | 0.2000 | 259200.00 | 253.13 | Unlimited | £82.00 | £590,400.00 | £196,514.00 | £393,886 |
| 20% | 130 | 600 | 0.2167 | 280800.00 | 274.22 | Unlimited | £82.00 | £590,400.00 | £196,584.00 | £393,816 |
| 20% | 140 | 600 | 0.2333 | 302400.00 | 295.31 | Unlimited | £82.00 | £590,400.00 | £196,651.00 | £393,749 |
| 20% | 150 | 600 | 0.2500 | 324000.00 | 316.41 | Unlimited | £82.00 | £590,400.00 | £196,716.00 | £393,684 |
| 20% | 160 | 600 | 0.2667 | 345600.00 | 337.50 | Unlimited | £82.00 | £590,400.00 | £196,779.00 | £393,621 |
| 20% | 170 | 600 | 0.2833 | 367200.00 | 358.59 | Unlimited | £82.00 | £590,400.00 | £196,839.00 | £393,561 |
| 20% | 180 | 600 | 0.3000 | 388800.00 | 379.69 | Unlimited | £82.00 | £590,400.00 | £196,898.00 | £393,502 |
| 20% | 190 | 600 | 0.3167 | 410400.00 | 400.78 | Unlimited | £82.00 | £590,400.00 | £196,956.00 | £393,444 |

**Table 18 Profitability table showing profit calculation for 9% and 91% uptake in 4G and 3G subscriptions respectively**

| 4G - 30% Uptake in Subscription Demand | | | | | | | | | | |
|---|---|---|---|---|---|---|---|---|---|---|
| Uptake of % | Rate of Traffic demand Mbps/km² | Subscribers uptake per km² | Mbps/subscriber | Mb/month/subscriber | GB/month/subscriber | Equivlent tarrif (GB) | Tarrif cost per month | (Charge per subscriber x number of subscribers) x 12 months | Cost for cells per km² | Profit/Loss (Before Retail Expenditure & Tax) |
| 30% | 5 | 900 | 0.0056 | 7200.00 | 7.03 | 7.00 | £61.50 | £664,200.00 | £195,152.00 | £469,048 |
| 30% | 10 | 900 | 0.0111 | 14400.00 | 14.06 | 14.00 | £71.00 | £766,800.00 | £195,296.00 | £571,504 |
| 30% | 20 | 900 | 0.0222 | 28800.00 | 28.13 | Unlimited | £82.00 | £885,600.00 | £195,501.00 | £690,099 |
| 30% | 30 | 900 | 0.0333 | 43200.00 | 42.19 | Unlimited | £82.00 | £885,600.00 | £195,658.00 | £689,942 |
| 30% | 40 | 900 | 0.0444 | 57600.00 | 56.25 | Unlimited | £82.00 | £885,600.00 | £195,790.00 | £689,810 |
| 30% | 50 | 900 | 0.0556 | 72000.00 | 70.31 | Unlimited | £82.00 | £885,600.00 | £195,907.00 | £689,693 |
| 30% | 60 | 900 | 0.0667 | 86400.00 | 84.38 | Unlimited | £82.00 | £885,600.00 | £196,012.00 | £689,588 |
| 30% | 70 | 900 | 0.0778 | 100800.00 | 98.44 | Unlimited | £82.00 | £885,600.00 | £196,109.00 | £689,491 |
| 30% | 80 | 900 | 0.0889 | 115200.00 | 112.50 | Unlimited | £82.00 | £885,600.00 | £196,200.00 | £689,400 |
| 30% | 90 | 900 | 0.1000 | 129600.00 | 126.56 | Unlimited | £82.00 | £885,600.00 | £196,284.00 | £689,316 |
| 30% | 100 | 900 | 0.1111 | 144000.00 | 140.63 | Unlimited | £82.00 | £885,600.00 | £196,365.00 | £689,235 |
| 30% | 110 | 900 | 0.1222 | 158400.00 | 154.69 | Unlimited | £82.00 | £885,600.00 | £196,441.00 | £689,159 |
| 30% | 120 | 900 | 0.1333 | 172800.00 | 168.75 | Unlimited | £82.00 | £885,600.00 | £196,514.00 | £689,086 |
| 30% | 130 | 900 | 0.1444 | 187200.00 | 182.81 | Unlimited | £82.00 | £885,600.00 | £196,584.00 | £689,016 |
| 30% | 140 | 900 | 0.1556 | 201600.00 | 196.88 | Unlimited | £82.00 | £885,600.00 | £196,651.00 | £688,949 |
| 30% | 150 | 900 | 0.1667 | 216000.00 | 210.94 | Unlimited | £82.00 | £885,600.00 | £196,716.00 | £688,884 |
| 30% | 160 | 900 | 0.1778 | 230400.00 | 225.00 | Unlimited | £82.00 | £885,600.00 | £196,779.00 | £688,821 |
| 30% | 170 | 900 | 0.1889 | 244800.00 | 239.06 | Unlimited | £82.00 | £885,600.00 | £196,839.00 | £688,761 |
| 30% | 180 | 900 | 0.2000 | 259200.00 | 253.13 | Unlimited | £82.00 | £885,600.00 | £196,898.00 | £688,702 |
| 30% | 190 | 900 | 0.2111 | 273600.00 | 267.19 | Unlimited | £82.00 | £885,600.00 | £196,956.00 | £688,644 |

| 3G - 70% Uptake in Subscription Demand | | | | | | | | | | | |
|---|---|---|---|---|---|---|---|---|---|---|---|
| Uptake of % | Rate of Traffic demand Mbps/km² | Subscribers uptake per km² | Mbps/subscriber | Mb/month/subscriber | GB/month/subscriber | Equivlent tarrif (GB) | Tarrif cost per month (to the closes .5 tarrif) | (Charge per subscriber x number of subscribers) x 12 months | Cost for cells per km² | Profit/Loss (Before Retail Expenditure & Tax) | Total Profitability for network operator |
| 70% | 5 | 2100 | 0.0024 | 3085.71 | 3.01 | 3.00 | £24.50 | £617,400.00 | £294,773.00 | £322,627 | £791,675 |
| 70% | 10 | 2100 | 0.0048 | 6171.43 | 6.03 | 6.00 | £30.00 | £756,000.00 | £295,169.00 | £460,831 | £1,032,335 |
| 70% | 20 | 2100 | 0.0095 | 12342.86 | 12.05 | 12.00 | £38.00 | £957,600.00 | £295,730.00 | £661,870 | £1,351,969 |
| 70% | 30 | 2100 | 0.0143 | 18514.29 | 18.08 | 18.00 | £45.00 | £1,134,000.00 | £296,160.00 | £837,840 | £1,527,782 |
| 70% | 40 | 2100 | 0.0190 | 24685.71 | 24.11 | 24.00 | £51.00 | £1,285,200.00 | £296,523.00 | £988,677 | £1,678,487 |
| 70% | 50 | 2100 | 0.0238 | 30857.14 | 30.13 | Unlimited | £52.00 | £1,310,400.00 | £296,824.00 | £1,013,576 | £1,703,269 |
| 70% | 60 | 2100 | 0.0286 | 37028.57 | 36.16 | Unlimited | £52.00 | £1,310,400.00 | £297,131.00 | £1,013,269 | £1,702,857 |
| 70% | 70 | 2100 | 0.0333 | 43200.00 | 42.19 | Unlimited | £52.00 | £1,310,400.00 | £297,397.00 | £1,013,003 | £1,702,494 |
| 70% | 80 | 2100 | 0.0381 | 49371.43 | 48.21 | Unlimited | £52.00 | £1,310,400.00 | £297,644.00 | £1,012,756 | £1,702,156 |
| 70% | 90 | 2100 | 0.0429 | 55542.86 | 54.24 | Unlimited | £52.00 | £1,310,400.00 | £297,876.00 | £1,012,524 | £1,701,840 |
| 70% | 100 | 2100 | 0.0476 | 61714.29 | 60.27 | Unlimited | £52.00 | £1,310,400.00 | £298,096.00 | £1,012,304 | £1,701,539 |
| 70% | 110 | 2100 | 0.0524 | 67885.71 | 66.29 | Unlimited | £52.00 | £1,310,400.00 | £298,305.00 | £1,012,095 | £1,701,254 |
| 70% | 120 | 2100 | 0.0571 | 74057.14 | 72.32 | Unlimited | £52.00 | £1,310,400.00 | £298,504.00 | £1,011,896 | £1,700,982 |
| 70% | 130 | 2100 | 0.0619 | 80228.57 | 78.35 | Unlimited | £52.00 | £1,310,400.00 | £298,696.00 | £1,011,704 | £1,700,720 |
| 70% | 140 | 2100 | 0.0667 | 86400.00 | 84.38 | Unlimited | £52.00 | £1,310,400.00 | £298,880.00 | £1,011,520 | £1,700,469 |
| 70% | 150 | 2100 | 0.0714 | 92571.43 | 90.40 | Unlimited | £52.00 | £1,310,400.00 | £299,058.00 | £1,011,342 | £1,700,226 |
| 70% | 160 | 2100 | 0.0762 | 98742.86 | 96.43 | Unlimited | £52.00 | £1,310,400.00 | £299,229.00 | £1,011,171 | £1,699,992 |
| 70% | 170 | 2100 | 0.0810 | 104914.29 | 102.46 | Unlimited | £52.00 | £1,310,400.00 | £299,396.00 | £1,011,004 | £1,699,765 |
| 70% | 180 | 2100 | 0.0857 | 111085.71 | 108.48 | Unlimited | £52.00 | £1,310,400.00 | £299,558.00 | £1,010,842 | £1,699,544 |
| 70% | 190 | 2100 | 0.0905 | 117257.14 | 114.51 | Unlimited | £52.00 | £1,310,400.00 | £299,715.00 | £1,010,685 | £1,699,329 |

**Table 19 Profitability table showing profit calculation for 30% and 70% uptake in 4G and 3G subscriptions respectively**

| 4G - 40% Uptake in Subscription Demand | | | | | | | | | |
|---|---|---|---|---|---|---|---|---|---|
| Uptake of % | Rate of Traffic demand Mbps/km² | Subscribers uptake per km² | Mbps/subscriber | Mb/month/subscriber | GB/month/subscriber | Equivlent tarrif (GB) | Tarrif cost per month | (Charge per subscriber x number of subscribers) x 12 months | Cost for cells per km² | Profit/Loss (Before Retail Expenditure & Tax) |
| 40% | 5 | 1200 | 0.0042 | 5400.00 | 5.27 | 5.50 | £57.50 | £828,000.00 | £195,152.00 | £632,848 |
| 40% | 10 | 1200 | 0.0083 | 10800.00 | 10.55 | 10.50 | £67.50 | £972,000.00 | £195,296.00 | £776,704 |
| 40% | 20 | 1200 | 0.0167 | 21600.00 | 21.09 | 21.00 | £78.00 | £1,123,200.00 | £195,501.00 | £927,699 |
| 40% | 30 | 1200 | 0.0250 | 32400.00 | 31.64 | Unlimited | £82.00 | £1,180,800.00 | £195,658.00 | £985,142 |
| 40% | 40 | 1200 | 0.0333 | 43200.00 | 42.19 | Unlimited | £82.00 | £1,180,800.00 | £195,790.00 | £985,010 |
| 40% | 50 | 1200 | 0.0417 | 54000.00 | 52.73 | Unlimited | £82.00 | £1,180,800.00 | £195,907.00 | £984,893 |
| 40% | 60 | 1200 | 0.0500 | 64800.00 | 63.28 | Unlimited | £82.00 | £1,180,800.00 | £196,012.00 | £984,788 |
| 40% | 70 | 1200 | 0.0583 | 75600.00 | 73.83 | Unlimited | £82.00 | £1,180,800.00 | £196,109.00 | £984,691 |
| 40% | 80 | 1200 | 0.0667 | 86400.00 | 84.38 | Unlimited | £82.00 | £1,180,800.00 | £196,200.00 | £984,600 |
| 40% | 90 | 1200 | 0.0750 | 97200.00 | 94.92 | Unlimited | £82.00 | £1,180,800.00 | £196,284.00 | £984,516 |
| 40% | 100 | 1200 | 0.0833 | 108000.00 | 105.47 | Unlimited | £82.00 | £1,180,800.00 | £196,365.00 | £984,435 |
| 40% | 110 | 1200 | 0.0917 | 118800.00 | 116.02 | Unlimited | £82.00 | £1,180,800.00 | £196,441.00 | £984,359 |
| 40% | 120 | 1200 | 0.1000 | 129600.00 | 126.56 | Unlimited | £82.00 | £1,180,800.00 | £196,514.00 | £984,286 |
| 40% | 130 | 1200 | 0.1083 | 140400.00 | 137.11 | Unlimited | £82.00 | £1,180,800.00 | £196,584.00 | £984,216 |
| 40% | 140 | 1200 | 0.1167 | 151200.00 | 147.66 | Unlimited | £82.00 | £1,180,800.00 | £196,651.00 | £984,149 |
| 40% | 150 | 1200 | 0.1250 | 162000.00 | 158.20 | Unlimited | £82.00 | £1,180,800.00 | £196,716.00 | £984,084 |
| 40% | 160 | 1200 | 0.1333 | 172800.00 | 168.75 | Unlimited | £82.00 | £1,180,800.00 | £196,779.00 | £984,021 |
| 40% | 170 | 1200 | 0.1417 | 183600.00 | 179.30 | Unlimited | £82.00 | £1,180,800.00 | £196,839.00 | £983,961 |
| 40% | 180 | 1200 | 0.1500 | 194400.00 | 189.84 | Unlimited | £82.00 | £1,180,800.00 | £196,898.00 | £983,902 |
| 40% | 190 | 1200 | 0.1583 | 205200.00 | 200.39 | Unlimited | £82.00 | £1,180,800.00 | £196,956.00 | £983,844 |

| 3G - 60% Uptake in Subscription Demand | | | | | | | | | | |
|---|---|---|---|---|---|---|---|---|---|---|
| Uptake of % | Rate of Traffic demand Mbps/km² | Subscribers uptake per km² | Mbps/subscriber | Mb/month/subscriber | GB/month/subscriber | Equivlent tarrif (GB) | Tarrif cost per month (to the closes .5 tarriff) | (Charge per subscriber x number of subscribers) x 12 months | Cost for cells per km² | Profit/Loss (Before Retail Expenditure & Tax) | Total Profitability for network operator |
| 60% | 5 | 1800 | 0.0028 | 3600.00 | 3.52 | 3.50 | £25.50 | £550,800.00 | £294,773.00 | £256,027 | £888,875 |
| 60% | 10 | 1800 | 0.0056 | 7200.00 | 7.03 | 7.00 | £32.00 | £691,200.00 | £295,169.00 | £396,031 | £1,172,735 |
| 60% | 20 | 1800 | 0.0111 | 14400.00 | 14.06 | 14.00 | £40.50 | £874,800.00 | £295,730.00 | £579,070 | £1,506,769 |
| 60% | 30 | 1800 | 0.0167 | 21600.00 | 21.09 | Unlimited | £48.00 | £1,036,800.00 | £296,160.00 | £740,640 | £1,725,782 |
| 60% | 40 | 1800 | 0.0222 | 28800.00 | 28.13 | Unlimited | £52.00 | £1,123,200.00 | £296,523.00 | £826,677 | £1,811,687 |
| 60% | 50 | 1800 | 0.0278 | 36000.00 | 35.16 | Unlimited | £52.00 | £1,123,200.00 | £296,824.00 | £826,376 | £1,811,269 |
| 60% | 60 | 1800 | 0.0333 | 43200.00 | 42.19 | Unlimited | £52.00 | £1,123,200.00 | £297,131.00 | £826,069 | £1,810,857 |
| 60% | 70 | 1800 | 0.0389 | 50400.00 | 49.22 | Unlimited | £52.00 | £1,123,200.00 | £297,397.00 | £825,803 | £1,810,494 |
| 60% | 80 | 1800 | 0.0444 | 57600.00 | 56.25 | Unlimited | £52.00 | £1,123,200.00 | £297,644.00 | £825,556 | £1,810,156 |
| 60% | 90 | 1800 | 0.0500 | 64800.00 | 63.28 | Unlimited | £52.00 | £1,123,200.00 | £297,876.00 | £825,324 | £1,809,840 |
| 60% | 100 | 1800 | 0.0556 | 72000.00 | 70.31 | Unlimited | £52.00 | £1,123,200.00 | £298,096.00 | £825,104 | £1,809,539 |
| 60% | 110 | 1800 | 0.0611 | 79200.00 | 77.34 | Unlimited | £52.00 | £1,123,200.00 | £298,305.00 | £824,895 | £1,809,254 |
| 60% | 120 | 1800 | 0.0667 | 86400.00 | 84.38 | Unlimited | £52.00 | £1,123,200.00 | £298,504.00 | £824,696 | £1,808,982 |
| 60% | 130 | 1800 | 0.0722 | 93600.00 | 91.41 | Unlimited | £52.00 | £1,123,200.00 | £298,696.00 | £824,504 | £1,808,720 |
| 60% | 140 | 1800 | 0.0778 | 100800.00 | 98.44 | Unlimited | £52.00 | £1,123,200.00 | £298,880.00 | £824,320 | £1,808,469 |
| 60% | 150 | 1800 | 0.0833 | 108000.00 | 105.47 | Unlimited | £52.00 | £1,123,200.00 | £299,058.00 | £824,142 | £1,808,226 |
| 60% | 160 | 1800 | 0.0889 | 115200.00 | 112.50 | Unlimited | £52.00 | £1,123,200.00 | £299,229.00 | £823,971 | £1,807,992 |
| 60% | 170 | 1800 | 0.0944 | 122400.00 | 119.53 | Unlimited | £52.00 | £1,123,200.00 | £299,396.00 | £823,804 | £1,807,765 |
| 60% | 180 | 1800 | 0.1000 | 129600.00 | 126.56 | Unlimited | £52.00 | £1,123,200.00 | £299,558.00 | £823,642 | £1,807,544 |
| 60% | 190 | 1800 | 0.1056 | 136800.00 | 133.59 | Unlimited | £52.00 | £1,123,200.00 | £299,715.00 | £823,485 | £1,807,329 |

Table 20 Profitability table showing profit calculation for 40% and 70% uptake in 4G and 3G subscriptions respectively

| 4G - 90% Uptake in Subscription Demand | | | | | | | | | | |
|---|---|---|---|---|---|---|---|---|---|---|
| Uptake of % | Rate of Traffic demand Mbps/km² | Subscribers uptake per km² | Mbps/subscriber | Mb/month/subscriber | GB/month/subscriber | Equivlent tarrif (GB) | Tarrif cost per month | (Charge per subscriber x number of subscribers) x 12 months | Cost for cells per km² | Profit/Loss (Before Retail Expenditure & Tax) |
| 90% | 5 | 2700 | 0.0019 | 2400.00 | 2.34 | 2.50 | £47.00 | £1,522,800.00 | £195,152.00 | £1,327,648 |
| 90% | 10 | 2700 | 0.0037 | 4800.00 | 4.69 | 5.00 | £56.00 | £1,814,400.00 | £195,296.00 | £1,619,104 |
| 90% | 20 | 2700 | 0.0074 | 9600.00 | 9.38 | 9.50 | £64.50 | £2,089,800.00 | £195,501.00 | £1,894,299 |
| 90% | 30 | 2700 | 0.0111 | 14400.00 | 14.06 | 14.00 | £70.50 | £2,284,200.00 | £195,658.00 | £2,088,542 |
| 90% | 40 | 2700 | 0.0148 | 19200.00 | 18.75 | 19.00 | £75.00 | £2,430,000.00 | £195,790.00 | £2,234,210 |
| 90% | 50 | 2700 | 0.0185 | 24000.00 | 23.44 | 23.50 | £80.50 | £2,608,200.00 | £195,907.00 | £2,412,293 |
| 90% | 60 | 2700 | 0.0222 | 28800.00 | 28.13 | Unlimited | £82.00 | £2,656,800.00 | £196,012.00 | £2,460,788 |
| 90% | 70 | 2700 | 0.0259 | 33600.00 | 32.81 | Unlimited | £82.00 | £2,656,800.00 | £196,109.00 | £2,460,691 |
| 90% | 80 | 2700 | 0.0296 | 38400.00 | 37.50 | Unlimited | £82.00 | £2,656,800.00 | £196,200.00 | £2,460,600 |
| 90% | 90 | 2700 | 0.0333 | 43200.00 | 42.19 | Unlimited | £82.00 | £2,656,800.00 | £196,284.00 | £2,460,516 |
| 90% | 100 | 2700 | 0.0370 | 48000.00 | 46.88 | Unlimited | £82.00 | £2,656,800.00 | £196,365.00 | £2,460,435 |
| 90% | 110 | 2700 | 0.0407 | 52800.00 | 51.56 | Unlimited | £82.00 | £2,656,800.00 | £196,441.00 | £2,460,359 |
| 90% | 120 | 2700 | 0.0444 | 57600.00 | 56.25 | Unlimited | £82.00 | £2,656,800.00 | £196,514.00 | £2,460,286 |
| 90% | 130 | 2700 | 0.0481 | 62400.00 | 60.94 | Unlimited | £82.00 | £2,656,800.00 | £196,584.00 | £2,460,216 |
| 90% | 140 | 2700 | 0.0519 | 67200.00 | 65.63 | Unlimited | £82.00 | £2,656,800.00 | £196,651.00 | £2,460,149 |
| 90% | 150 | 2700 | 0.0556 | 72000.00 | 70.31 | Unlimited | £82.00 | £2,656,800.00 | £196,716.00 | £2,460,084 |
| 90% | 160 | 2700 | 0.0593 | 76800.00 | 75.00 | Unlimited | £82.00 | £2,656,800.00 | £196,779.00 | £2,460,021 |
| 90% | 170 | 2700 | 0.0630 | 81600.00 | 79.69 | Unlimited | £82.00 | £2,656,800.00 | £196,839.00 | £2,459,961 |
| 90% | 180 | 2700 | 0.0667 | 86400.00 | 84.38 | Unlimited | £82.00 | £2,656,800.00 | £196,898.00 | £2,459,902 |
| 90% | 190 | 2700 | 0.0704 | 91200.00 | 89.06 | Unlimited | £82.00 | £2,656,800.00 | £196,956.00 | £2,459,844 |

| 3G - 10% Uptake in Subscription Demand | | | | | | | | | | | |
|---|---|---|---|---|---|---|---|---|---|---|---|
| Uptake of % | Rate of Traffic demand Mbps/km² | Subscribers uptake per km² | Mbps/subscriber | Mb/month/subscriber | GB/month/subscriber | Equivlent tarrif (GB) | Tarrif cost per month (to the closes .5 tarriff) | (Charge per subscriber x number of subscribers) x 12 months | Cost for cells per km² | Profit/Loss (Before Retail Expenditure & Tax) | Total Profitability for network operator |
| 10% | 5 | 300 | 0.0167 | 21600.00 | 21.09 | 21.00 | £48.00 | £172,800.00 | £294,773.00 | -£121,973 | £1,205,675 |
| 10% | 10 | 300 | 0.0333 | 43200.00 | 42.19 | Unlmited | £52.00 | £187,200.00 | £295,169.00 | -£107,969 | £1,511,135 |
| 10% | 20 | 300 | 0.0667 | 86400.00 | 84.38 | Unlmited | £52.00 | £187,200.00 | £295,730.00 | -£108,530 | £1,785,769 |
| 10% | 30 | 300 | 0.1000 | 129600.00 | 126.56 | Unlmited | £52.00 | £187,200.00 | £296,160.00 | -£108,960 | £1,979,582 |
| 10% | 40 | 300 | 0.1333 | 172800.00 | 168.75 | Unlmited | £52.00 | £187,200.00 | £296,523.00 | -£109,323 | £2,124,887 |
| 10% | 50 | 300 | 0.1667 | 216000.00 | 210.94 | Unlmited | £52.00 | £187,200.00 | £296,824.00 | -£109,624 | £2,302,669 |
| 10% | 60 | 300 | 0.2000 | 259200.00 | 253.13 | Unlmited | £52.00 | £187,200.00 | £297,131.00 | -£109,931 | £2,350,857 |
| 10% | 70 | 300 | 0.2333 | 302400.00 | 295.31 | Unlmited | £52.00 | £187,200.00 | £297,397.00 | -£110,197 | £2,350,494 |
| 10% | 80 | 300 | 0.2667 | 345600.00 | 337.50 | Unlmited | £52.00 | £187,200.00 | £297,644.00 | -£110,444 | £2,350,156 |
| 10% | 90 | 300 | 0.3000 | 388800.00 | 379.69 | Unlmited | £52.00 | £187,200.00 | £297,876.00 | -£110,676 | £2,349,840 |
| 10% | 100 | 300 | 0.3333 | 432000.00 | 421.88 | Unlmited | £52.00 | £187,200.00 | £298,096.00 | -£110,896 | £2,349,539 |
| 10% | 110 | 300 | 0.3667 | 475200.00 | 464.06 | Unlmited | £52.00 | £187,200.00 | £298,305.00 | -£111,105 | £2,349,254 |
| 10% | 120 | 300 | 0.4000 | 518400.00 | 506.25 | Unlmited | £52.00 | £187,200.00 | £298,504.00 | -£111,304 | £2,348,982 |
| 10% | 130 | 300 | 0.4333 | 561600.00 | 548.44 | Unlmited | £52.00 | £187,200.00 | £298,696.00 | -£111,496 | £2,348,720 |
| 10% | 140 | 300 | 0.4667 | 604800.00 | 590.63 | Unlmited | £52.00 | £187,200.00 | £298,880.00 | -£111,680 | £2,348,469 |
| 10% | 150 | 300 | 0.5000 | 648000.00 | 632.81 | Unlmited | £52.00 | £187,200.00 | £299,058.00 | -£111,858 | £2,348,226 |
| 10% | 160 | 300 | 0.5333 | 691200.00 | 675.00 | Unlmited | £52.00 | £187,200.00 | £299,229.00 | -£112,029 | £2,347,992 |
| 10% | 170 | 300 | 0.5667 | 734400.00 | 717.19 | Unlmited | £52.00 | £187,200.00 | £299,396.00 | -£112,196 | £2,347,765 |
| 10% | 180 | 300 | 0.6000 | 777600.00 | 759.38 | Unlmited | £52.00 | £187,200.00 | £299,558.00 | -£112,358 | £2,347,544 |
| 10% | 190 | 300 | 0.6333 | 820800.00 | 801.56 | Unlmited | £52.00 | £187,200.00 | £299,715.00 | -£112,515 | £2,347,329 |

**Table 21 Profitability table showing profit calculation for 90% and 10% uptake in 4G and 3G subscriptions respectively**

# 10.0 Glossary of Terms

**3G** - The third generation cellular networks were developed with the aim of offering high speed data and multimedia connectivity to subscribers. The International Telecommunication Union (ITU) under the initiative IMT-2000 has defined 3G systems as being capable of supporting high speed data ranges of 144 kbps to greater than 2 Mbps. A few technologies are able to fulfil the International Mobile Telecommunications (IMT) standards, such as CDMA, UMTS and some variation of GSM such as EDGE.

**4G -** 4G standards, named the International Mobile Telecommunications Advanced (IMT-Advanced) specification, setting peak speed requirements for 4G service at 100 megabits per second (Mbit/s) for high mobility communication (such as from trains and cars) and 1 gigabit per second (Gbit/s) for low mobility communication (such as pedestrians and stationary users).

**Bandwidth -** In wireless communications, the word 'bandwidth' is used to refer to analogue signal bandwidth measured in hertz. The connection is that according to Hartley's law, the digital data rate limit (or channel capacity) of a physical communication link is proportional to its bandwidth in hertz.

**Capacity-** In electrical engineering, computer science and information theory, **channel capacity** is the tightest upper bound on the amount of information that can be reliably transmitted over a communications channel. By the noisy-channel coding theorem, the channel capacity of a given channel is the limiting information rate (in units of information per unit time) that can be achieved with arbitrarily small error probability.

**Cellular network** - A **cellular network** or **mobile network** is a radio network distributed over land areas called cells, each served by at least one fixed-location transceiver, known as a cell site or base station. In a cellular network, each cell uses a different set of frequencies (frequency reuse) from neighbouring cells, to avoid interference and provide guaranteed bandwidth within each cell.

**Fading** – Fading or more specifically small scale fading is often described by a mathematical model that describes the distortion that a carrier-modulated communication signal experiences during its propagation through an operating environment. Thus, fading describes the rapid fluctuations of the amplitudes, phases, or multi-path delays of a radio signal over a short period of time or distance. Fading is caused when two or more versions of a transmitted signal arrive at the receiver with small amounts of time in between them. Therefore these signals, or multi-path waves combine at the receiver antenna, resulting in a signal that varies widely in phase as well as in amplitude, which depends on the distribution of the intensity, relative propagation time of the waves, and the bandwidth of the transmitted signal.

**HSPA –** High-speed packet access (HSPA) refers to a set of technologies that came about as a result of enhancements to wideband code division multiple access (WCDMA) systems. HSPA is composed of the High-Speed Downlink Packet Access (HSDPA) protocol and the High-Speed Uplink Packet Access (HSUPA) protocol. It features peak data rates of up to 14 Mbps downlink and up to 5.7 Mbps uplink. HSPA+, or Evolved High-Speed Packet Access, is a technical standard for wireless, broadband telecommunication. HSPA+ enhances the widely used WCDMA (UMTS) based 3G networks with higher speeds for the end user that are comparable to the newer LTE networks. HSPA+ was first defined in the technical standard 3GPP release 7 and expanded further in later releases.

**Latency** is a measure of time delay experienced in a system, the precise definition of which depends on the system and the time being measured. In communications, the lower limit of latency is determined by the medium being used for communications. In reliable two-way communication systems, latency limits the maximum rate that information can be transmitted, as there is often a limit on the amount of information that is "in-flight" at any one moment. In the field of human-machine interaction, perceptible latency has a strong effect on user satisfaction and usability.

**LTE –** is the acronym of **long-term evolution**, marketed as **4G LTE**, is a standard for wireless communication of high-speed data for mobile phones and data terminals. It is based on the GSM/EDGE and UMTS/HSPA network technologies, increasing the capacity and speed using a different radio interface together with core network improvements.

**Network capacity** is the maximum capacity of a link or network path to convey data from one location in the network to another.

**Shadowing** - Shadowing has a medium scale effect on the signal. The shadowing effect is observed when field strength variations occur, and this happens if the antenna is displaced over distances larger than a few tens or hundreds of meters or the signal passes through an obstruction.